\documentclass[preprint2]{aastex}

\usepackage{amsmath}
\usepackage{amsfonts}
\usepackage{color}

\shorttitle{Mapping Spectral Line Survey toward W51}
\shortauthors{Watanabe et al.}

\begin{document}


\title{Molecular-Cloud-Scale Chemical Composition I: \\Mapping Spectral Line Survey toward W51 in the 3~mm Band}


\author{Yoshimasa~Watanabe\altaffilmark{1,2,3}, Yuri~Nishimura\altaffilmark{1}, Nanase~Harada\altaffilmark{4}, Nami~Sakai\altaffilmark{5}, Takashi~Shimonishi\altaffilmark{6,7}, Yuri~Aikawa\altaffilmark{8}, Akiko~Kawamura\altaffilmark{9}}
%\affil{Department of Physics, The University of Tokyo}

\and

\author{Satoshi Yamamoto\altaffilmark{1,10}}


\altaffiltext{1}{Department of Physics, The University of Tokyo, 7-3-1 Hongo, Bunkyo-ku, Tokyo, 113-0033, Japan}
\altaffiltext{2}{Division of Physics, Faculty of Pure and Applied Sciences, University of Tsukuba,  Tsukuba, Ibaraki 305-8571, Japan}
\altaffiltext{3}{Center for Integrated Research in Fundamental Science and Engineering (CiRfSE), Faculty of Pure and Applied Sciences, University of Tsukuba, Tsukuba, Ibaraki 305-8571, Japan   }
\altaffiltext{4}{Academia Sinica Institute of Astronomy and Astrophysics, No.1, Sec. 4, Roosevelt Rd, 10617, Taipei, R.O.C., Taiwan}
\altaffiltext{5}{RIKEN, 2-1, Hirosawa, Wako, Saitama 351-0198, Japan}
\altaffiltext{6}{Frontier Research Institute for Interdisciplinary Sciences, Tohoku University, Aramakiazaaoba 6-3, Aoba-ku, Sendai, Miyagi, 980-8578, Japan}
\altaffiltext{7}{Astronomical Institute, Tohoku University, Aramakiazaaoba 6-3, Aoba-ku, Sendai, Miyagi, 980-8578, Japan}
\altaffiltext{8}{Center for Computational Sciences, The University of Tsukuba, 1-1-1, Tennodai, Tsukuba, Ibaraki 305-8577, Japan}
\altaffiltext{9}{National Astronomical Observatory of Japan, Osawa, Mitaka, Tokyo, 181-8588, Japan}
\altaffiltext{10}{Research Center for the Early Universe, The University of Tokyo, 7-3-1 Hongo, Bunkyo-ku, Tokyo, 113-0033, Japan}

%%\newpage

\begin{abstract}
We have conducted a mapping spectral line survey toward the Galactic giant molecular cloud W51 in the 3~mm band with the Mopra 22~m telescope in order to study an averaged chemical composition of the gas extended over a molecular cloud scale in our Galaxy.  We have observed the area of $25' \times 30'$, which corresponds to 39~pc $\times$ 47~pc.  The frequency ranges of the observation are 85.1 - 101.1~GHz and 107.0 - 114.9~GHz.  In the spectrum spatially averaged over the observed area, spectral lines of 12 molecular species and 4 additional isotopologues are identified.  An intensity pattern of the spatially-averaged spectrum is found to be similar to that of the spiral arm in the external galaxy M51, indicating that these two sources have similar chemical compositions.  The observed area has been classified into 5 sub-regions according to the integrated intensity of $^{13}$CO($J=1-0$) ($I_{\rm ^{13}CO}$), and contributions of the fluxes of 11 molecular lines from each sub-region to the averaged spectrum have been evaluated.  For most of molecular species, 50~\% or more of the flux come from the sub-regions with $I_{\rm ^{13}CO}$ from 25~K~km~s$^{-1}$ to 100~K~km~s$^{-1}$, which does not involve active star forming regions.  Therefore, the molecular-cloud-scale spectrum observed in the 3~mm band  hardly represents the chemical composition of star forming cores, but mainly represents the chemical composition of an extended quiescent molecular gas.  The present result constitutes a sound base for interpreting the spectra of external galaxies at a resolution of a molecular cloud scale ($\sim$10~pc) or larger.  
\end{abstract}





\section{Introduction}
Chemical compositions of dense molecular cloud cores in our Galaxy have been studied since the early era of radio astronomy.  On the basis of observations of some representative molecules such as CCS, NH$_3$, and N$_2$H$^+$ in combination with chemical models, a systematic change in chemical composition along cloud evolution has been established, which has substantially contributed to our understanding of a starless-core phase of star formation \citep[e.g.][]{Suzuki1992,Aikawa2001,Caselli2003}.  On the other hand, chemical composition of molecular gas largely extended over a molecular cloud scale ($>$10~pc) has attracted less attention, although it would define an initial condition for chemical evolution toward dense molecular cloud cores.  Moreover, an understanding of the molecular-cloud-scale chemical composition is of crucial importance in extragalactic astrochemistry, because only such a large-scale chemical composition without resolving dense cores can be observed in most of nearby galaxies even in high spatial resolution observations with ALMA (Atacama Large Millimeter/submillimeter Array).  For example, the angular resolution of 0.2$''$ readily achieved with ALMA corresponds to the linear resolution of 10~pc for nearby galaxies at a distance of 10~Mpc.

Recently, chemical compositions have extensively been studied in extragalactic objects such as active galactic nuclei (AGN), starburst galaxies, and ultra luminous infrared galaxies (ULIRGs) by multi-molecular-line observations and unbiased spectral line surveys with single-dish telescopes \citep[e.g.][]{Martin2006,Aladro2015} and interferometers \citep[e.g.][]{Martin2011,Costagliola2015,Meier2015}.  In these studies, chemical compositions are discussed in relation to physical environments specific to target sources.  In addition, chemical compositions have been studied in quiescent regions of some external galaxies, such as spiral arms and bars \citep[e.g.][]{Meier2005,Meier2012,Watanabe2014,Watanabe2016}.  \citet{Watanabe2014} conducted a spectral line survey in the 3~mm and 2~mm bands toward two positions in the spiral arm of M~51 with the IRAM~30~m telescope, and revealed chemical compositions averaged over a 1~kpc scale.  They found that the chemical compositions of the two positions are similar to each other in spite of a difference in star formation activities.  Moreover, an effect of star formation activities cannot be seen on the chemical compositions in the spiral arm even at a spatial resolution of 300~pc \citep{Watanabe2016}.  \citet{Nishimura2016} observed seven molecular clouds in the Large Magellanic Cloud (LMC), and found that chemical compositions at a 10~pc scale are similar to one another regardless of their star formation activities.  These results suggest that molecular-cloud-scale chemical compositions observed in the 3~mm band, particularly for the relatively quiescent regions, are not seriously affected by local star formation activities, and are mostly dominated by contributions from a largely extended molecular gas.  Moreover, similarity of the molecular-cloud-scale chemical composition among various extragalactic sources implies that it would reflect fundamental physical and chemical properties of molecular clouds without influences of starburst activities and AGNs.  According to the above results, it is revealed that high excitation lines in the submillimeter-wave band with high critical densities have to be observed to trace star forming regions.

One practical way to {\rm investigate} the dominant factor which determines the molecular-cloud-scale chemical composition is to conduct a large-scale mapping observation of a Galactic molecular cloud in various molecular lines.  By averaging the spectra over the observed area, we can simulate the spectrum observed toward external galaxies.  Moreover, contributions of line fluxes from particular parts of the observed region can be studied.  In this way, we interpret the molecular-cloud-scale chemical compositions in terms of astrochemical processes of smaller regions, which have been established so far.  Although large-scale mapping observations have extensively been done in the CO and its isotopologue lines \citep[e.g.][]{Dame2001,Jackson2006,Fukui2008}, observations in other molecular lines are sparse, except for the Galactic center clouds and Orion~B \cite[e.g.][]{Jones2008,Pety2017}.  Large-scale mapping observations in various molecular lines, which we call `mapping spectral line survey', require a high sensitivity receiver and a large set of backend spectrometers to cover a wide frequency range simultaneously.  Thanks to recent advances of radio astronomy, the mapping spectral line surveys are now possible with a reasonable observation time.  

In the present study, we conducted the mapping spectral line survey toward the Galactic molecular cloud complex W51, which is well known as one of vigorous star forming giant molecular clouds (GMCs) \citep[e.g.][]{Mehringer1994,Carpenter1998,Bieging2010,Ginsburg2016}.  It is located at a distance of about $5.1-5.4$~kpc \citep{Xu2009,Sato2010} in the Sagittarius arm.  CO mapping observations have extensively been conducted toward W51, the molecular gas mass has been evaluated to be higher than $10^6 M_{\odot}$ \citep[e.g.][]{Carpenter1998,Jackson2006,Bieging2010}.  W51 is known to harbor the hot cores such as W51 e1/e2 and W51N \citep[e.g.][]{Ho1983,Zhang1998,Liu2001,Remijan2004}.  More than 50 molecular species including complex organic molecules have been identified toward the hot core e1/e2 \citep[e.g.][]{Liu2001, Ikeda2001, Remijan2002, Demyk2008, Kalenskii2010, Lykke2015, Rivilla2016a, Rivilla2016b}.  We observed the W51 cloud in order to investigate how the above vigorous star-formation activities are reflected on the spectrum averaged over the molecular cloud. 

\section{Observation}
Observations were carried out with the Mopra 22~m telescope in 2013 October and 2014 August and September.  An on-the-fly (OTF) mapping method was employed to cover the $25' \times 30'$ area of W51 centered at ($l$, $b$) = ($49.^{\circ}4902$, $-0.^{\circ}2622$) in the Galactic coordinate.  The area corresponds to $39 \, {\rm pc} \times 47 \, {\rm pc}$ at the distance of W51.  The off-source position was: ($l$, $b$) = ($49^{\circ}.8775$, $+0^{\circ}.2622$), which is known to be a $^{13}$CO free position \citep{Bieging2010}.  The OTF map was obtained with scans along the Galactic longitude.  Three frequency settings were observed to cover the frequency ranges of 85.2 - 101.1~GHz and 107.0 - 114.9~GHz (Table~\ref{tab01}).  The 3~mm HEMT MMIC receiver, which can simultaneously observe two orthogonal polarizations, was used as a front end.  A typical system noise temperature ranged from 140 to 800~K, depending on the frequency and the weather condition.  The beam sizes are $38''$ and $30''$ in the 90~GHz and 115~GHz, respectively.  Backends were the Mopra Spectrometer (MOPS) in a wideband mode, whose band width and frequency resolution are 8.3~GHz and 0.27~MHz, respectively.  Telescope pointing was checked every hour by observing the SiO maser source V1111 Oph.  Pointing accuracy was confirmed to be better than $6''$.  An intensity scale was calibrated to the antenna temperature scale ($T^{*}_{\rm a}$) by a chopper wheel method.  A daily fluctuation of the intensity was monitored by observing the hot cores e1/e2 in W51 at ($l$, $b$) = ($49.^{\circ}4898$, $-0.^{\circ}3874$), and was evaluated to be less than 14~\%.  The observed data were reduced with LIVEDATA/GRID-ZILLA \citep{Barnes2001}.  Bandpass correction and intensity calibration were processed by the LIVEDATA.  Spectral baselines were subtracted by fitting line-free channels to a seventh-order polynomial in a frequency range of 2~GHz in the LIVEDATA.  The processed spectra were re-sampled with a grid size of $15''$, and were integrated by the GRID-ZILLA.  After these procedures, the antenna temperature was converted to the main beam temperature as $T_{\rm mb} = T_{\rm a}^{*}/\eta_{\rm mb}$.  Here, $\eta_{\rm mb}$ is the main beam efficiency, which is 0.5 and 0.4 at 86~GHz and 115~GHz, respectively. The final rms noise levels of the maps were summarized in Table~\ref{tab01}.

\begin{figure*}
\epsscale{1.99}
\plotone{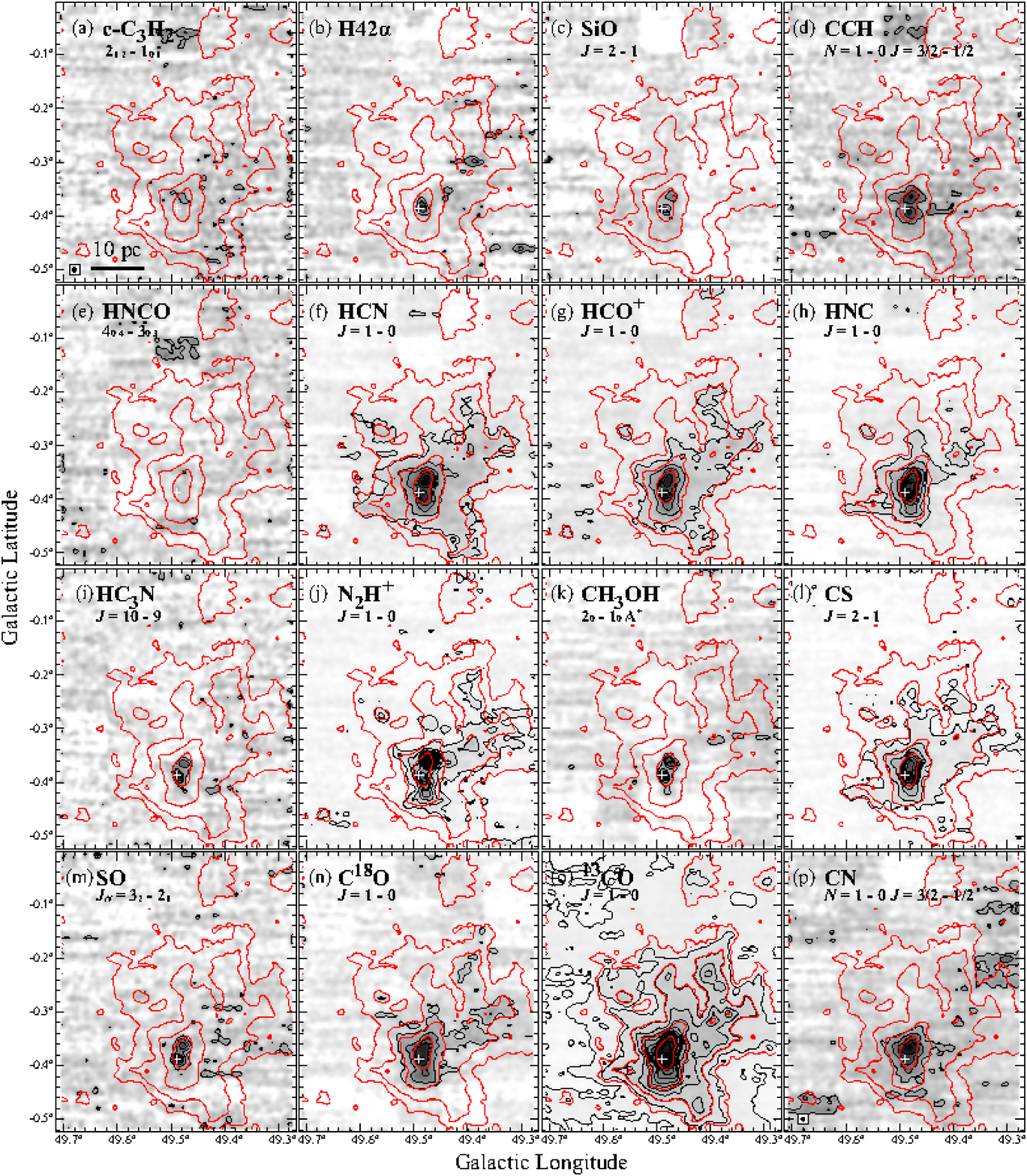}
\caption{\small Integrated intensity maps of (a) c-C$_3$H$_2$, (b) H42$\alpha$, (c) SiO, (d) CCH, (e) HNCO, (f) HCN, (g) HCO$^{+}$, (h) HNC, (i) HC$_3$N, (j) N$_2$H$^+$, (k) CH$_3$OH, (l) CS, (m) SO, (n) C$^{18}$O, (o) $^{13}$CO, and (p) CN.  A white cross indicates the position of hot cores W51~e1/e2.  Contour levels of the c-C$_3$H$_2$, H42$\alpha$, SiO, CCH, HNCO, HCN, HCO$^+$, HNC, HC$_3$N, N$_2$H$^+$, CH$_3$OH, CS, SO, C$^{18}$O, $^{13}$CO, and CN integrated intensities are $1.5 \times (3, 5)$~K~km~s$^{-1}$, $2.6 \times (3, 5, 7)$~K~km~s$^{-1}$, $2.0 \times (3, 5)$~K~km~s$^{-1}$,  $3.4 \times (3, 5, 7)$~K~km~s$^{-1}$, 5.1~K~km~s$^{-1}$, $3.1 \times (3, 7, ..., 23)$~K~km~s$^{-1}$, $3.0 \times (3, 7, ..., 23)$~K~km~s$^{-1}$, $2.2 \times (3, 6, ..., 24)$~K~km~s$^{-1}$, $1.8 \times (3, 5, ..., 9)$~K~km~s$^{-1}$, $1.5 \times (3, 7, ..., 43)$~K~km~s$^{-1}$, $2.6 \times (3, 5, ..., 11)$~K~km~s$^{-1}$, $2.6 \times (3, 8, ..., 48)$~K~km~s$^{-1}$, $3.8 \times (3, 5, ..., 11)$~K~km~s$^{-1}$, $3.2 \times (3, 6, ..., 15)$~K~km~s$^{-1}$, $4.0 \times (3, 8, ..., 78)$~K~km~s$^{-1}$, and $5.2 \times (3, 5, ..., 11)$, respectively.  Black closed circles shown in left-bottom corner of (a) and (p) are maximum ($38''$) and minimum ($32''$) beam sizes of the observation, respectively.  Red contours indicate sub-regions defined in Figure~\ref{fig04}.}
\label{fig01}
\end{figure*}

Figures~\ref{fig01} (a)--(p) reveal the integrated intensity maps of representative emission lines, c-C$_3$H$_2$($2_{1\,2}-1_{0\,1}$), SiO($J=2-1$), CCH($N=1-0, J=3/2-1/2$), HNCO($4_{0\,4}-3_{0\,3}$), HCN($J=1-0$), HCO$^+$($J=1-0$), NHC($J=1-0$), HC$_3$N($J=10-9$), N$_2$H$^+$($J=1-0$), CH$_3$OH($2_0-1_0, {\rm A}^+$), CS($J=2-1$), SO($J_N=3_2-2_1$), C$^{18}$O($J=1-0$), $^{13}$CO($J=1-0$), CN($N=1-0$), and H42$\alpha$ in order to show a typical quality of the imaging data, where the velocity ranges for integration is from 40~km~s$^{-1}$ to 90~km~s$^{-1}$.  All these emission lines are successfully imaged, although the sensitivity is not good enough to study their small-scale distribution in detail.  Molecular distributions are mostly concentrated in the vicinity of the hot cores W51~e1/e2 (a white cross in Figure~\ref{fig01}), with which a large H~II region is associated.  In addition to the component around the hot core, the HCN, HCO$^+$, HNC, N$_2$H$^+$, CS, C$^{18}$O, and $^{13}$CO emission have a spatially extended component.  

\section{Chemical Composition at a Molecular-Cloud Scale }
\subsection{Spatially-Averaged Spectrum}
\begin{figure*}
\epsscale{1.99}
\plotone{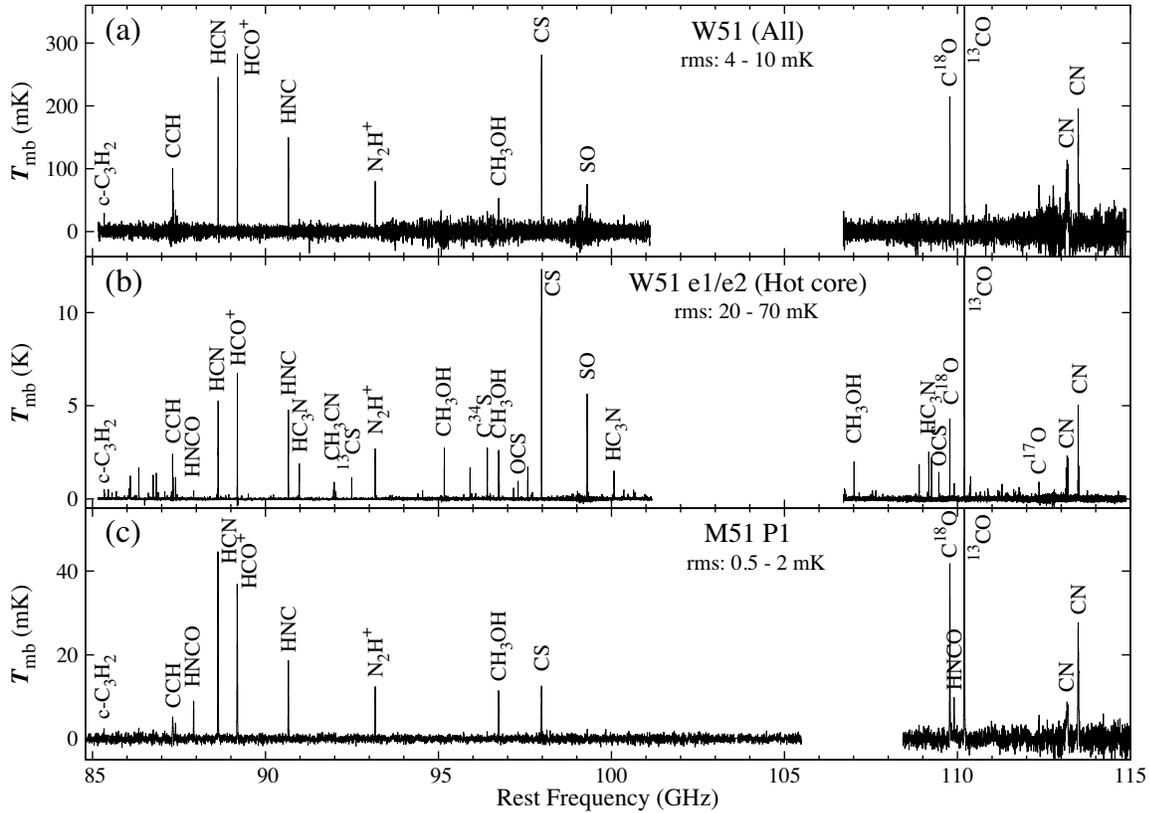}
\caption{Spectra of (a) averaged over all the observed area in W51, (b) W51~e1/e2 (hot cores), and (c) M51~P1 \citep{Watanabe2014}.  $V_{\rm LSR}$ is assumed to be 60~km~s$^{-1}$ and 55~km~s$^{-1}$ for W51 (All) and W51~e1/e2, respectively.  The angular resolution in the observations with the Mopra telescope is $38''$ -- $31''$.  The data of the spectra (a) and (b) are available in the online version of the journal.}
\label{fig02}
\end{figure*}

We prepare a spectrum averaged over the full observed region in W51 (39~pc $\times$ 47~pc), as shown in Figure~\ref{fig02}(a).  This is a spectrum of a GMC-scale molecular gas in our Galaxy, which can be compared with the spectrum observed toward external galaxies using interferometers such as ALMA.  In the preparation, we notice that the averaged spectrum suffers from a strong baseline ripple, which is caused by a standing wave between the main reflector and the subreflector of the Mopra 22~m telescope.  The baseline ripple is not recognized in the individual spectrum before averaging due to overwhelming receiver noise, while it is evident in the averaged spectrum.  Since the baseline ripple cannot be subtracted by using a polynomial function, we employ a baseline function consisting of several sinusoidal functions with different wavenumbers.  After the baseline subtraction, the spectrum is smoothed to the frequency resolution of 1.1~MHz by summing-up 4 successive channels.  A range of rms noise level of the spatially-averaged spectrum is from 12~mK to 48~mK depending on frequency. 

\begin{figure*}
\epsscale{1.99}
\plotone{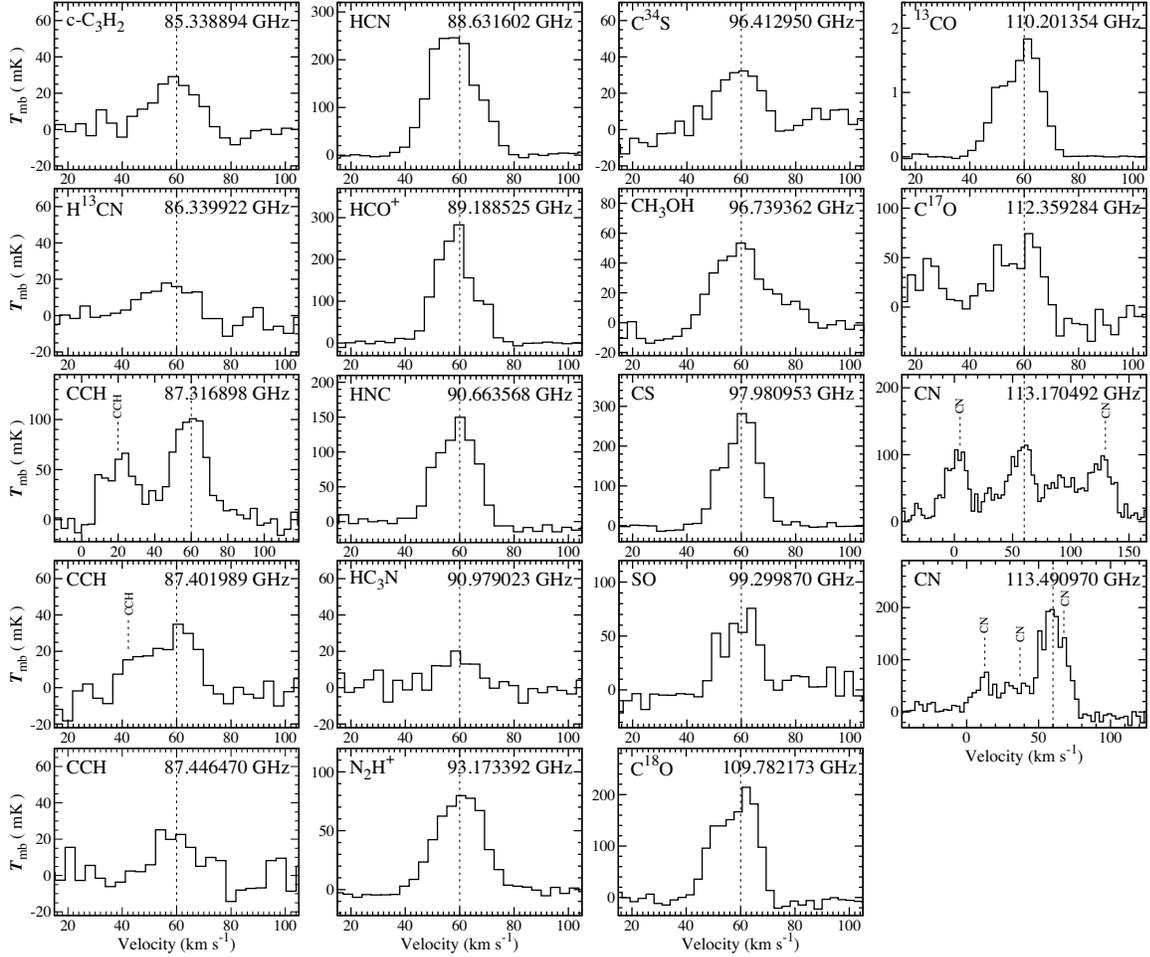}
\caption{Profiles of individual molecular lines averaged over all the observed area.  Vertical dashed-lines indicate $V_{\rm LSR}$ of 60~km~s$^{-1}$, which is the typical systemic velocity of W51 estimated by the Gaussian fitting to the averaged spectra (Table~\ref{tab02}).  }
\label{fig03}
\end{figure*}

In the spatially-averaged spectrum, 24 emission lines are detected.  These lines are assigned to 12 molecular species ($^{13}$CO, HCO$^+$, HCN, HNC, CN, SO, CS, CCH, N$_2$H$^+$, c-C$_3$H$_2$, HC$_3$N, and CH$_3$OH) and 4 additional isotopologues (C$^{17}$O, C$^{18}$O, H$^{13}$CN, and C$^{34}$S) on the basis of the spectral line databases, the Cologne Database for Molecular Spectroscopy (CDMS) managed by University of Cologne \citep{muller01,muller05} and the Submillimeter, Millimeter and Microwave Spectral Line Catalog provided by Jet Propulsion Laboratory \citep{pickett98}.  All the emission lines are simple molecular species, which consist of three heavy atoms or less, with low upper-state energies ($E_{\rm u} < 24$~K).  The line parameters and line profiles are summarised in Table~\ref{tab02} and Figure~\ref{fig03}, respectively.    

Figure~\ref{fig02} shows the comparison of the averaged spectrum with the spectrum of the hot cores W51~e1/e2 (Figure~\ref{fig02}~(b)) and that of the spiral arm of the external galaxy (M51~P1: Figure~\ref{fig02}(c)) \citep{Watanabe2014}.  The hot core e1/e2 was observed with a single pointing as an intensity calibration source (Section~2).  A spectral pattern of the averaged spectrum is found to be different from the spectrum of the hot cores W51~e1/e2, indicating that the spatially-averaged chemical composition is different from the chemical composition of the hot core e1/e2.  Although the spectral lines of fundamental species such as HCO$^+$, HCN, HNC, and CS are bright in both spectra, the spectrum of the hot core reveals much more emission lines than the averaged spectrum.  Emission lines of higher upper state energies ($E_{\rm u} > 100$~K), those of complex organic molecules such as HCOOCH$_3$ and CH$_3$OCH$_3$, and hydrogen recombination lines can be seen in the spectrum of the hot core in addition to the fundamental molecular species identified in the averaged spectrum (Appendix).  This comparison suggests that the spectral features specific to the hot core are smeared out in the averaged spectrum by the overwhelming contribution of an extended molecular gas.  This is a natural consequence of a small size of hot cores \citep[e.g. $\sim 2.4$~arcsec:][]{Zhang1998}.  If a hot core size is as large as 0.04--0.06~pc \citep{Zhang1998,Hernandez2014}, the molecular emission lines from the hot core are diluted by a factor of $10^{-6}$ in the spectrum averaged over the 39~pc $\times$ 47~pc area.  If we assume the gas kinetic temperature of the hot cores to be 200~K, which is typical for hot core \citep[e.g.][]{Zapata2009}, the intensity of the thermal emission does not exceed 200~K.  Hence, the expected intensity in the spectrum is less than $2 \times 10^{-4}$~K assuming the dilution factor of 10$^{-6}$.  This intensity is indeed much below the typical rms noise level of the averaged spectrum.  Conversely, the beam size smaller than about 3~pc is necessary to detect the hot core emission with the $3\sigma$ noise level of 60~mK in the 3~mm band.  For instance, the HCOOCH$_3$ emission ($8_{0\,8}-7_{0\,7}$ A,E: 90.23~GHz) is indeed detected with the intensity of 0.2~K at a $35''$ ($\sim0.9$~pc) resolution toward hot core W51~e1/e2 in the present study (Appendix).

Here, we assess the sensitivities of the averaged spectrum and the line intensities estimated from those observed in the W51~e1/e2 and M51~P1, in order to examine the reasons of non-detection of the molecular lines of complex organic molecules, OCS, SiO, and HNCO, as well as the CH$_3$OH line with higher $E_{\rm u}$, in the averaged spectrum.  A HNC line is detected with a reasonable signal-to-noise ratio (S/N) of 23, 240, and 57 in the averaged spectrum of W51, the W51~e1/e2 spectrum, and the M51~P1 spectrum, respectively.  Therefore, we estimate the molecular line intensities expected for the averaged spectrum of W51 and the M51~P1 spectrum, by applying the intensity ratios relative to HNC in the W51~e1/e2 spectrum to the HNC intensity of each source.  

In W51~e1/e2, many lines of complex organic molecules are detected with a typical intensity of 0.2~K (Appendix), while none of these lines is detected in the averaged spectrum, except for the CH$_3$OH lines.  If the intensity ratio between the complex organic molecules and the HNC intensity in the averaged spectrum were the same as that in the W51~e1/e2 spectrum, the intensity of complex organic molecule would be 0.006~K in the averaged spectrum.  This intensity is lower than the rms noise of the averaged spectrum.  Therefore, the non-detection of complex organic molecules in the averaged spectrum is partially due to the insufficient observation sensitivity.  However, the sensitivity difference does not explain non-detection of a few moderately intense lines such as OCS ($J=9-8,10-9$), SiO ($J=2-1$), and CH$_3$OH ($8_0-7_1,{\rm A}^+$) with high $E_{\rm u}$ in the averaged spectrum.  For example, the OCS ($J=9-8$) line is expected to be detected with the S/N ratio of 6 under the above estimation.  The deficiency of these molecular lines would be due to a heavier beam dilution effect, since the distributions of these molecular lines are thought to be more compact than that of HNC.  The beam dilution effect also contributes to the non-detection of the complex organic molecule.

On the other hand, the difference of observation sensitivities does not seriously affect the differences between the averaged spectrum of W51 and that of M51~P1.  For example, one notable difference is non-detection of HNCO($4_{0\,4}-3_{0\,3}$) in the averaged spectrum of W51.  If HNCO/HNC intensity ratio in the averaged spectrum were the same as that of M51~P1, the HNCO would be detected with the S/N ratio of 9 in the averaged spectrum.  The difference could not be explained by the beam dilution effect, because HNCO is detected in M51~P1 with a much larger beam size ($\sim 1$~kpc) than that of W51 ($\sim 50$~pc).  Therefore, non-detection of HNCO is due to deficiency of the HNCO abundance in the averaged spectrum of W51. 

\subsection{Contribution of Extended Molecular Gas}
\begin{figure*}
\epsscale{1.00}
\plotone{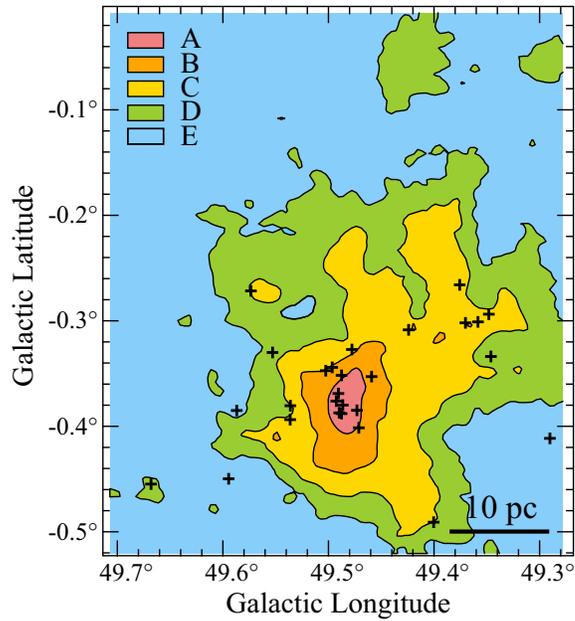}
\caption{Definitions of the sub-regions A to E.  A, B, C, D, and E are the sub-regions where $I_{\rm ^{13}CO} > 200 \,{\rm K\,km\,s^{-1}}$, $200 \,{\rm K\,km\,s^{-1}} > I_{\rm ^{13}CO} > 100 \,{\rm K\,km\,s^{-1}}$, $100 \,{\rm K\,km\,s^{-1}} > I_{\rm ^{13}CO} > 50 \,{\rm K\,km\,s^{-1}}$, $50 \,{\rm K\,km\,s^{-1}} > I_{\rm ^{13}CO} > 25 \,{\rm K\,km\,s^{-1}}$, and $25 \,{\rm K\,km\,s^{-1}} > I_{\rm ^{13}CO}$, respectively.  Crosses indicate positions of the zero-age main-sequence OB stars \citep{Mehringer1994}. }
\label{fig04}
\end{figure*}

\begin{figure*}
\epsscale{1.99}
\plotone{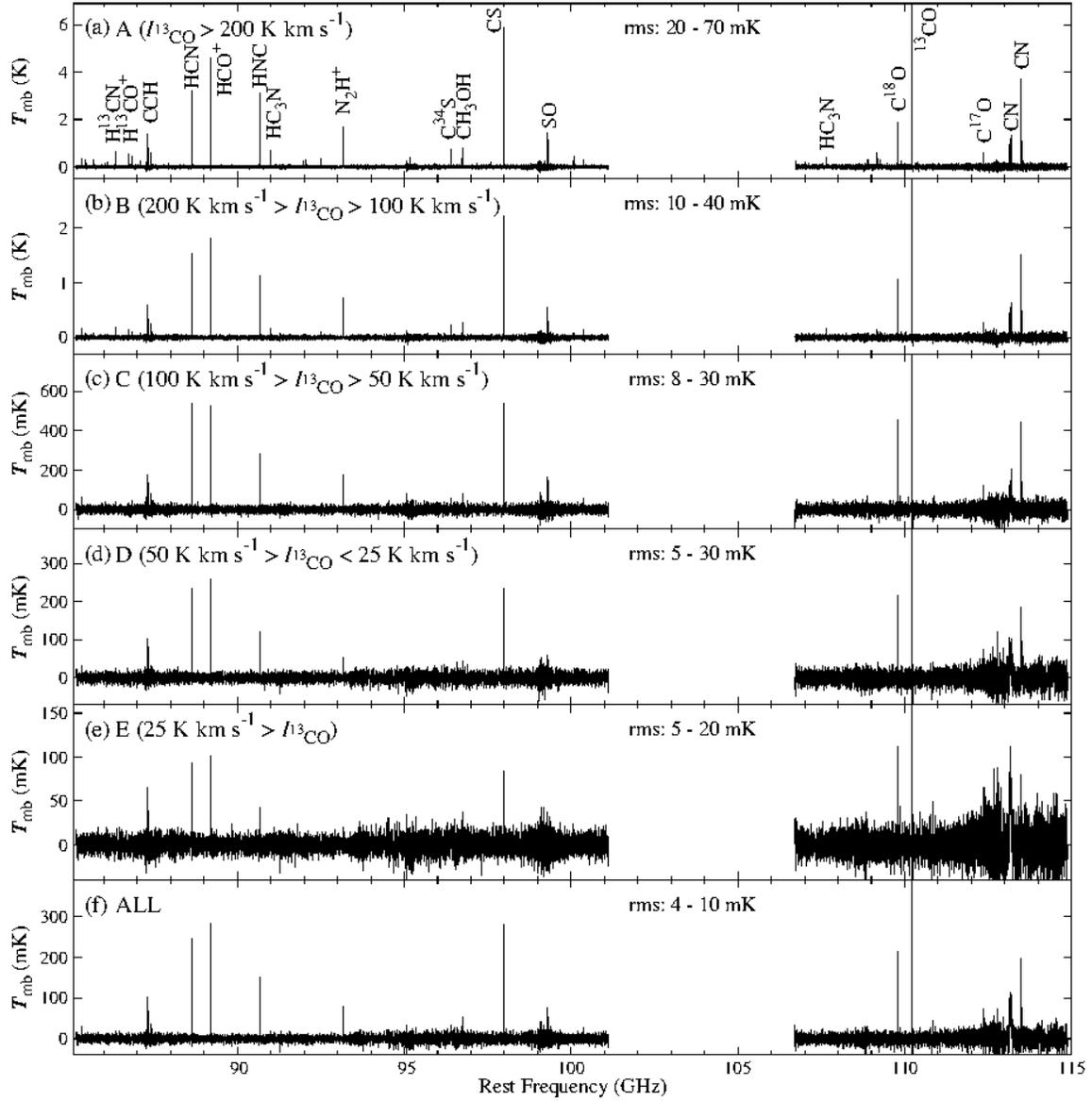}
\caption{(a) $\sim$ (e) Averaged spectra in the sub-regions A, B, C, D, and E, respectively.  (f) The averaged spectrum over the whole observed area for comparison, which is the same as Figure~\ref{fig01}(a).  The data of the spectra (a) $\sim$ (e) are available in the online version of the journal.}
\label{fig05}
\end{figure*}

\begin{figure*}
\epsscale{1.00}
\plotone{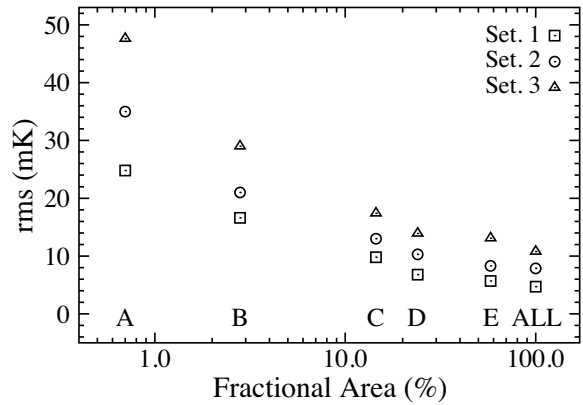}
\caption{Rms noises of the spectra of the sub-regions (A, B, C, D, and E) and that of the full spectrum (ALL) as a function of the fractional areas.  Squares, circles, and triangles indicate the plots for the frequency settings of 1, 2, and, 3 (Table~\ref{tab01}), respectively.  The rms noises are evaluated in the frequency ranges of 89.5 -- 90.5~GHz, 96.85 -- 97.85~GHz, and 110.5 -- 111.5~GHz, which are the emission-line-free ranges, for the frequency settings 1, 2, and 3, respectively. }
\label{fig11}
\end{figure*}

We here evaluate the contribution of a widely extended molecular gas to the average spectrum.  For this purpose, we classify the observed area into 5 sub-regions (A--E) according to the integrated intensities of $^{13}$CO($J=1-0$), as shown in Table~\ref{tab04} and Figure~\ref{fig04}.  Here, we use the $^{13}$CO($J=1-0$) emission as a proxy of the line-of-sight column density of the molecular gas.  Then, we derive the averaged spectrum for each sub-region, and evaluate a fraction of the flux from each sub-region to the total flux for each emission line.  While the sub-regions A and B involve the H~II regions and many main-sequence OB stars (Figure~\ref{fig04}), the sub-regions C, D, and E show relatively mild star-formation activities.  Figure~\ref{fig05} shows the averaged spectrum for each sub-region.  The averaged spectrum is prepared by the same method described in Section 3.1.  In the sub-region A, we identify 18 molecular species, 10 isotopologues, and 3 hydrogen recombination lines on the basis of the spectral line databases.  A summary of the detected molecular species in the spectrum of each sub-region and the averaged spectrum is given in Table~\ref{tab04}.  Integrated intensities of molecules and their upper-limits for the 5 sub-regions are summarized in Table~\ref{tab05}.  Table~\ref{tab05} lists all the emission lines detected with the $3\sigma$ or higher confidence level in the spectrum averaged over the sub-region A.  In order to show the noise in the sub-regions, we prepared a plots of the rms noise as a function of a fraction of area to the all observed area (fractional area) (Figure~\ref{fig11}).  The rms noise decreases as increasing the fractional area for each observation setting. 

\begin{figure*}
\epsscale{1.99}
\plotone{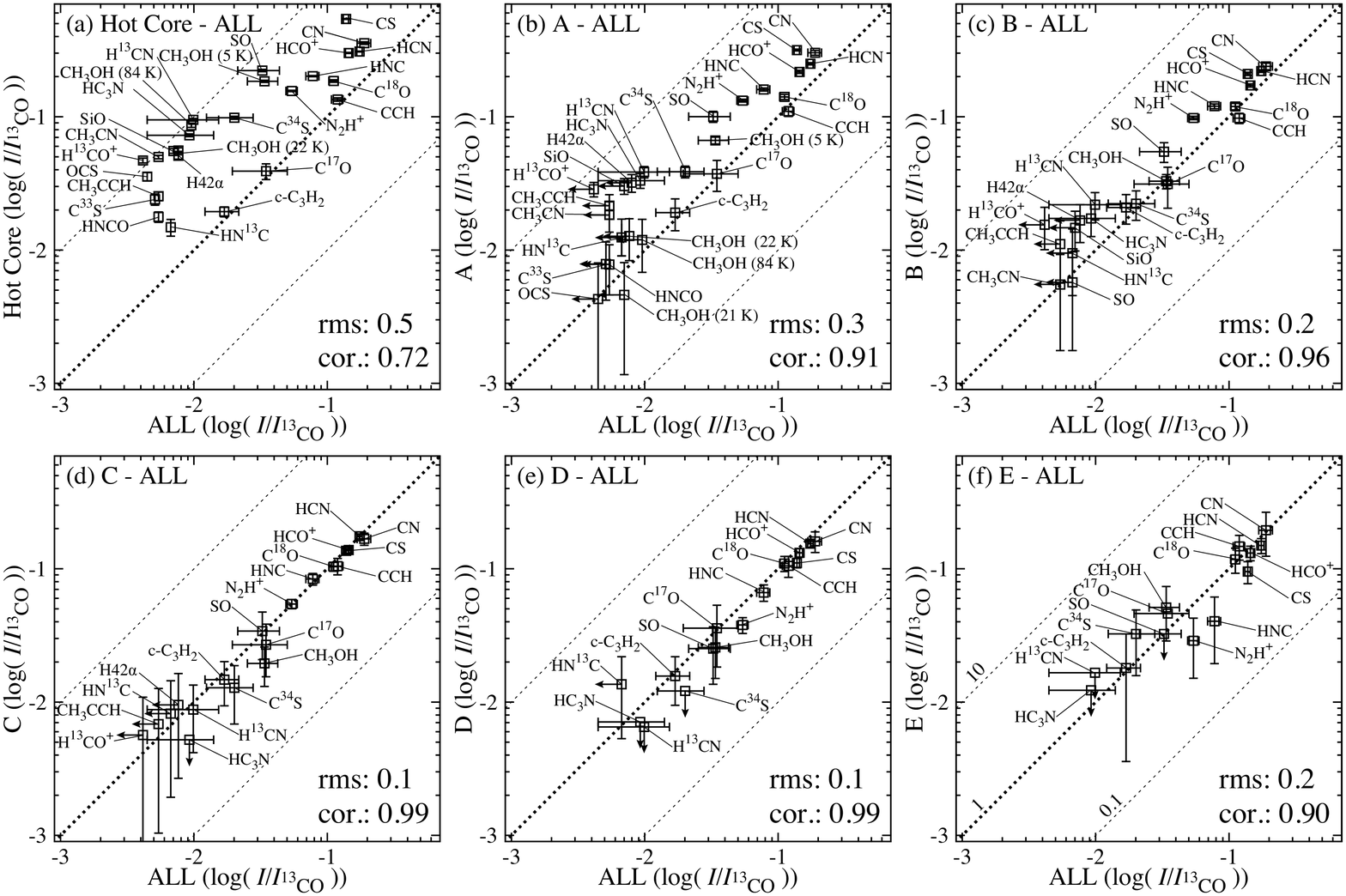}
\caption{Correlation diagrams of the integrated intensity normalized by the integrated intensity of $^{13}$CO between the spectrum averaged over all the region of W51 and (a) hot core, (b) the sub-region A, (c) the sub-region B, (d) the sub-region C, (e) the sub-region D, and (f) the sub-region E.  Arrows indicate upper limits.  The upper state energies are given in parentheses for CH$_3$OH.  Thin dotted lines indicate the ratios of the two axis of 0.1 and 10, while thick dotted line indicates the ratio of 1.  The rms values are calculated as $(1/n \sum_{i}( \log(I_{i,{\rm sub-region}}/I_{\rm ^{13}CO, sub-region} ) - \log(I_{i,{\rm ALL}}/I_{\rm ^{13}CO, ALL}))^2)^{1/2}$, where $n$, $I_{i,{\rm sub-region}}$, $I_{\rm ^{13}CO, sub-region}$, $I_{i,{\rm ALL}}$, and $I_{\rm ^{13}CO, ALL}$ are the number of molecular species, the integrated intensities of a particular molecular species in the sub-region of W51, the integrated intensity of $^{13}$CO in the sub-region, the integrated intensities of molecule in the spectrum averaged over all the region of W51, and the integrated intensity of $^{13}$CO in the spectrum averaged over all the region of W51, respectively.  The correlation coefficients between $\log(I_{i,{\rm sub-region}}/I_{\rm ^{13}CO, sub-region})$ and $\log(I_{i,{\rm ALL}}/I_{\rm ^{13}CO, ALL})$ are evaluated and shown in the panels (cor.). } 
\label{fig10}
\end{figure*}

Spectral patterns of all the sub-regions (A-E) look similar to that of the full region, particularly when we focus on bright lines.  In order to see in more detail how they are similar or different, we prepare correlation diagrams of integrated intensities normalized by the integrated intensity of $^{13}$CO between the full region and each sub-region (Figure~\ref{fig10}).  In addition, we also prepare the similar diagram between the full region and the hot core region (Figure~\ref{fig10}a).  If the spectral pattern of a sub-region is the same as that of the full region, the plots for all the spectral lines are on the straight line indicated by the thick dashed line.  This trend can be seen for the sub-regions C, D, and E.  This result indicates the similarity of the spectra of the sub-regions (C-E) and the full region.  On the other hand, the plots tend to be shifted to the upper-left direction for the sub-regions A and B.  This trend is significant for the sub-region A and also for the hot core e1/e2.  Although some spectral lines such as C$^{18}$O, C$^{17}$O, CCH and c-C$_3$H$_2$ are almost on the thick dashed line, the intensities of most spectral lines normalized by the $^{13}$CO intensity tend to be higher in the sub-regions A and B and the hot core e1/e2.  For quantitative comparison, we calculate the dispersion from the thick dashed line (the same normalized intensities between a sub-region and the full region).  Indeed, the dispersion is larger for the sub-regions A and B than the sub-regions C and D, as shown in Figure~\ref{fig10}.  Although the dispersion for the sub-region E is a bit large, this is due to the low signal-to-noise ratio of the spectrum of the sub-region E.  Above all, the spectra of the sub-regions (C-E) are similar to that of the full region, while those of the sub-regions A and B are rather different.  A widely extended molecular gas in the sub-regions (C-E) mainly contributes the spectrum of the full region. 

The spectrum of sub-region A (Figure~\ref{fig05}a) shows the emission lines of HC$_3$N, CH$_3$CCH, CH$_3$CN, HNCO, SiO, and OCS, as well as isotopologues of the simple molecular species (H$^{13}$CN, H$^{13}$CO$^+$, C$^{34}$S, C$^{18}$O, and C$^{17}$O).  High excitation lines of CH$_3$OH and the hydrogen recombination lines are also visible.  Many of them can also be seen in the spectrum of the sub-region B.  If these emissions come from the star forming regions associated with OB stars (Figure~\ref{fig04}), they would be significantly diluted by spatial averaging.  Since the area of the sub-region A is about 14~pc$^2$, the dilution factor for the hot core size (0.06~pc) is the order of 10$^{-4}$.  If the high excitation line of CH$_3$OH ($E_{\rm u} = 83.6$ K) at 95.169 GHz originates only from the hot core, its peak temperature is roughly estimated to be as high as 1500~K.  This is too high for the brightness temperature of the thermal emission from the hot core.  The same situation holds for most of the other molecules listed above.  Hence, the lines of these molecules do not come only from the hot core itself, but most likely from a dense and warm gas distributed around the hot core.  In this case, the emission is less affected by the dilution in the sub-region A. In the other sub-regions, such warm and dense regions may not be as large as the sub-region A case, considering their lower star-formation activities.  Hence, the emission from the warm and dense component affected by star-formation activities would intrinsically be weak, and would further be affected by the heavy dilution effect.

To show the contribution from each region to the full spectrum more quantitatively, we evaluate the fluxes of c-C$_3$H$_2$, CCH, HCN, HCO$^+$, HNC, N$_2$H$^+$, CH$_3$OH, CS, SO, C$^{18}$O, $^{13}$CO, and CN for each sub-region, and calculate their fractions to the total flux (fractional flux) as:
\begin{equation}
{\rm Fractional \, Flux} = \frac{\int_{\rm S} I_{\rm mol}(x,y) dxdy}{\int_{\rm All} I_{\rm mol}(x,y) dxdy} \times 100 \, \%,
\end{equation}
%\begin{equation}
%{\rm Fractional \, Flux\,(\%)} = \frac{\sum_{i \in {\rm R}^{}} I_{i}(X)}{\sum_{i \in {\rm All}^{}} I_{i}(X)} \times 100 \, \%,
%\end{equation}
where $I_{\rm mol}(x,y)$ stands for the integrated intensity of a molecule.  The integration area is designated by S (A, B, C, D, or E of Table~\ref{tab04}), where `All' means all the observed area.  A fractional flux of SO is estimated only in the sub-regions A, B, C, and D, since only the $3\sigma$ upper limit of SO is available in the sub-region E.  Table~\ref{tab07} summarizes the results.  Figure~\ref{fig06} shows the fractional area of each sub-region, along with the fractional fluxes of the 11 molecular species except for SO. 

For most molecules, 50~\% or more of the flux are found to come from the sub-regions C and D, as shown in Figure~\ref{fig06} and Table~\ref{tab07}.  The fractional fluxes are 60~\% or higher, if the sub-region E is included.  On the other hand, contributions from the sub-regions A and B are relatively small ($\lesssim30$~\%), because the fractions of the areas are only 0.7~\% and 2.8~\% for the sub-regions A and B, respectively.  Therefore, chemical compositions of a widely extended molecular gas in the sub-regions C, D, and E mainly contribute to the spectrum of the full region.  This effect would especially be significant in the 3~mm band observation, because many of the bright emission lines in this band are transitions to the ground rotational state.  Although the critical densities of these lines are 10$^5$--10$^6$~cm$^{-3}$ except for CO isotopologues \citep[e.g.][]{Evans1999,Yamamoto2017}, they can be sub-thermally excited even in less dense regions ($\sim$10$^4$~cm$^{-3}$) including cloud peripheries \citep[e.g.][see also Section 3.4]{Nishimura2016}.  On the other hand, emission lines from higher energy levels can be seen under higher temperature and higher density conditions.  In fact, emission lines with high $E_{\rm u}$ ($> 50$~K) are observed only in the vicinity of the hot cores, thereby in the sub-region A of W51 (Table~\ref{tab05}).  

It is interesting to examine a variation of the excitation condition from sub-region to sub-region for molecules whose multiple lines are detected.  CH$_3$OH is a good case for such an analysis, because we detected the high excitation line of CH$_3$OH (95.169 GHz, $E_{\rm u} = 83.6$~K) in the sub-region A.  On the other hand, the high excitation line is not detected in the other sub-regions.  This indicates an insufficient excitation condition in these sub-regions than the sub-region A.  In addition, the high excitation line of CH$_3$OH will be more affected by the beam dilution effect, because the high excitation line is expected to trace more compact molecular gas in the vicinity of the star forming regions.  We can thus evaluate the intensity ratio relative to the three blending low-excitation lines of CH$_3$OH (96.741 GHz, $E_{\rm u} = 7.0$ K; 96.739 GHz, $E_{\rm u} = 4.6$ K; 96.745 GHz, $E_{\rm u} = 13.6 $ K) to be 0.18 in the sub-region A, according to Table~\ref{tab05}.  However, this line was not detected in the other sub-regions, and we only derive the $3\sigma$ upper limit to the ratio.  It is mostly higher than 0.2, and hence, we cannot see any definitive variation of the excitation conditions in this analysis.

\subsection{Distributions of Molecular Emissions}
\begin{figure*}
\epsscale{1.99}
\plotone{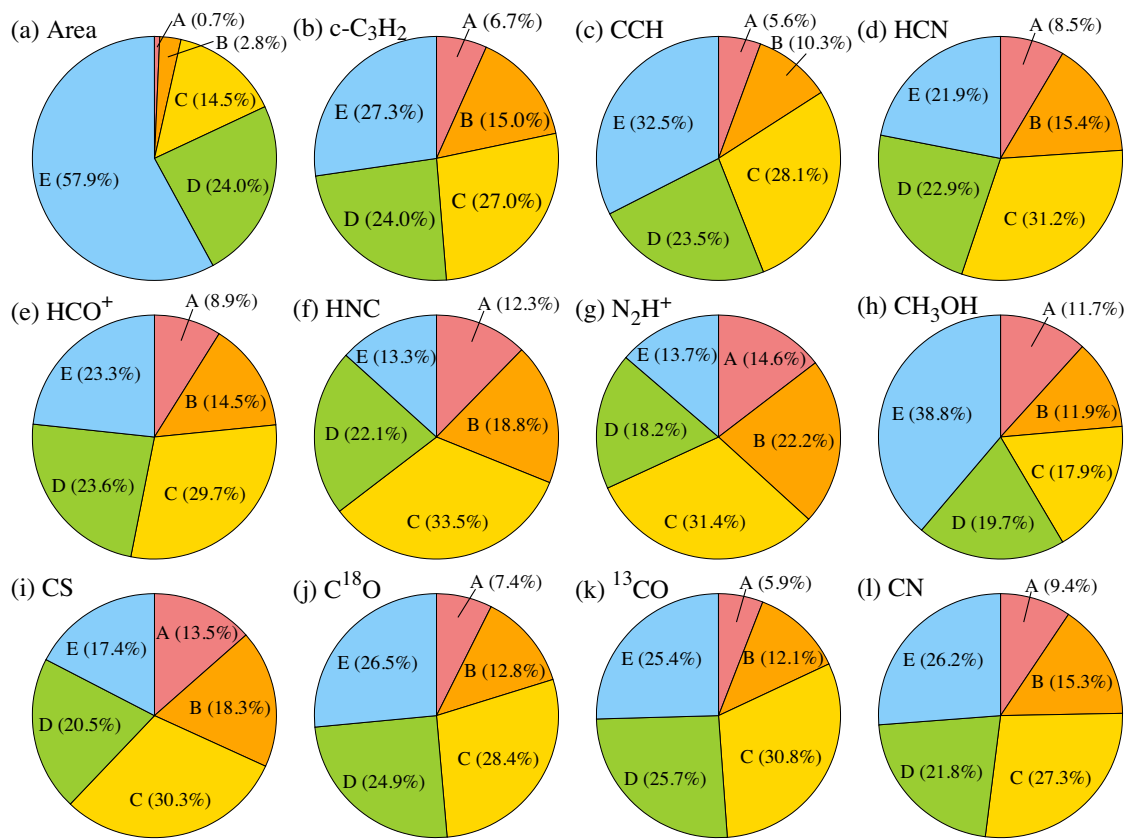}
\caption{(a) A fractional area of each sub-region and (b)$\sim$(m) fractional fluxes of c-C$_3$H$_2$, CCH, HCN, HCO$^+$, HNC, N$_2$H$^+$, CH$_3$OH, CS, SO, C$^{18}$O, $^{13}$CO, and CN for the sub-regions.  The transition used for each molecule is given in the footnote of Table~\ref{tab07}.} 
\label{fig06}
\end{figure*}

Contributions from the sub-regions to the total flux are different from molecule to molecule, as shown in Figure~\ref{fig06}.  These would also be different from transition to transition within the same molecular species due to different excitation conditions.  In particular the contributions from the sub-regions C and D are relatively small for HNC, N$_2$H$^+$, and CS in comparison with the other molecules.  For HNC, N$_2$H$^+$, and CS, the contributions from the sub-region A and B share 30~\% of the total flux or higher.  For CH$_3$OH, the contribution from the sub-region E looks higher than those from the other sub-regions.  However, this higher contribution would be affected by a large uncertainty due to a poor signal-to-noise ratio in the spectrum of the sub-region E.  Except for these molecules, the contribution from the sub-region C, D, and E are dominant.  

In order to examine the dependences of molecular line intensities on the total line-of-sight column density of the molecular gas in more detail, we divide the mapping area into 99 ($9 \times 11$) grid points, where the grid spacing is $2.'5$.  Then, the spectrum averaged over a circular area with a radius of $2.'5$ is prepared for each grid point.  The baseline of each averaged spectrum is subtracted by the same method described in the section 3.1.  We calculate integrated intensities of molecular lines at each point.  

Figure~\ref{fig07} shows correlation diagrams of the integrated intensities of 11 molecules against that of $^{13}$CO, a proxy of the line-of-sight column density of the molecular gas.  For HCN, CCH, CN, and C$^{18}$O, the integrated intensities almost linearly increase as increasing $^{13}$CO integrated intensity.  This trend indicates that the abundances of these molecules are almost constant over a wide range of the column density of the molecular gas.  On the other hand, the integrated intensities of HNC, N$_2$H$^+$, and CS intensities are found to increase rapidly as increasing $^{13}$CO integrated intensity in a non-linear way.  The similar trend can be seen for the SO and CH$_3$OH.  In particular, there seems to be a threshold value of the $^{13}$CO integrated intensity for appearance of the SO and CH$_3$OH lines (maybe N$_2$H$^+$ as well).  This behavior indicates that the abundances of these molecules are enhanced in the sub-region with the higher molecular column density, namely in the vicinity of the star forming regions in W51.  However, it should be stressed that this enhancement does not mean that the contribution of the sub-regions A and B is dominated in the full spectrum averaged over the large area.  Rather, the emission from the extended region, the sub-regions (C-E), makes a dominant contribution in the full spectrum because of the larger emitting area, as shown in Figure~\ref{fig06}. 

%The distribution of CH$_3$OH is also different from other molecules.  But this would be due to poor signal-to-noise ration of spectrum in the region E.  

\subsection{Column Densities and Fractional Abundances}
Column densities averaged over the full region are evaluated by the local thermodynamic equilibrium (LTE) analysis and the non-LTE analysis.  In order to evaluate the column densities by the LTE and non-LTE analyses, the temperature of molecular gas is necessary.  However, no rotation temperature can be derived from our data, because only one transition is observed for each species in the averaged spectrum for the full region except for hyperfine components.  Hence, we have to assume the temperature in the both analyses.  In the LTE analysis, column densities are evaluated under the assumption of the optically thin condition by using the following formula:
\begin{equation}
W_{\nu} = \frac{8 \pi^3 S\mu_0^2 \nu N}{3 k U(T_{\rm rot})} \left\{ 1 - \frac{\exp(h \nu/kT_{\rm rot})-1}{\exp(h \nu/kT_{\rm bg})-1} \right\}\exp\left(-\frac{E_{\rm u}}{k T_{\rm rot}}\right),
\label{eq01}
\end{equation}
where $W_{\nu}$, $S$, $\mu_0$, $\nu$, $N$, $k$, $U$, $T_{\rm rot}$, $h$, $T_{\rm bg}$, and $E_{\rm u}$ are integrated intensity, line strength, dipole moment, transition frequency, total column density, the Boltzmann constant, partition function, rotation temperature, the Planck constant, the cosmic microwave background temperature, and upper state energy, respectively.  Table~\ref{tab03} summarizes the column densities derived for the rotation temperatures of 10~K, 15~K, and 20~K.  The errors are evaluated by taking into account of the rms noise of the averaged spectrum and the calibration error of the chopper wheel method (20~\%).  In addition to the column densities (Table~\ref{tab03}), the fractional abundances relative to the H$_2$ column density are calculated by dividing the molecular column densities by the H$_2$ column density.  The latter is obtained from the column density of C$^{18}$O, where the [C$^{18}$O]/[H$_2$] ratio is assumed to be $1.7 \times 10^{-7}$ \citep[e.g.][]{Frerking1982,Goldsmith1997}.

\begin{figure*}
\epsscale{1.99}
\plotone{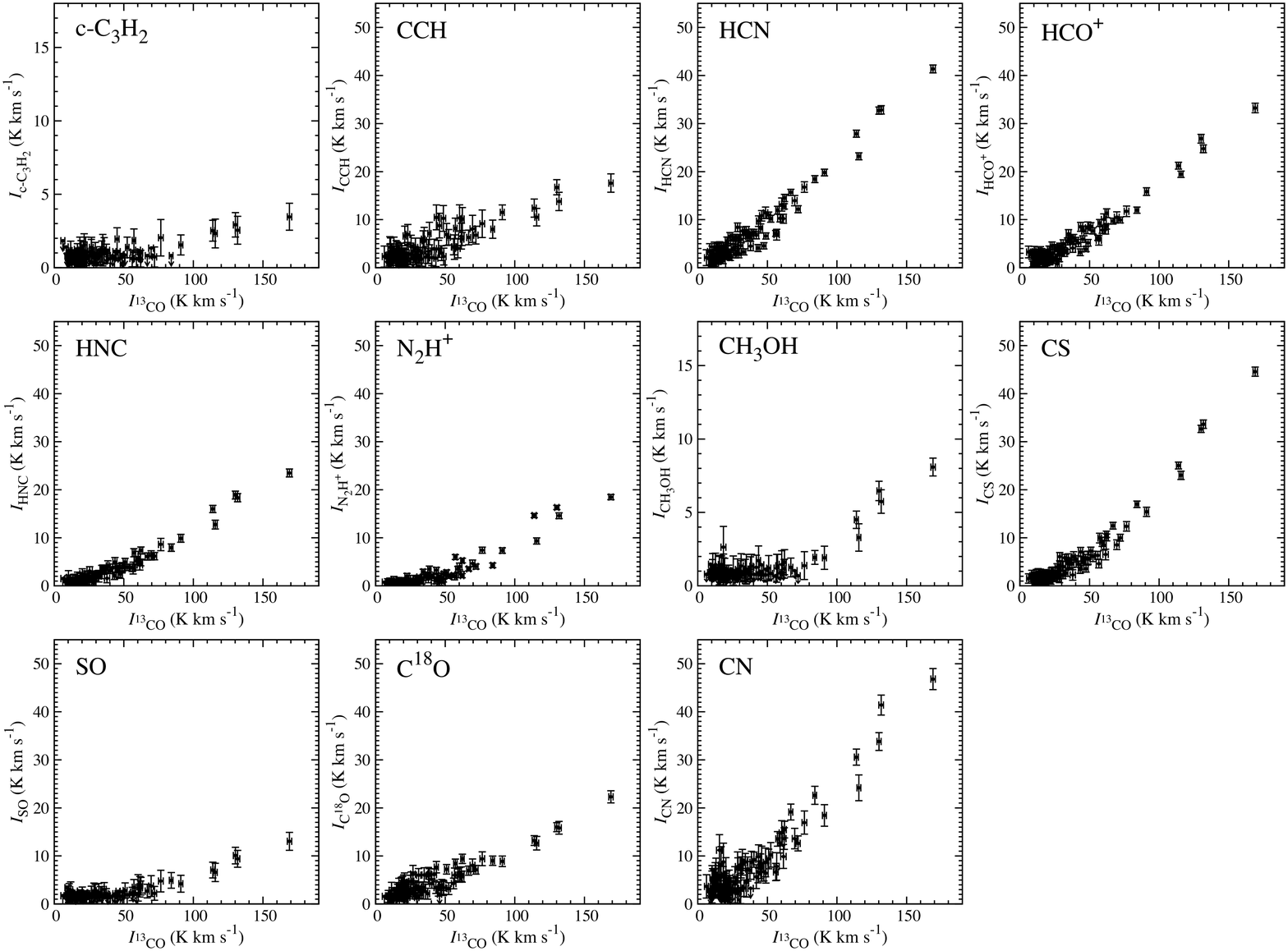}
\caption{Plots of the integrated intensities of c-C$_3$H$_2$, CCH, HCN, HCO$^+$, HNC, N$_2$H$^+$, CH$_3$OH, CS, SO, C$^{18}$O, and CN as a function of the integrated intensity of $^{13}$CO at a resolution of 5~arcmin.  Arrows indicate upper-limits.  The vertical axis is expanded for c-C$_3$H$_2$ and CH$_3$OH, because the integrated intensities of these molecules are weaker than 10~k~km~s$^{-1}$.   }
\label{fig07}
\end{figure*}

In the above analysis, we assume the optically thin condition.  However, this assumption is not the case for some relatively intense lines.  We therefore evaluate the optical depth for HCN and CS by comparing their intensities with the isotopologue lines, H$^{13}$CN and C$^{34}$S, respectively.  The HCN/H$^{13}$CN and CS/C$^{34}$S ratios are calculated to be $16 \pm 10$ and $7 \pm 3$, respectively.   These values are lower than the elemental isotope ratios in the Solar vicinity ($^{12}$C/$^{13}$C $\sim 60-70$; $^{32}$S/$^{34}$S $\sim22$) \citep{Wilson1999,Milam2005}.  The optical depths of the HCN and CS lines are estimated from these ratios to be about 4 and 3, respectively.  Hence, the column densities of HCN and CS are calculated by using their isotopologue lines, assuming the isotope ratios in the Solar vicinity (Table~\ref{tab03}).  While the H$^{13}$CO$^+$ line is not detected in the spectrum of the full region, the HCO$^+$ line may be also optically thick.  The column density of HCO$^+$ should thus be regarded as the lower limit.  For other lines, the optically thin assumption would be held, because the line intensities are lower than those of the HCN, CS, and HCO$^+$ lines.  Note that the $^{13}$CO and C$^{18}$O lines can be regarded as optically thin, because they are well correlated with each other over a wide range of the $^{13}$CO intensity (Figure~\ref{fig07}).  The $^{13}$CO/C$^{18}$O ratio is 10, which is consistent with the elemental $^{13}$C/$^{18}$O ratio of 9 \citep{Lucas1998}.

A non-LTE excitation effect may be important in the estimation of column densities, especially when a critical density of a molecular line is lower than a number density of molecular hydrogen ($n_{\rm H_2}$).  Hence, we derive the column densities by using the statistical equilibrium radiative transfer code RADEX \citep{Tak2007}.  Here, we assume the $n_{\rm H_2}$ values of $10^4$~cm$^{-3}$ and $10^5$~cm$^{-3}$ in our calculations, since the dense gas ($n_{\rm H_2} > 10^{4}\,{\rm cm^{-3}}$) fraction relative to the total molecular gas is reported to be more than 70~\% in the W51 cloud \citep{Ginsburg2015}.   The gas kinetic temperatures is assumed to be 10~K, 15~K, and 20~K as in the case of the LTE analysis.  Table~\ref{tab06} shows the results of the non-LTE analyses for molecules whose the collisional coefficients are available.  For the $n_{\rm H_2}$ value of $10^4$~cm$^{-3}$, the column densities are higher than the LTE values by an order of magnitude for the transitions with high critical densities ($\gtrsim 10^5$~cm$^{-3}$) such as HCN($J=1-0$), H$^{13}$CN($J=1-0$), and CS($J=2-1$).  These transitions are thought to be in a sub-thermal excitation conditions and have lower excitation temperatures than the assumed kinetic temperature.  As the result, the higher column densities are required in the none-LTE analysis than in the LTE analysis to reproduce the observed intensities.  On the other hand, the column densities of CO isotoplogues are found to be similar to the LTE value, because the critical density of CO itotopologues ($\sim 10^3$~cm$^{-3}$) is lower than the assumed molecular hydrogen densities.  For the $n_{\rm H_2}$ value of $10^5$~cm$^{-3}$, the column densities of most molecules are similar to the LTE values within a factor of two.  This result indicates that the LTE approximation is reasonable for most of transitions, if the number density of molecular hydrogen is as high as $10^5$~cm$^{-3}$. 

In order to determine the column densities with the non-LTE method, more accurate values of the number density of molecular hydrogen and the kinetic temperature are necessary.  For this purpose, observations of other transition lines are awaited.

\section{Comparison with Spectra of External Galaxies}
In this section, we compare the molecular-cloud-scale chemical composition of W51 with those of the spiral arm region of M51 \citep{Watanabe2014}, the starburst region of NGC~253 \citep{Aladro2015}, the AGN region of NGC~1068 \citep{Aladro2015}, the nuclear region of Luminous Infrared Galaxy (LIRG) NGC~4418 \citep{Costagliola2015}, and the star forming cloud (N44C) in the Large Magellanic Cloud \citep{Nishimura2016}.  Although the spatial-scale of the chemical compositions for M51, NGC~253, and NGC~1068 is larger than that for the W51 case by an order of magnitude, those of the last two cases are similar to the W51 case.

\subsection{Comparison with Spiral Arm of M51}
We here compare the molecular-cloud-scale chemical compositions of W51 with those of the spiral arm of M51 \citep{Watanabe2014}, because the spiral arm is expected to consist of molecular clouds similar to W51.  In spite of large size-scale difference between W51 ($\sim 50$~pc) and M51 ($\sim 1$~kpc), we find that the averaged spectral pattern of the full region of W51 (Figure~\ref{fig02}) is similar to the spectrum observed toward the spiral arm in the external galaxy M51, except for HNCO, CS, and SO.  This result indicates similar chemical compositions for the two sources.

Figure~\ref{fig08}(a) is a correlation diagram of fractional abundances of various molecules relative to H$_2$ between W51 and a spiral arm position of M51~P1 \citep{Watanabe2014}.  Here we employ the column densities derived under the LTE approximation both for W51 and M51 for simplicity.  Note that the elemental abundances of M51 is similar to that of the solar neighborhood \citep{Bresolin2004,Garnett2004}.   As expected from the spectra of these two sources (Figure~\ref{fig02}), the fractional abundances in W51 well correlate with those in the M51~P1  with a very small scatter (rms: 0.3).  The correlation coefficient in the log-scale is 0.97.  The correlation coefficient is similar to that between the sub-region C and the full region (Figure~\ref{fig10}d) and larger than that between the hot core and the full region (Figure~\ref{fig10}a), although these correlation coefficients are derived not in the fractional abundances but in the integrated intensity ratios.  Thus, the molecular-cloud-scale chemical composition of W51 is almost similar to the chemical composition observed toward the spiral arm in M51, although the size-scale of W51 is much smaller than that of M51.  On the other hand, slight differences can be seen in several molecules in Figure~\ref{fig08}(a).  For example, the abundance of HNCO is higher in the M51~P1 than the W51.  S-bearing species are found to be slightly more abundant in the W51, as seen in the Figure~\ref{fig02}.   However, the difference of the fractional abundances between W51 and M51 are within a factor of two.  \citet{Watanabe2014,Watanabe2016} reported that the star-formation activities do not significantly affect the chemical composition of the molecular gas at the scale of 0.3--1~kpc in M51 on the basis of the observation with the IRAM~30~m telescope and the CARMA.  They suggested that the observed spectral pattern would represent a chemical composition of an extended molecular gas in the spiral arm of M51.  The similarity between the averaged spectrum of W51 and the M51 spectrum supports their suggestion.

The similarity of chemical compositions between W51 and the spiral arm of M51 suggests that the chemical composition of molecular clouds in galactic disks would be similar among galaxies with similar elemental abundances, although this result has to be examined by more samples of molecular clouds in the Galaxy and the external galaxies.  In addition, the origin of the molecular-cloud-scale scale chemical composition should be investigated by chemical models.  We will discuss these points in the forthcoming papers.

\subsection{Comparison with AGN and Starburst}
Figures~\ref{fig08b}a and b are correlation diagrams of the fractional abundances relative to C$^{18}$O between the whole region of W51 and the starburst region in NGC~253 and between the whole region of W51 and the AGN in NGC~1068, respectively \citep{Aladro2015}.  The star formation rate of NGC~253 is estimated to be 3.6~M$_{\odot}$~yr$^{-1}$ within a few 100~pc area of the galactic center \citep{Karachentsev2004}, which is higher than the Galactic value of the whole disk by a factor of three \citep[e.g.][]{Murray2010,Robitaille2010}.  NGC~1068 is a prototypical Seyfert 2 type AGN.  The chemical compositions of a circumnuclear disk suggest effects of X-ray dominated regions \citep[e.g.][]{Krips2011,Garcia-Burillo2014,Nakajima2015}.  The elemental abundances of oxygen and nitrogen of both galaxies are similar to those of the solar neighborhood \citep[e.g.][]{Evans1987,Pilyugin2014}.  These diagrams show that the chemical compositions of the starburst region and the AGN roughly correlate with that of W51, although the correlation coefficients, which are 0.88 and 0.88 for NGC~253 and NGC~1068, respectively, are smaller than the M51 case (0.97).  A scatter in the correlation diagram may originate from the effects of the starburst region and the AGN.  It is noted that the chemical compositions in a central region of a galaxy would be affected by the star formation activities and/or the active galactic nuclei at a scale of a few 10~pc.  For example, complex organic molecules, which are expected to relate with the star formation activities, have been detected in various positions of the central molecular zone of our Galaxy with similar abundances \citep[e.g.][]{Requena2006,Requena2008}. 

Nevertheless, the overall correlations in the abundances are rather good, although the observation beams, which corresponds to 300~pc and 1.5~kpc at distances of 3.4~Mpc \citep[NGC~253:][]{Dalcanton2009} and 14.4~Mpc \citep[NGC~1068:][]{Bland-Hawthorn1997}, cover the central molecular zones of the galaxies and do not the spiral arm.  The result suggests that the effect of nuclear activities would be almost smeared out by that of the surrounding extended molecular gas, when they are observed by the large observation beam of $\sim 1$~kpc scale in the 3~mm band.  The effect of the large observation beam is revealed in the comparison of the chemical compositions between W51 and LIRG NGC~4418 at a spatial resolution of $\sim 330$~pc scale (Section~4.3).

\subsection{Comparison with LIRG NGC~4418}
Here, we compare W51 and the nuclear region of NGC~4418 to see differences of molecular-cloud-scale chemical compositions between the molecular cloud in the disk and a molecular clouds in the nuclear region at a similar spatial scale ($\sim 100$~pc).  NGC~4418 is a LIRG (infrared luminosity ($L_{\rm}$) of $\sim 10^{11}L_{\odot}$) at a distance of 34~Mpc \citep{Sakamoto2013}.  Toward the nuclear region of the galaxy, \citet{Costagliola2015} have conducted a spectral line scan with ALMA in Bands 3, 6, and 7.  Because an angular resolution of $\sim2''$ at the Band~3 corresponds to a spatial resolution of $\sim 330$~pc, we can compare the W51 and NGC~4418 at a roughly similar spatial scale.  Figure~\ref{fig08c} is a correlation diagram of the column densities derived under the LTE approximation relative to that of C$^{18}$O.  In contrast to the NGC~253 and NGC~1068 plots (Figure~\ref{fig08}a and b), Figure~\ref{fig08c} shows a larger scatter.  The rms value of the scatter for NGC~4418 (1.4) is larger than those of NGC~253 (0.6) and NGC~1068 (0.7).  One may think that these differences would come from the resolving-out effect of the radio interferometer, because the extended component, which is similar to the component mainly contributing to the full spectrum of W51, is resolved out in the observation of NGC~4418.  However, the missing flux of the NGC~4418 observation is evaluated to be small according to the comparison with the spectra obtained with the IRAM~30~m telescope \citep{Costagliola2015}.  Therefore, these differences would likely reflect the different chemical compositions between NGC~4418 and W51 in the full spectrum. 

The column density ratios of molecules are generally higher in NGC~4418 than in W51.  In particular, the ratios of H$^{13}$CN, $^{13}$CO$^+$, CH$_3$CN, c-C$_3$H$_2$, and HC$_3$N are higher in NGC~4418 than those in W51 by two orders of magnitudes.  We do not include HCN and HCO$^+$ in Figure~\ref{fig08c}, because these lines are optically thick, as discussed by \citet{Costagliola2015}.  \citet{Costagliola2015} pointed out the higher abundances of HC$_3$N and c-C$_3$H$_2$ as well as lower the abundance of CH$_3$OH in NGC~4418 than those in the other extragalactic and Galactic sources.  Our results are consistent with theirs. 

In W51, we find that the observed spectrum of the full region is dominated by the contribution from the extended quiescent molecular gas.  On the other hand, the observation of NGC~4418 traces a very compact ($< 5$~pc) gas with the large observation beam of $\sim 330$~pc \citep{Costagliola2015}.  In fact, high excitation lines of vibrationally excited HCN, HNC, and HC$_3$N are detected \citep[e.g.][]{Costagliola2010,Sakamoto2010}.  One possible explanation for the difference is that a fraction of the high temperature and high density region is much larger in NGC~4418 than in W51 due to the starburst/AGN activities.  In contrast, the vibrationally excited CH$_3$OH lines are detected only in the hot core e1/e2 ($\sim 0.05$~pc) in W51.  By comparing the NGC~4418 spectrum with the W51 spectrum, we can thus delineate the influence of the extreme environment on the chemical composition at a molecular cloud scale in NGC~4418.

\subsection{Comparison with the Large Magellanic Cloud (LMC)}
Figure~\ref{fig09} is a correlation diagram of the integrated intensities relative to the HCO$^+$ intensity between the {\rm full} region of W51 and the molecular cloud N44C in the LMC \citep{Nishimura2016}.  Since C$^{18}$O was not observed in N44C, we use HCO$^+$ as a reference molecule for the comparison.  N44C has an embedded high-mass young stellar object ST2 \citep{Shimonish2010}.  \citet{Nishimura2016} carried out a spectral line survey toward N44C with the Mopra telescope in the 3~mm band.  The beam size of their observation ($38''$) corresponds to $\sim 10$~pc at the distance of LMC \citep[49.97~kpc:][]{Pietrzynski2013}.  Therefore, the spectral pattern of their observation represents a molecular-cloud scale chemical composition.  Figure~\ref{fig09} shows that emission lines of the N-bearing species (HCN, HNC, CN, and N$_2$H$^+$), C$^{18}$O, and CH$_3$OH are relatively weaker in N44C than W51, indicating deficiency of these molecules.  As discussed by \citet{Nishimura2016}, the deficiency of the N-bearing species are due to the low elemental abundance of nitrogen in the LMC.  The deficiency of CH$_3$OH would be due to warmer dust temperature caused by strong UV radiation, which reduces the production efficiency of CH$_3$OH on dust grains \citep{Nishimura2016,Shimonishi2016}.

\section{Summary}
We carried out a mapping spectral line survey toward the Galactic giant molecular cloud W51 in the 3~mm band with the Mopra 22~m telescope.  Our mapping observation approximately covers the $39\,{\rm pc} \times 47\,{\rm pc}$ area of the W51 molecular cloud in the frequency ranges of 85.1 - 101.1~GHz and 107.0 - 114.9~GHz.  The main results are summarized as follows:

\noindent (1) We prepare the spectrum averaged over all the observed area, in which we identify 12 molecular species and 4 additional isopopologues.  All the identified molecules are fundamental molecular species, which consist of four heavy atoms or less.

\noindent (2) The intensity pattern of the averaged spectrum is different from the spectrum of the hot core W51~e1/e2.  The hydrogen recombination lines, the CH$_3$OH lines with higher upper state energies ($> 100$~K), and the lines of complex organic molecules are not detected in the averaged spectrum.  These emission lines from the hot core are thought to be heavily diluted in the averaged spectrum. 
 
\noindent (3) We classify the observed area into 5 sub-regions according to the integrated intensity of $^{13}$CO ($I_{\rm ^{13}CO}$), and calculate the fractional flux of 11 molecules of each sub-region relative to the total flux.  For most of the observed molecules, 50~\% or more of the flux come from the region with the range of $I_{\rm ^{13}CO}$ from 25~K~km~s$^{-1}$ to 100~K~km~s$^{-1}$, which does not involve active star forming regions.  This analysis clearly shows that the molecular-cloud-scale chemical composition mainly represents the chemical composition of an extended molecular gas.

\noindent (4) The spectrum averaged over all observed area is similar to the spectra of the spiral arm in the external galaxy M51 observed with a large telescope beam of $\sim 1$~kpc.  The chemical composition of W51 is found to be similar to that of the spiral arm region of M51.  Thus, the observation of M51 in the 3~mm band traces the molecular-cloud-scale chemical composition.

\noindent (5) The molecular abundances in the AGN and starburst based on the 3~mm observation roughly correlate with those in W51, although high density, high temperature, and extreme radiation field are expected in these objects.  These results suggest that the contribution of the nuclear region would mostly be smeared out by an extended molecular gas around the nucleus in the 3~mm band observations.  Moreover, it is striking that the molecular-cloud scale chemical compositions are similar to each other among the three galaxies, i.e. the Galaxy, NGC~253, and NGC~1068, which have similar elemental abundances.

\noindent (5) The molecular abundances of the LIRG, NGC~4418, observed at a 330~pc scale with ALMA are rather different from those of W51, which would reflect the extreme environment of the nuclear region.


%% Included in this acknowledgments section are examples of the

%% Included in this acknowledgments section are examples of the
%% AASTeX hypertext markup commands. Use \url without the optional [HREF]
%% argument when you want to print the url directly in the text. Otherwise,
%% use either \url or \anchor, with the HREF as the first argument and the
%% text to be printed in the second.

\acknowledgments
The authors thank the anonymous reviewer for many helpful comments and suggestions.  The authors are grateful to the Mopra staff for their excellent support.  The Mopra radio telescope is operated by funding from the National Astronomical Observatory of Japan, the University of New South Wales, the University of Adelaide, and the Commonwealth of Australia through CSIRO.  This study is supported by a Grant-in-Aid from the Ministry of Education, Culture, Sports, Science, and Technology of Japan (No. 25108005 and 16K17657).


%% in parentheses, after the keyword, but they will not be verified.

{\it Facilities:} \facility{Mopra 22~m}.

%%%%%%%%%%%%%%%%%%%%%%%%
%% The reference list %%
%%%%%%%%%%%%%%%%%%%%%%%%

\clearpage

%%%%%%%%%%%%%
%% Figures %%
%%%%%%%%%%%%%

\begin{figure*}
\epsscale{1.99}
\plotone{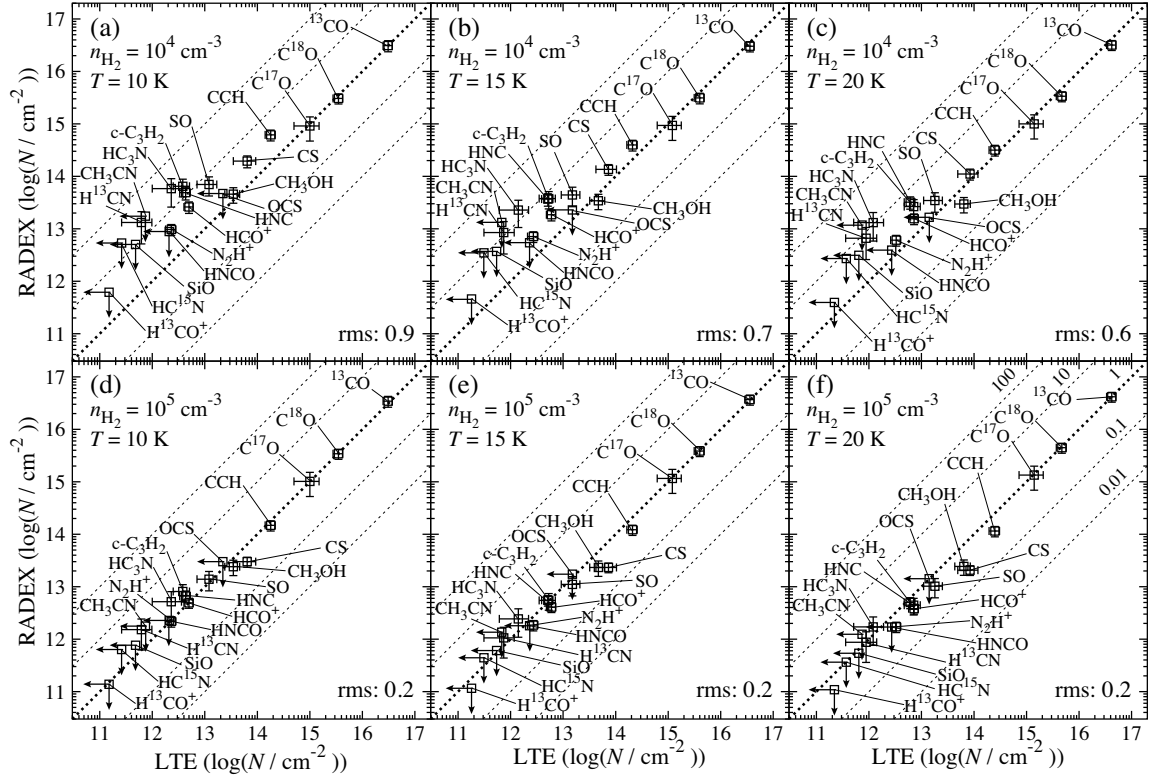}
\caption{Correlation plots between the column densities derived under the LTE approximation and those by the RADEX.  The assumed number density of molecular hydrogen ($n_{\rm H_2}$) and temperature ($T$) are shown in the left top corner of each panel.  Thin dotted lines indicate the ratios of the two axis of 0.01, 0.1, 10 and 100, while thick dotted lines indicate the ratio of 1.  The rms values are calculated as $(1/n \sum_{i}( \log(N_{i,{\rm LTE}}) - \log(N_{i,{\rm RADEX}}))^2)^{1/2}$, where $n$, $N_{i,{\rm LTE}}$, and $X_{i,{\rm RADEX}}$ are the number of molecular species, the column density derived under the LTE approximation, and the column density derived by the RADEX, respectively. }
\label{fig12}
\end{figure*}

\begin{figure*}
\epsscale{1.00}
\plotone{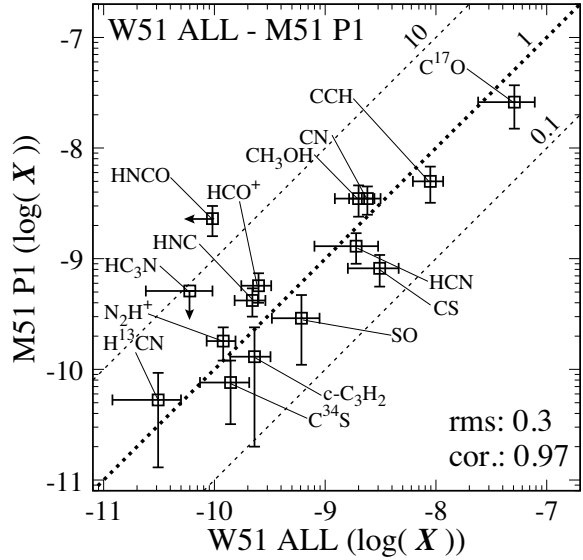}
\caption{Plots of fractional abundances between the whole observed area in W51 and a spiral arm of M51~P1 \citep{Watanabe2014}.  Fractional abundances are relative to H$_2$.  The fractional abundances of W51 are estimated with a rotation temperature of 15~K under the LTE approximation.  The fractional abundances of M51~P1 are estimated by \citet{Watanabe2014} under assumptions of a source size of $10''$, a rotation temperature of 5~K, and the LTE condition.  Dashed lines indicate the fractional abundance ratios of 10, 1, and 0.1.  The rms value is calculated as $(1/n \sum_{i}(\log(X_{i,{\rm M51\,P1}}) - \log(X_{i,{\rm ALL}}))^2)^{1/2}$, where $n$, $X_{i,{\rm M51\,P1}}$, and $X_{i,{\rm ALL}}$ are the number of molecular species, the fractional abundance of a particular molecular species in M51~P1, and the fractional of molecule in the spectrum averaged over all the region of W51, respectively.  The correlation coefficient between $\log(X_{i,{\rm M51\,P1}}$ and $\log(X_{i,{\rm ALL}})$ is evaluated and shown in the bottom-right corner (cor.). } 
\label{fig08}
\end{figure*}

\begin{figure*}
\epsscale{1.99}
\plotone{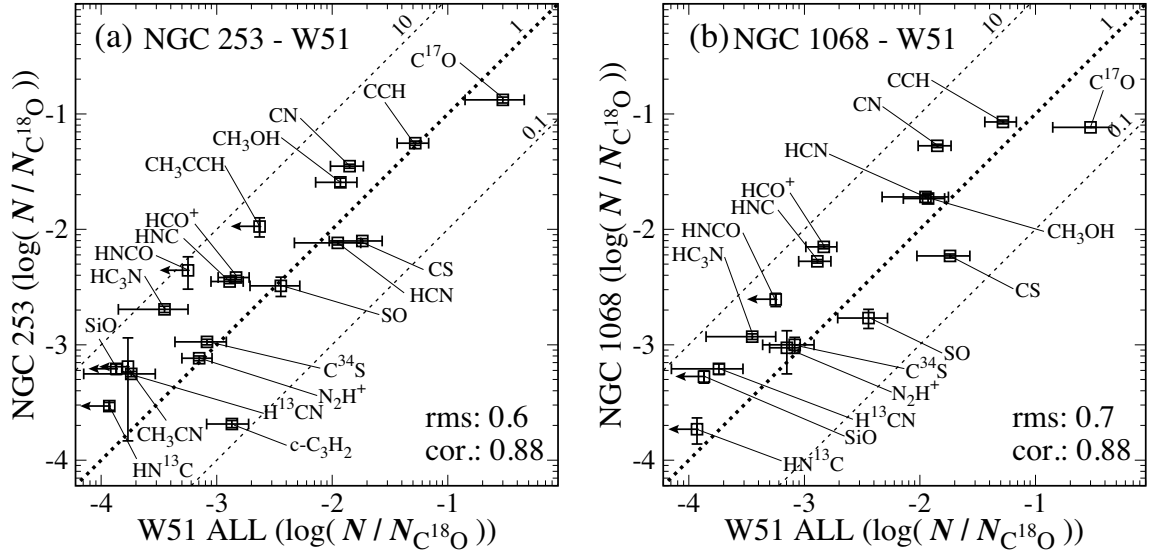}
\caption{Plots of fractional abundances between the whole observed area in W51 and nearby external galaxies: (a) a starburst in NGC~253 and (b) an AGN in NGC~1068.  Fractional abundances are relative to the column density of C$^{18}$O.  The fractional abundances of W51 are estimated by using the rotation temperature of 15~K under the LTE approximation.  The data for NGC~253 and NGC~1068 are taken from \citet{Aladro2015}.   Dashed lines indicate the fractional abundance ratios of 10, 1, and 0.1.  The rms values are calculated as $(1/n \sum_{i}( \log(N_{i}/N_{\rm C^{18}O}) - \log(N_{i,{\rm ALL}}/N_{\rm ^{13}CO, ALL}))^2)^{1/2}$, where $n$, $N_{i}$, $N_{\rm ^C{18}O}$, $N_{i,{\rm ALL}}$, and $N_{\rm C^{18}O, ALL}$ are the number of molecular species, the column density of a particular molecular species in NGC~253 or NGC~1068, the column density of C$^{18}$O in NGC~253 or NGC~1068, the column density of molecule in the spectrum averaged over all the region of W51, and the column density of C$^{18}$O in the spectrum averaged over all the region of W51, respectively.  The correlation coefficients between $\log(N_{i}/N_{\rm C^{18}O})$ and $\log(N_{i,{\rm ALL}}/N_{\rm C^{18}O, ALL})$ are evaluated and shown in the panels (cor.).} 
\label{fig08b}
\end{figure*}

\begin{figure*}
\epsscale{1.00}
\plotone{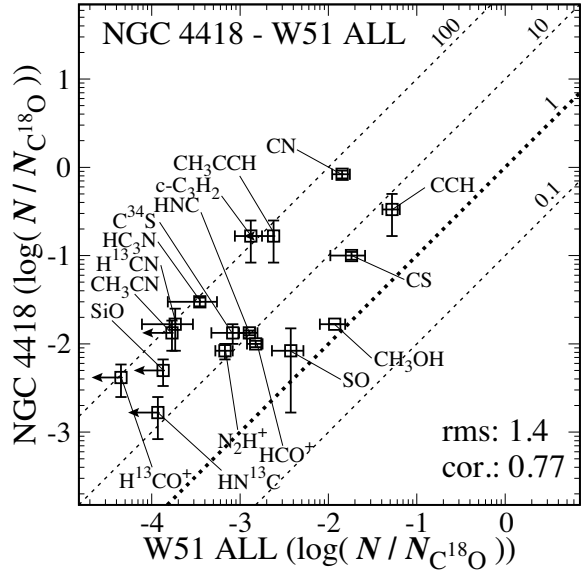}
\caption{Plots of intensity ratios relative to that of C$^{18}$O between the full region of W51 and the nuclear region of NGC~4418 \citep{Costagliola2015}.  The data for NGC~4418 are taken from \citep{Costagliola2015}.  Dashed lines indicate the ratio of column density ratios of 100, 10, 1, and 0.1.  The rms values are calculated as $(1/n \sum_{i}(\log(N_{i,{\rm NGC\,4418}}/N_{\rm C^{18}O, NGC\,4418} ) - \log(N_{i,{\rm ALL}}/N_{\rm C^{18}O, ALL}))^2)^{1/2}$, where $n$, $N_{i,{\rm NGC\,4418}}$, $N_{\rm C^{18}O, NGC\,4418}$, $N_{i,{\rm ALL}}$, and $N_{\rm C^{18}O, ALL}$ are the number of molecular species, the column density of a particular molecular species in NGC~4418, the column density of C$^{18}$O in NGC~4418, the column density of molecule in the spectrum averaged over all the region of W51, and the column density of C$^{18}$O in the spectrum averaged over all the region of W51.  The correlation coefficient between $\log(N_{i,{\rm NGC\,4418}}/N_{\rm C^{18}O, NGC\,4418})$ and $\log(N_{i,{\rm ALL}}/N_{\rm C^{18}O, ALL})$ is evaluated and shown in the bottom-right corner (cor.). } 
\label{fig08c}
\end{figure*}

\begin{figure*}
\epsscale{1.00}
\plotone{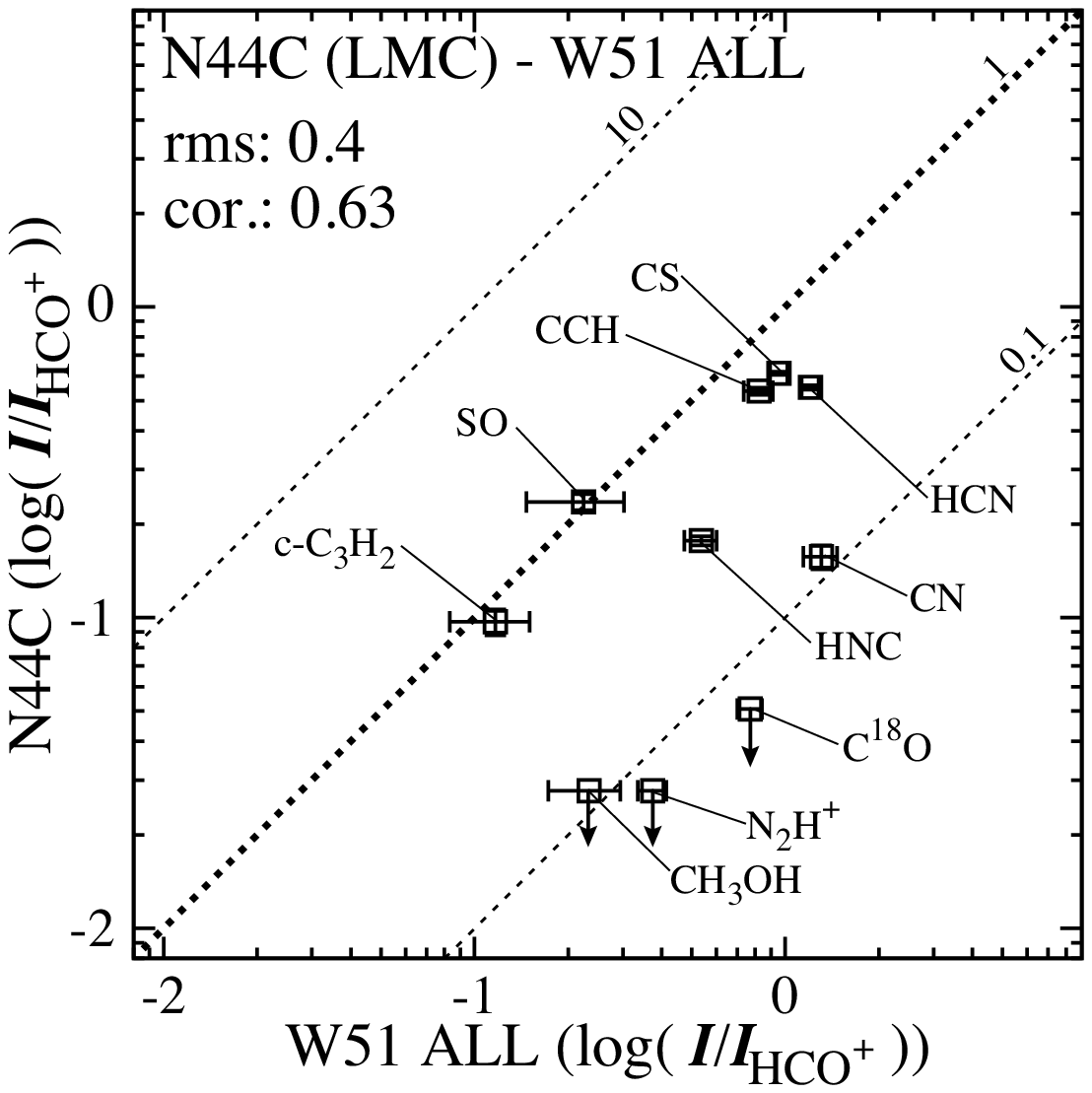}
\caption{Plots of intensity ratios relative to that of HCO$^+$ between the whole observed region of W51 and N44C in the LMC.  The data for N44C are taken from \citet{Nishimura2016}.   Dashed lines indicate the ratio of intensity ratios of 10, 1, and 0.1.  The rms value is calculated as $(1/n \sum_{i}(\log(I_{i,{\rm N44C}}/I_{\rm HCO^+, N44C} ) - \log(I_{i,{\rm ALL}}/I_{\rm HCO^+, ALL}))^2)^{1/2}$, where $n$, $I_{i,{\rm N44C}}$, $I_{\rm HCO^+, N44C}$, $I_{i,{\rm ALL}}$, and $I_{\rm HCO^+, ALL}$ are the number of molecular species, the integrated intensities of a particular molecular species in N44C, the integrated intensity of HCO$^+$ in N44C, the integrated intensities of molecule in the spectrum averaged over all the region of W51, and the integrated intensity of HCO$^{+}$ in the spectrum averaged over all the region of W51, respectively.  The correlation coefficient between $\log(I_{i,{\rm N44C}}/I_{\rm HCO^{+}, N44C})$ and $\log(I_{i,{\rm ALL}}/I_{\rm HCO^{+}, ALL})$ is evaluated and shown in the top-left corner (cor.). } 
\label{fig09}
\end{figure*}

%%%%%%%%%%%%
%% Tables %%
%%%%%%%%%%%%

\begin{table*}[h]
\caption{Summary of Observations \label{tab01}}
\begin{tabular}{ccccc}
\tableline\tableline
Setting     & Frequency\tablenotemark{a} & Frequency range\tablenotemark{b} & Beam Size & Sensitivity \tablenotemark{c} \\
            & (GHz)                      & (GHz) & (arcsec) & (K) \\
\tableline
1 &  89.240  &  85.096 -- 93.397  & 38 & 0.14 -- 0.5 \\
2 &  97.050  &  92.930 -- 101.100 & 35 & 0.16 -- 0.5 \\
3 & 110.800  & 106.980 -- 114.900 & 31 & 0.3 -- 0.7 \\
\tableline
\end{tabular}
\tablenotetext{a}{Central frequency of the MOPS spectrometer.}
\tablenotetext{b}{Observed frequency ranges.}
\tablenotetext{c}{Rms noise level in the $T_{\rm mb}$ scale at the frequency resolution of 0.27~MHz.}
\end{table*}

\begin{deluxetable}{lrlrrrrrr}
\tabletypesize{\tiny}
\tablecolumns{9}
\tablewidth{0pt}
\tablecaption{Line parameters of the averaged spectrum \tablenotemark{a} \label{tab02}}
\tablehead{
\colhead{Name} & \colhead{Frequency} & \colhead{Transition} & \colhead{$E_{\rm u}$} & \colhead{S$\mu^2$} & \colhead{$T_{\rm mb}$ Peak \tablenotemark{b}} & \colhead{$\int T_{\rm mb} dv$ \tablenotemark{b,c}} & \colhead{$V_{\rm LSR}$\tablenotemark{d}} & \colhead{FWHM\tablenotemark{d}} \\
\colhead{} & \colhead{(GHz)} & \colhead{} & \colhead{(K)} & \colhead{(Debye$^2$)} & \colhead{(K)} & \colhead{(K~km~s$^{-1}$)} & \colhead{(km~s$^{-1}$)} &\colhead{(km~s$^{-1}$)} 
}

\startdata
c-C$_3$H$_2$   & 85.338894 &$2_{1\,2}-1_{0\,1}$               &  6.4 &  16.1 & 0.03 (0.01) &  0.5 (0.2) & 59.0 (0.9) & 18 (2)     \\
CH$_3$CCH      & 85.457300 &$5_0-4_0$                         & 12.3 &  6.14 & $< 0.01$    & $< 0.2 $   & \nodata    & \nodata    \\
H42$\alpha$    & 85.688390 &                                  &      &       & $< 0.02 $   & $< 0.2 $   & \nodata    & \nodata    \\
HC$^{15}$N     & 86.054966 &$1-0$                             &  4.1 &  8.91 & $< 0.02 $   & $< 0.1 $   & \nodata    & \nodata    \\
SO             & 86.093950 &$J_N=2_2-1_1$                     & 19.3 &  3.53 & $< 0.02 $   & $< 0.2 $   & \nodata    & \nodata    \\
H$^{13}$CN     & 86.339922 &$1-0$                             &  4.1 &  8.91 & 0.02 (0.01) &  0.3 (0.2) & 57 (1)     & 19 (3)     \\
H$^{13}$CO$^+$ & 86.754288 &$1-0$                             &  4.2 &  15.2 & $< 0.02 $   &  $< 0.1 $  & \nodata    & \nodata    \\
SiO            & 86.846960 &$2-1$                             &  6.3 &  19.2 & $< 0.02 $   &  $< 0.2 $  & \nodata    & \nodata    \\
HN$^{13}$C     & 87.090825 &$1-0$                             &  4.2 &  9.30 & $< 0.02 $   &  $< 0.2 $  & \nodata    & \nodata    \\
CCH \tablenotemark{e} & 87.316898 &$N=1-0,J=3/2-1/2,F=2-1$    &  4.2 &  0.99 & 0.10 (0.02) &  3.7 (0.4) & ...\tablenotemark{f} & ...\tablenotemark{f} \\
CCH \tablenotemark{e} & 87.328585 &$N=1-0,J=3/2-1/2,F=1-0$    &  4.2 &  0.49 & 0.07 (0.02) &  \nodata   & \nodata    & \nodata    \\
CCH \tablenotemark{e} & 87.401989 &$N=1-0,J=1/2-1/2,F=1-1$    &  4.2 &  0.49 & 0.04 (0.02) &  0.7 (0.3) & ...\tablenotemark{f} & ...\tablenotemark{f} \\
CCH \tablenotemark{e} & 87.407165 &$N=1-0,J=1/2-1/2,F=0-1$    &  4.2 &  0.20 & \nodata     &  \nodata   & \nodata    & \nodata    \\
CCH            & 87.446470 &$N=1-0,J=1/2-1/2,F=1-0$           &  4.2 &  0.10 & 0.03 (0.02) &  0.4 (0.3) & 60 (2)     & 17 (4)     \\
HNCO           & 87.925237 &$4_{0\,4}-3_{0\,3}$               & 10.5 &  10.0 & $< 0.016$   &  $< 0.17$  & \nodata    & \nodata    \\
HCN            & 88.631602 &$1-0$                             &  4.3 &  8.91 & 0.25 (0.02) &  5.5 (0.2) & 58.1 (0.2) & 20.0 (0.6) \\
HCO$^+$        & 89.188525 &$1-0$                             &  4.3 &  15.2 & 0.28 (0.02) &  4.5 (0.2) & 57.9 (0.2) & 16.0 (0.5) \\
HNC            & 90.663568 &$1-0$                             &  4.4 &  9.30 & 0.15 (0.02) &  2.4 (0.3) & 58.9 (0.4) & 16.6 (0.9) \\
HC$_3$N        & 90.979023 &$10-9$                            & 24.0 & 139.3 & 0.02 (0.01) &  0.3 (0.2) & 59 (1)     & 16 (3)     \\
CH$_3$CN       & 91.987088 &$5_0-4_0$                         & 13.2 & 153.8 & $< 0.01 $   &  $< 0.1 $  & \nodata    & \nodata    \\
H41$\alpha$    & 92.034430 &                                  &      &       & $< 0.02 $   &  $< 0.3 $  & \nodata    & \nodata    \\
$^{13}$CS      & 92.494308 &$2-1$                             &  6.7 &  15.3 & $< 0.02 $   &  $< 0.2 $  & \nodata    & \nodata    \\
N$_2$H$^+$     & 93.173392 &$1-0$                             &  4.5 & 104.1 & 0.08 (0.01) &  1.7 (0.2) & 61.6 (0.3) & 19.4 (0.7) \\
CH$_3$OH       & 95.169463 &$8_0-7_1,{\rm A}^+$               & 83.6 &  7.22 & $< 0.03 $   &  $< 0.3 $  & \nodata    & \nodata    \\
CH$_3$OH       & 95.914309 &$2_1-1_1,{\rm A}^+$               & 21.4 &  1.21 & $< 0.03 $   &  $< 0.2 $  & \nodata    & \nodata    \\
C$^{34}$S      & 96.412950 &$2-1$                             &  6.9 &  7.67 & 0.03 (0.02) &  0.6 (0.2) & 59 (1)     & 18 (2)     \\
CH$_3$OH \tablenotemark{e} & 96.739362 &$2_{-1}-1_{-1},{E}$   &  4.6 &  1.21 & 0.05 (0.02) &  1.1 (0.3) & 65.7 (0.7) & 21 (2)     \\
CH$_3$OH \tablenotemark{e} & 96.741375 &$2_0-1_0,{\rm A}^+$   &  7.0 &  1.62 & \nodata     &  \nodata   & \nodata    & \nodata    \\
CH$_3$OH \tablenotemark{e} & 96.744550 &$2_{0}-1_{0},{E}$     & 12.2 &  1.62 & \nodata     &  \nodata   & \nodata    & \nodata    \\
C$^{33}$S      & 97.171840 &$2-1$                             &  7.0 &  30.8 & $< 0.02$    &  $< 0.2$   & \nodata    & \nodata    \\
OCS            & 97.301209 &$8-7$                             & 21.0 &  4.09 & $< 0.02$    &  $< 0.2$   & \nodata    & \nodata    \\
CH$_3$OH       & 97.582804 &$2_1-1_1, {\rm A}^-$              & 21.6 &  1.21 & $< 0.02$    &  $< 0.2$   & \nodata    & \nodata    \\
CS             & 97.980953 &$2-1$                             &  7.1 &  7.67 & 0.28 (0.01) &  4.3 (0.2) & 60.1 (0.3) & 15.4 (0.6) \\
H40$\alpha$    & 99.022950 &                                  &      &       & $< 0.029$   &  $< 0.37$  & \nodata    & \nodata    \\
SO             & 99.299870 &$J_N=3_2-2_1$                     &  9.2 &  6.91 & 0.08 (0.04) &  1.0 (0.4) & 60.1 (0.8) & 15 (2)     \\
HC$_3$N        &100.076392 &$11-10$                           & 28.8 & 153.2 & $<0.02$     &  $<0.2$    & \nodata    & \nodata    \\
HC$_3$N        &109.173634 &$12-11$                           & 34.1 & 167.1 & $< 0.02$    &  $< 0.2$   & \nodata    & \nodata    \\
SO             &109.252220 &$J_N=2_3-1_3$                     & 21.1 &  3.56 & $< 0.03$    &  $< 0.2$   & \nodata    & \nodata    \\  
OCS            &109.463063 &$9-8$                             & 26.3 &  4.60 & $< 0.03$    &  $< 0.2$   & \nodata    & \nodata    \\  
C$^{18}$O      &109.782173 &$1-0$                             &  5.3 & 0.012 & 0.21 (0.02) &  3.5 (0.3) & 58.4 (0.5) & 18 (1)     \\
HNCO           &109.905749 &$5_{0\,5}-4_{0\,4}$               & 15.8 &  12.5 & $< 0.03$    &  $< 0.3$   & \nodata    & \nodata    \\  
$^{13}$CO      &110.201354 &$1-0$                             &  5.3 & 0.012 & 1.83 (0.06) & 31.3 (0.7) & 59.0 (0.3) & 18.1 (0.7) \\
C$^{17}$O      &112.359284 &$1-0$                             &  5.4 & 0.012 & 0.07 (0.05) &  1.1 (0.5) & 58 (2)     & 18 (4)     \\
CN \tablenotemark{e} &113.123370 &$N=1-0,J=1/2-1/2,F=1/2-1/2$ &  5.4 &  0.15 & 0.06 (0.04) & 11 (1)     & ...\tablenotemark{f} & ...\tablenotemark{f} \\
CN \tablenotemark{e} &113.144157 &$N=1-0,J=1/2-1/2,F=1/2-3/2$ &  5.4 &  1.25 & 0.10 (0.04) & \nodata    & \nodata    & \nodata    \\
CN \tablenotemark{e} &113.170492 &$N=1-0,J=1/2-1/2,F=3/2-1/2$ &  5.4 &  1.22 & 0.11 (0.04) & \nodata    & \nodata    & \nodata    \\
CN \tablenotemark{e} &113.191279 &$N=1-0,J=1/2-1/2,F=3/2-3/2$ &  5.4 &  1.58 & 0.11 (0.04) & \nodata    & \nodata    & \nodata    \\
CN \tablenotemark{e} &113.490970 &$N=1-0,J=3/2-1/2,F=5/2-3/2$ &  5.4 &  4.20 & 0.20 (0.04) &  5.9 (0.7) & ...\tablenotemark{f} & ...\tablenotemark{f} \\
CN \tablenotemark{e} &113.488120 &$N=1-0,J=3/2-1/2,F=3/2-1/2$ &  5.4 &  1.58 & \nodata     & \nodata    & \nodata    & \nodata    \\
CN \tablenotemark{e} &113.499644 &$N=1-0,J=3/2-1/2,F=1/2-1/2$ &  5.4 &  1.25 & 0.08 (0.04) & \nodata    & \nodata    & \nodata    \\
\enddata

\tablenotetext{a}{Upper limits to the peak temperature and the integrated intensities are given for the lines detected in the sub-region A but not in the full region (see Table~\ref{tab05}).}
\tablenotetext{b}{The numbers in parentheses represent $3\sigma$ errors.}
\tablenotetext{c}{The upper limit to the integrated intensity is calculated as: $\int T_{\rm mb} dv < 3 \sigma \times \sqrt{\Delta V \times \Delta v_{\rm res}}$, where $\Delta V$ is the assumed line width (40~km~s$^{-1}$) and $\Delta v_{\rm res}$ is the velocity resolution per channel.}
\tablenotetext{d}{The numbers in parentheses represent $1\sigma$ errors.}
\tablenotetext{e}{The line is blended with other lines.}
\tablenotetext{f}{Gaussian fitting is not successful due to blending with other lines.}
\end{deluxetable}

\begin{table}
\caption{Definition of sub-regions and detected molecules \label{tab04}}
\begin{tabular}{|l|c|c|c|c|c|c|}
\tableline\tableline
     & \multicolumn{6}{|c|}{Sub-region Name} \\ \cline{2-7}
     & A & B & C & D & E & ALL \\ \tableline
$I_{\rm ^{13}CO}$ range (K~km~s$^{-1}$) $^{\rm a}$ & $> 200$ & $200 - 100$ & $100 - 50$ & $50 - 25$ & $25 >$ & -- \\
\tableline
H42$\alpha$    & Y & Y & Y & N & N & N \\ 
CCH            & Y & Y & Y & Y & Y & Y \\ 
CN             & Y & Y & Y & Y & Y & Y \\ 
HCN            & Y & Y & Y & Y & Y & Y \\ 
H$^{13}$CN     & Y & Y & Y & N & N & Y \\ 
HC$^{15}$N     & Y & Y & N & N & N & N \\ 
HNC            & Y & Y & Y & Y & Y & Y \\ 
HN$^{13}$C     & Y & Y & Y & Y & N & N \\ 
N$_2$H$^+$     & Y & Y & Y & Y & Y & Y \\ 
$^{13}$CO      & Y & Y & Y & Y & Y & Y \\ 
C$^{17}$O      & Y & Y & Y & Y & N & Y \\ 
C$^{18}$O      & Y & Y & Y & Y & Y & Y \\ 
HCO$^+$        & Y & Y & Y & Y & Y & Y \\ 
H$^{13}$CO$^+$ & Y & Y & Y & N & N & N \\ 
CH$_3$OH       & Y & Y & Y & Y & Y & Y \\ 
c-C$_3$H$_2$   & Y & Y & Y & Y & Y & Y \\ 
CH$_3$CCH      & Y & Y & Y & N & N & N \\ 
CH$_3$CN       & Y & Y & N & N & N & N \\ 
HNCO           & Y & N & N & N & N & N \\ 
SiO            & Y & Y & N & N & N & N \\ 
CS             & Y & Y & Y & Y & Y & Y \\ 
$^{13}$CS      & Y & Y & N & N & N & N \\ 
C$^{33}$S      & Y & N & N & N & N & N \\ 
C$^{34}$S      & Y & Y & Y & N & N & Y \\ 
SO             & Y & Y & Y & Y & N & Y \\ 
HC$_3$N        & Y & Y & N & N & N & Y \\ 
OCS            & Y & N & N & N & N & N \\ \tableline
\end{tabular}
\tablenotetext{a}{The integrated intensity range of $^{13}$CO($J=1-0$).}
\end{table}

\begin{deluxetable}{lrrrrrrrr}
%\rotate
\tabletypesize{\tiny}
\tablecolumns{9}
%\tablewidth{500pt}
\tablecaption{Integrated intensities of molecular lines in the averaged spectra for the sub-regions \label{tab05}}
\tablehead{
\colhead{Molecule} & \colhead{Frequency} & \colhead{E$_{\rm u}$} & \colhead{S$\mu^2$} & \multicolumn{5}{c}{Integrated Intensity in Sub-regions \tablenotemark{a}} \\
\colhead{} & \colhead{} &\colhead{} & \colhead{} &  \colhead{A} & \colhead{B} & \colhead{C} & \colhead{D} & \colhead{E} \\
\colhead{} & \colhead{(GHz)} & \colhead{(K)} & \colhead{(Debye$^2$)} & \colhead{(K~km~s$^{-1}$)} & \colhead{(K~km~s$^{-1}$)} & \colhead{(K~km~s$^{-1}$)} & \colhead{(K~km~s$^{-1}$)} & \colhead{(K~km~s$^{-1}$)} 
}
\startdata
c-C$_3$H$_2$   & 85.338894 &  6.4 &  16.1  & 5 (1)        & 2.5 (0.7)    & 1.0 (0.4)    & 0.5 (0.2)    & 0.3 (0.2)    \\
CH$_3$CCH      & 85.457300 & 12.3 &  6.15  & 5 (1)        & 1.5 (0.7)    & 0.5 (0.4)    & $< 0.34$     & $< 0.2$      \\
H42$\alpha$    & 85.688390 &      &        & 8 (1)        & 2.7 (0.7)    & 0.6 (0.5)    & $< 0.3$      & $< 0.3$      \\
HC$^{15}$N     & 86.054966 &  4.1 &  8.91  & 1.7 (0.7)    & 0.5 (0.5)    & $< 0.3$      & $< 0.2$      & $< 0.2$      \\
SO             & 86.093950 & 19.3 &  3.53  & 3.1 (0.8)    & 0.8 (0.5)    & $< 0.4$      & $< 0.3$      & $< 0.2$      \\
H$^{13}$CN     & 86.339922 &  4.1 &  8.91  & 9.6 (0.9)    & 3.0 (0.7)    & 0.6 (0.3)    & $< 0.2$      & $< 0.2$      \\
H$^{13}$CO$^+$ & 86.754288 &  4.2 &  15.2  & 7.1 (0.9)    & 2.1 (0.7)    & 0.4 (0.4)    & $< 0.2$      & $< 0.1$      \\
SiO            & 86.846960 &  6.3 &  19.2  & 8 (1)        & 2.0 (0.7)    & $< 0.5$      & $< 0.3$      & $< 0.3$      \\
HN$^{13}$C     & 87.090825 &  4.2 &  9.30  & 3 (1)        & 1.3 (0.7)    & 0.6 (0.4)    & 0.5 (0.3)    & $< 0.2$      \\
CCH \tablenotemark{c} & 87.316898 &  4.2 &  0.99  & 27 (2)       & 13 (1)       & 7 (1)        & 3.6 (0.7)    & 2.0 (0.4)    \\
CCH \tablenotemark{c} & 87.328585 &  4.2 &  0.49  & ...          & ...          & ...          & ...          & ...          \\
CCH \tablenotemark{c} & 87.401989 &  4.2 &  0.49  & 9 (1)        & 4.5 (0.9)    & 1.6 (0.6)    & 0.5 (0.5)    & $< 0.3$      \\
CCH \tablenotemark{c} & 87.407165 &  4.2 &  0.20  & ...          & ...          & ...          & ...          & ...          \\
CCH            & 87.446470 &  4.2 &  0.10  & $< 2$        & $< 1$        & 1.0 (0.6)    & 0.4 (0.4)    & $< 0.3$      \\
HNCO           & 87.925237 & 10.6 &  10.0  & 1.9 (0.8)    & $< 0.7$      & $< 0.5$      & $< 0.3$      & $< 0.2$      \\
HCN            & 88.631602 &  4.3 &  8.91  & 62 (1)       & 30.0 (0.6)   & 11.8 (0.5)   & 5.2 (0.3)    & 2.1 (0.3)    \\
HCO$^+$        & 89.188525 &  4.3 &  15.2  & 54 (2)       & 23.4 (0.6)   & 9.3 (0.5)    & 4.5 (0.3)    & 1.8 (0.2)    \\
HNC            & 90.663568 &  4.4 &  9.30  & 40 (1)       & 16.3 (0.6)   & 5.6 (0.6)    & 2.2 (0.3)    & 0.6 (0.3)    \\
HC$_3$N        & 90.979023 & 24.0 & 139.3  & 8 (1)        & 2.4 (0.63)   & $< 0.4$      & $< 0.2$      & $< 0.2$      \\
CH$_3$CN       & 91.987088 & 13.2 & 153.8  & 4.6 (1.2)    & 0.8 (0.5)    & $< 0.3$      & $< 0.3$      & $< 0.2$      \\
H41$\alpha$    & 92.034430 &      &        & 8.0 (1.2)    & 1.8 (0.7)    & $< 0.5$      & $< 0.4$      & $< 0.3$      \\
$^{13}$CS      & 92.494308 &  6.7 &  15.3  & 4.70 (0.90)  & 1.3 (0.6)    & $< 0.4$      & $< 0.3$      & $< 0.2$      \\
N$_2$H$^+$     & 93.173392 &  4.5 & 111.8  & 32.95 (0.75) & 13.3 (0.3)   & 3.7 (0.3)    & 1.3 (0.2)    & 0.4 (0.2)    \\
CH$_3$OH       & 95.169463 & 83.6 &  7.22  & 3 (1)        & $< 0.9$      & $< 0.5$      & $< 0.4$      & $< 0.4$      \\
CH$_3$OH       & 95.914309 & 21.4 &  1.21  & 1.2 (0.9)    & $< 0.5$      & $< 0.4$      & $< 0.3$      & $< 0.2$      \\
C$^{34}$S      & 96.412950 &  6.9 &  7.67  & 10 (1)       & 3.0 (0.8)    & 0.9 (0.4)    & $< 0.4$      & 0.5 (0.2)    \\
CH$_3$OH \tablenotemark{c} & 96.741375 &  7.0 &  1.62  & 17 (1)       & 4.5 (0.6)    & 1.3 (0.4)    & 0.9 (0.4)    & 0.7 (0.3)    \\
CH$_3$OH \tablenotemark{c} & 96.739362 &  4.6 &  1.21  & ...          & ...          & ...          & ...          & ...          \\
CH$_3$OH \tablenotemark{c} & 96.744550 & 13.6 &  1.62  & ...          & ...          & ...          & ...          & ...          \\
C$^{33}$S      & 97.171840 &  7.0 &  30.3  & 2.0 (0.9)    & $< 0.5$      & $< 0.4$      & $< 0.3$      & $< 0.2$      \\
OCS            & 97.301209 & 21.0 &  4.09  & 1.1 (0.9)    & $< 0.6$      & $< 0.2$      & $< 0.2$      & $< 0.2$      \\
CH$_3$OH       & 97.582804 & 21.6 &  1.21  & 3 (1)        & $< 0.9$      & $< 0.3$      & $< 0.4$      & $< 0.3$      \\
CS             & 97.980953 &  7.1 &  7.67  & 79 (1)       & 28.6 (0.7)   & 9.2 (0.3)    & 3.7 (0.3)    & 1.3 (0.2)    \\
H40$\alpha$    & 99.022950 &      &        & 6 (2)        & 2 (2)        & $< 0.7$      & $< 0.6$      & $< 0.4$      \\
SO             & 99.299870 &  9.2 &  6.91  & 25 (2)       & 8 (1)        & 2.3 (0.9)    & 0.9 (0.4)    & $< 0.5$      \\
HC$_3$N        &100.076392 & 28.8 & 153.2  & 5 (1)        & 0.7 (0.6)    & $< 0.4$      & $< 0.3$      & $< 0.2$      \\
HC$_3$N        &109.173634 & 34.1 & 167.1  & 7 (1)        & 2.0 (0.8)    & $< 0.5$      & $< 0.4$      & $< 0.3$      \\
SO             &109.252220 & 21.1 &  3.56  & 4 (1)        & $< 1$        & $< 0.6$      & $< 0.3$      & $< 0.3$      \\  
OCS            &109.463063 & 26.3 &  4.60  & 2 (1)        & $< 0.8$      & $< 0.4$      & $< 0.3$      & $< 0.3$      \\  
C$^{18}$O      &109.782173 &  5.3 & 0.012  & 35 (2)       & 16.2 (0.9)   & 7.0 (0.5)    & 3.7 (0.5)    & 1.6 (0.3)    \\
HNCO           &109.905749 & 15.8 &  12.5  & 4 (2)        & 1.6 (0.8)    & $< 0.5$      & $< 0.4$      & $< 0.4$      \\  
$^{13}$CO      &110.201354 &  5.3 & 0.012  & 294 (2)      & 136 (1)      & 67 (1)       & 33.8 (0.9)   & 13.9 (0.6)   \\
C$^{17}$O      &112.359284 &  5.4 & 0.012  & 9.3 (2.4)    & 4.3 (1.4)    & 1.81 (0.77)  & 1.21 (0.59)  & $< 0.64$     \\
CN \tablenotemark{c} &113.123370 &  5.4 &  0.15  & 49 (6)       & 22 (3)       & 12 (2)       & 11 (2)       & 10 (1)       \\
CN \tablenotemark{c} &113.144157 &  5.4 &  1.25  & ...          & ...          & ...          & ...          & ...          \\
CN \tablenotemark{c} &113.170492 &  5.4 &  1.22  & ...          & ...          & ...          & ...          & ...          \\
CN \tablenotemark{c} &113.191279 &  5.4 &  1.58  & ...          & ...          & ...          & ...          & ...          \\
CN \tablenotemark{c} &113.490970 &  5.4 &  4.21  & 75 (3)       & 33 (2)       & 11 (1)       & 5.4 (0.9)    & 3 (1)        \\
CN \tablenotemark{c} &113.488120 &  5.4 &  1.58  & ...          & ...          & ...          & ...          & ...          \\
CN \tablenotemark{c} &113.499644 &  5.4 &  1.25  & ...          & ...          & ...          & ...          & ...          \\
\enddata

%CH$_3$CN       &110.38350 & 18.5 & 184.6 0 &  &  &  &  &  \\
\tablenotetext{a}{The numbers in parentheses represent $3\sigma$ errors.}
\tablenotetext{b}{The upper limit to the integrated intensity is calculated as: $\int T_{\rm mb} dv < 3 \sigma \times \sqrt{\Delta V \times \Delta v_{\rm res}}$, where $\Delta V$ is the assumed line width (40~km~s$^{-1}$) and $\Delta v_{\rm res}$ is the velocity resolution per channel.}
\tablenotetext{c}{The line is blended with other lines.}
\end{deluxetable}

\clearpage
\begin{table*}
\caption{Fractions of the area and the line fluxes from each sub-region}
\begin{tabular}{lrrrrr}
\tableline\tableline
Name         & A & B & C & D & E \\
             & (\%) & (\%) & (\%) & (\%) & (\%) \\
\tableline
Area                           &  0.7 &  2.8 & 14.5 & 24.0 & 57.9 \\
c-C$_3$H$_2$ \tablenotemark{a} &  6.7 $\pm$ 1.8 & 15.0 $\pm$ 3.8 & 27.0 $\pm$  9.8 & 24.0 $\pm$  9.5 & 27.3 $\pm$ 21.8 \\
CCH \tablenotemark{a}          &  5.6 $\pm$ 0.5 & 10.3 $\pm$ 0.9 & 28.1 $\pm$  3.9 & 23.5 $\pm$  4.2 & 32.5 $\pm$  6.7 \\
HCN \tablenotemark{a}          &  8.5 $\pm$ 0.1 & 15.4 $\pm$ 0.3 & 31.2 $\pm$  1.2 & 22.9 $\pm$  1.3 & 21.9 $\pm$  2.6 \\
HCO$^+$ \tablenotemark{a}      &  8.9 $\pm$ 0.2 & 14.5 $\pm$ 0.4 & 29.7 $\pm$  1.5 & 23.6 $\pm$  1.6 & 23.3 $\pm$  2.7 \\
HNC \tablenotemark{a}          & 12.3 $\pm$ 0.3 & 18.8 $\pm$ 0.7 & 33.5 $\pm$  3.3 & 22.1 $\pm$  3.1 & 13.1 $\pm$  6.9 \\
N$_2$H$^+$ \tablenotemark{a}   & 14.6 $\pm$ 0.3 & 22.2 $\pm$ 0.6 & 31.4 $\pm$  2.4 & 18.2 $\pm$  2.4 & 13.7 $\pm$  6.5 \\
CH$_3$OH \tablenotemark{a}     & 11.7 $\pm$ 0.7 & 11.9 $\pm$ 1.5 & 17.9 $\pm$  5.9 & 19.7 $\pm$  8.2 & 38.8 $\pm$ 16.9 \\
CS \tablenotemark{a}           & 13.5 $\pm$ 0.2 & 18.3 $\pm$ 0.5 & 30.3 $\pm$  1.1 & 20.5 $\pm$  1.6 & 17.4 $\pm$  3.2 \\
SO \tablenotemark{a}           & 20.1 $\pm$ 1.9 & 22.5 $\pm$ 3.8 & 35.6 $\pm$ 13.8 & 21.9 $\pm$ 10.0 & 0    $^{+20}_{0}$ \\
C$^{18}$O \tablenotemark{a}    &  7.4 $\pm$ 0.3 & 12.8 $\pm$ 0.7 & 28.4 $\pm$  2.2 & 24.9 $\pm$  3.2 & 26.5 $\pm$  5.5 \\
$^{13}$CO \tablenotemark{a}    & 5.91 $\pm$ 0.05 & 12.1 $\pm$ 0.1 & 30.8 $\pm$  0.5 & 25.7 $\pm$  0.7 & 25.4 $\pm$  1.1 \\
CN \tablenotemark{a}           &  9.4 $\pm$ 0.4 & 15.3 $\pm$ 0.9 & 27.3 $\pm$  2.8 & 21.8 $\pm$  3.8 & 26.2 $\pm$  9.4 \\
\tableline
\end{tabular}
\label{tab07}
\tablenotetext{a}{c-C$-3$H$_2$ ($2_{1\,2}-1_{0\,1}$: 85.338894~GHz), CCH ($N=1-0,J=3/2-1/2,F=2-1$: 87.316898~GHz), HCN ($J=1-0$: 88.631602~GHz), HCO$^+$ ($J=1-0$: 89.188525~GHz), HNC ($J=1-0$: 90.663568~GHz), N$_2$H$^+$ ($J=1-0$: 93.173392~GHz), CH$_3$OH ($2_{-1}-1_{-1},{E}$: 96.739362~GHz), CS ($J=2-1$: 97.980953~GHz), SO ($J_N=3_2-2_1$: 99.299870~GHz), C$^{18}$O ($J=1-0$: 109.782173~GHz), $^{13}$CO ($J=1-0$: 110.201354~GHz), and CN ($N=1-0,J=3/2-1/2,F=5/2-3/2$: 113.490970~GHz) are used in this analysis. }
\end{table*}

%\newpage
\begin{deluxetable}{lllllll}
\tablecolumns{7}
%\tablewidth{0pt}
\tabletypesize{\tiny}
\tablecaption{Column densities and fractional abundances of molecules averaged over whole the area of W51.$^{\rm a}$ \label{tab03}}
\tablehead{
\colhead{Molecule} & \colhead{$^{\rm b}N$ ($T$=10~K)$^{\rm c}$} & \colhead{$X^{\rm d}$ ($T$=10~K)$^{\rm c}$} & \colhead{$N^{\rm b}$ ($T$=15~K)$^{\rm c}$} & \colhead{$X^{\rm d}$ ($T$=15~K)$^{\rm c}$} & \colhead{$N^{\rm b}$ ($T$=20~K)$^{\rm c}$} & \colhead{$X^{\rm d}$ ($T$=20~K)$^{\rm c}$} \\
\colhead{}         & \colhead{(cm$^{-2}$)}              & \colhead{} & \colhead{(cm$^{-2}$)}              & \colhead{} & \colhead{(cm$^{-2}$)} & \colhead{} }
\startdata
CCH                         &
 $1.8 (0.4) \times 10^{14}$ & $9   (2)   \times 10^{-9}$ &
 $2.1 (0.5) \times 10^{14}$ & $9   (3)   \times 10^{-9}$ &
 $2.5 (0.6) \times 10^{14}$ & $9   (3)   \times 10^{-9}$   \\
CN                          &
 $5   (1)   \times 10^{13}$ & $2.4 (0.6) \times 10^{-9}$ &
 $6   (1)   \times 10^{13}$ & $2.4 (0.8) \times 10^{-9}$ &
 $7   (2)   \times 10^{13}$ & $2.4 (0.8) \times 10^{-9}$   \\
HCN $^{\rm e}$              &
 $4   (2)   \times 10^{13}$ & $2   (1)   \times 10^{-9} $ &
 $4   (3)   \times 10^{13}$ & $2   (1)   \times 10^{-9} $ &
 $5   (3)   \times 10^{13}$ & $2   (1)   \times 10^{-9} $   \\
H$^{13}$CN                  &
 $6   (4)   \times 10^{11}$ & $3   (2)   \times 10^{-11}$ &
 $7   (4)   \times 10^{11}$ & $3   (2)   \times 10^{-11}$ &
 $9   (5)   \times 10^{11}$ & $3   (2)   \times 10^{-11}$   \\
HC$^{15}$N $^{\rm f}$       &
 $< 3   \times 10^{11}$     & $< 1   \times 10^{-11}$    &
 $< 3   \times 10^{11}$     & $< 1   \times 10^{-11}$    &
 $< 4   \times 10^{11}$     & $< 1   \times 10^{-11}$      \\
HNC                         &
 $4   (1)   \times 10^{12}$ & $2.1 (0.5) \times 10^{-10}$ &
 $5   (1)   \times 10^{12}$ & $2.2 (0.7) \times 10^{-10}$ &
 $6   (1)   \times 10^{12}$ & $2.2 (0.7) \times 10^{-10}$   \\
HN$^{13}$C                  &
 $< 4       \times 10^{11}$ & $< 2       \times 10^{-11}$ &
 $< 5       \times 10^{11}$ & $< 2       \times 10^{-11}$ &
 $< 6       \times 10^{11}$ & $< 2       \times 10^{-11}$   \\
$^{13}$CO                   &
 $3.1 (0.6) \times 10^{16}$ & $1.5 (0.3) \times 10^{-6}$ &
 $3.6 (0.7) \times 10^{16}$ & $1.5 (0.5) \times 10^{-6}$ &
 $4.2 (0.9) \times 10^{16}$ & $1.5 (0.4) \times 10^{-6}$   \\
C$^{17}$O                   &
 $1.0 (0.5) \times 10^{15}$ & $5.1 (0.3) \times 10^{-8}$ &
 $1.2 (0.6) \times 10^{15}$ & $5.1 (0.3) \times 10^{-8}$ &
 $1.4 (0.7) \times 10^{15}$ & $5.1 (0.3) \times 10^{-8}$   \\
C$^{18}$O                   &
 $3.5 (0.7) \times 10^{15}$ &  &
 $4.0 (0.9) \times 10^{15}$ &  &
 $5   (1)   \times 10^{15}$ &    \\
HCO$^+$                     &
 $5   (1)   \times 10^{12}$ & $2.5 (0.5) \times 10^{-10}$ &
 $6   (1)   \times 10^{12}$ & $2.5 (0.8) \times 10^{-10}$ &
 $7   (2)   \times 10^{12}$ & $2.6 (0.8) \times 10^{-10}$   \\
H$^{13}$CO$^+$ $^{\rm f}$   &
 $< 2   \times 10^{11}$     & $< 7   \times 10^{-12}$ &
 $< 2   \times 10^{11}$     & $< 8   \times 10^{-12}$ &
 $< 2   \times 10^{11}$     & $< 8   \times 10^{-12}$   \\
CH$_3$OH                    &
 $4   (1)   \times 10^{13}$ & $1.7 (0.6) \times 10^{-9}$ &
 $5   (2)   \times 10^{13}$ & $2.0 (0.8) \times 10^{-9}$ &
 $6   (2)   \times 10^{13}$ & $2.3 (0.9) \times 10^{-9}$   \\
N$_2$H$^+$                  &
 $2.3 (0.5) \times 10^{12}$ & $1.1 (0.3) \times 10^{-10}$ &
 $2.7 (0.6) \times 10^{12}$ & $1.2 (0.4) \times 10^{-10}$ &
 $3.3 (0.7) \times 10^{12}$ & $1.2 (0.4) \times 10^{-10}$   \\
c-C$_3$H$_2$ $^{\rm g}$     &
 $4   (1)   \times 10^{12}$ & $1.9 (0.7) \times 10^{-10}$ &
 $5   (2)   \times 10^{12}$ & $2.3 (0.9) \times 10^{-10}$ &
 $7   (2)   \times 10^{12}$ & $3   (1)   \times 10^{-10}$   \\
CH$_3$CCH $^{\rm f, h}$     &
 $< 1   \times 10^{13}$     & $< 5   \times 10^{-10}$ &
 $< 1   \times 10^{13}$     & $< 4   \times 10^{-10}$ &
 $< 1   \times 10^{13}$     & $< 4   \times 10^{-10}$   \\
CH$_3$CN $^{\rm f, h}$      &
 $< 7   \times 10^{11}$     & $< 4   \times 10^{-11}$ &
 $< 7   \times 10^{11}$     & $< 3   \times 10^{-11}$ &
 $< 7   \times 10^{11}$     & $< 3   \times 10^{-11}$   \\
HNCO $^{\rm f}$             &
 $< 2   \times 10^{12}$     & $< 1   \times 10^{-10}$ &
 $< 2   \times 10^{12}$     & $< 1   \times 10^{-10}$ &
 $< 3   \times 10^{12}$     & $< 1   \times 10^{-10}$   \\
SiO $^{\rm f}$              &
 $< 5   \times 10^{11}$     & $< 2   \times 10^{-11}$ &
 $< 5   \times 10^{11}$     & $< 2   \times 10^{-11}$ &
 $< 6   \times 10^{11}$     & $< 2   \times 10^{-11}$   \\
CS $^{\rm i}$               &
 $6   (3)   \times 10^{13}$ & $3   (1)   \times 10^{-9}$ &
 $7   (3)   \times 10^{13}$ & $3   (2)   \times 10^{-9}$ &
 $8   (4)   \times 10^{13}$ & $3   (1)   \times 10^{-9}$   \\
$^{13}$CS $^{\rm f}$        &
 $< 9   \times 10^{11}$     & $< 5   \times 10^{-11}$ &
 $< 1   \times 10^{12}$     & $< 4   \times 10^{-11}$ &
 $< 1   \times 10^{12}$     & $< 4   \times 10^{-11}$        \\
C$^{33}$S $^{\rm f}$        &
 $< 9   \times 10^{11}$     & $< 5   \times 10^{-11}$ &
 $< 1   \times 10^{12}$     & $< 4   \times 10^{-11}$ &
 $< 1   \times 10^{12}$     & $< 4   \times 10^{-11}$        \\
C$^{34}$S                   &
 $3   (1)   \times 10^{12}$ & $1.4 (0.6) \times 10^{-10}$ &
 $3   (1)   \times 10^{12}$ & $1.4 (0.7) \times 10^{-10}$ &
 $4   (2)   \times 10^{12}$ & $1.4 (0.7) \times 10^{-10}$   \\
SO                        &
 $1.2 (0.5) \times 10^{13}$ & $6   (2)   \times 10^{-10}$ &
 $1.5 (0.6) \times 10^{13}$ & $6   (3)   \times 10^{-10}$ &
 $1.8 (0.7) \times 10^{13}$ & $6   (3)   \times 10^{-10}$   \\
OCS $^{\rm f}$              &
 $< 2   \times 10^{13}$     & $< 1   \times 10^{-9}$ &
 $< 2   \times 10^{13}$     & $< 7   \times 10^{-10}$ &
 $< 1   \times 10^{13}$     & $< 5   \times 10^{-10}$   \\
HC$_3$N                     &
 $2   (1)   \times 10^{12}$ & $1.1 (0.6) \times 10^{-10}$ &
 $1.4 (0.8) \times 10^{12}$ & $6.0 (0.4) \times 10^{-11}$ &
 $1.2 (0.7) \times 10^{12}$ & $4.5 (0.3) \times 10^{-11}$   \\
\enddata
\tablenotetext{a}{Errors of the column densities are estimated by taking into account the rms noise and calibration uncertainties of the chopper-wheel method (20~\%). }
\tablenotetext{b}{Column density.}
\tablenotetext{c}{Assumed excitation temperatures.  }
\tablenotetext{d}{Fractional abundance relative to the H$_2$.  The column density of H$_2$ is calculated from the column density of C$^{18}$O, where the $N({\rm C^{18}O})/N({\rm H_2}) = 1.7 \times 10^{-7}$ is assumed.}
\tablenotetext{e}{Obtained from the H$^{13}$CN data assuming the $^{12}$C/$^{13}$C ratio of 60.}
\tablenotetext{f}{The upper limit to the column density is estimated from the 3$\sigma$ upper limit of the integrated intensity assuming the line width of 40~km/s.}
\tablenotetext{g}{An ortho-to-para ratio of 3 is assumed.}
\tablenotetext{h}{The column density is calculated from the A species ($K=0$) on the assumption that the column density of the E species is the same as that of the A species.}
\tablenotetext{i}{Obtained from the C$^{34}$S data assuming the $^{32}$S/$^{34}$S ratio of 22.}
\end{deluxetable}

%\newpage
\begin{table*}
\tiny
\rotate
\caption{Column densities and fractional abundances of molecules averaged over whole the area of W51 estimated by the RADEX.$^{\rm a}$ \label{tab06}}
\begin{tabular}{lllllll}
\tableline\tableline
Molecule & \multicolumn{6}{c}{ $n_{\rm H_2} = 10^4$~cm$^{-3}$ $^{\rm b}$ } \\ \cline{2-7}
         & $N^{\rm c}$ ($T_{\rm kin}$=10~K)$^{\rm d}$ & $X^{\rm e}$ ($T_{\rm kin}$=10~K)$^{\rm d}$ & $N^{\rm c}$ ($T_{\rm kin}$=15~K)$^{\rm d}$ & $X^{\rm e}$ ($T_{\rm kin}$=15~K)$^{\rm d}$ & $N^{\rm c}$ ($T_{\rm kin}$=20~K)$^{\rm d}$ & $X^{\rm e}$ ($T_{\rm kin}$=20~K)$^{\rm d}$ \\ 
         & (cm$^{-2}$) &  & (cm$^{-2}$) &  & (cm$^{-2}$) & \\
\tableline
CCH                         &
 $6   (1)   \times 10^{14}$ & $3   (1)   \times 10^{-8}$ &
 $4.0 (0.9) \times 10^{14}$ & $2.1 (0.7) \times 10^{-8}$ &
 $3.2 (0.7) \times 10^{14}$ & $1.6 (0.5) \times 10^{-8}$ \\
HCN $^{\rm f}$              &
 $8   (5)   \times 10^{14}$ & $4   (3)   \times 10^{-8}$ &
 $5   (3)   \times 10^{14}$ & $3   (2)   \times 10^{-8}$ &
 $4   (2)   \times 10^{14}$ & $2   (1)   \times 10^{-8}$ \\
H$^{13}$CN                  &
 $1.3 (0.8) \times 10^{13}$ & $7   (5)   \times 10^{-10}$ &
 $9   (5)   \times 10^{12}$ & $5   (3)   \times 10^{-10}$ &
 $7   (4)   \times 10^{12}$ & $3   (2)   \times 10^{-10}$ \\
HC$^{15}$N $^{\rm g}$       &
 $< 5   \times 10^{12}$     & $< 3   \times 10^{-10}$    &
 $< 3   \times 10^{12}$     & $< 2   \times 10^{-10}$    &
 $< 3   \times 10^{12}$     & $< 1   \times 10^{-10}$    \\
HNC                         &
 $5   (1)   \times 10^{13}$ & $2.7 (0.9) \times 10^{-9}$ &
 $4   (1)   \times 10^{13}$ & $2.0 (0.7) \times 10^{-9}$ &
 $3.2 (0.8) \times 10^{13}$ & $1.6 (0.5) \times 10^{-9}$ \\
$^{13}$CO                   &
 $3.1 (0.7) \times 10^{16}$ & $1.7 (0.6) \times 10^{-6}$ &
 $3.1 (0.7) \times 10^{16}$ & $1.6 (0.5) \times 10^{-6}$ &
 $3.2 (0.7) \times 10^{16}$ & $1.6 (0.5) \times 10^{-6}$ \\
C$^{17}$O                   &
 $9   (4)   \times 10^{14}$ & $5   (3)   \times 10^{-8}$ &
 $9   (5)   \times 10^{14}$ & $5   (3)   \times 10^{-8}$ &
 $1.3 (0.5) \times 10^{15}$ & $5   (3)   \times 10^{-8}$ \\
C$^{18}$O                   &
 $3.1 (0.7) \times 10^{15}$ & \nodata &
 $3.1 (0.7) \times 10^{15}$ & \nodata &
 $3.4 (0.7) \times 10^{15}$ & \nodata \\
HCO$^+$                     &
 $2.6 (0.6) \times 10^{13}$ & $1.5 (0.5) \times 10^{-9}$ &
 $1.9 (0.4) \times 10^{13}$ & $1.0 (0.3) \times 10^{-9}$ &
 $1.6 (0.4) \times 10^{13}$ & $8   (2)   \times 10^{-10}$ \\
H$^{13}$CO$^+$ $^{\rm g}$   &
 $< 6   \times 10^{11}$     & $< 3   \times 10^{-11}$ &
 $< 5   \times 10^{11}$     & $< 2   \times 10^{-11}$ &
 $< 4   \times 10^{11}$     & $< 2   \times 10^{-11}$ \\
CH$_3$OH                    &
 $5   (1)   \times 10^{13}$ & $3   (1)   \times 10^{-9}$ &
 $3   (1)   \times 10^{13}$ & $1.9 (0.7) \times 10^{-9}$ &
 $3   (1)   \times 10^{13}$ & $9   (4)   \times 10^{-10}$ \\
N$_2$H$^+$                  &
 $1.0 (0.2) \times 10^{13}$ & $5   (2)   \times 10^{-10}$ &
 $7   (2)   \times 10^{12}$ & $4   (1)   \times 10^{-10}$ &
 $6   (1)   \times 10^{12}$ & $3   (1)   \times 10^{-10}$ \\
c-C$_3$H$_2$                &
 $6   (2)   \times 10^{13}$ & $4   (2)   \times 10^{-9}$ &
 $4   (1)   \times 10^{13}$ & $2.0 (0.9) \times 10^{-9}$ &
 $2.7 (0.9) \times 10^{13}$ & $1.4 (0.6) \times 10^{-9}$ \\
CH$_3$CN $^{\rm g}$         &
 $< 2   \times 10^{13}$     & $< 1   \times 10^{-9}$ &
 $< 1   \times 10^{13}$     & $< 7   \times 10^{-10}$ &
 $< 1   \times 10^{13}$     & $< 6   \times 10^{-10}$ \\
HNCO $^{\rm g}$             &
 $< 9   \times 10^{12}$     & $< 5   \times 10^{-10}$ &
 $< 5   \times 10^{12}$     & $< 3   \times 10^{-10}$ &
 $< 4   \times 10^{12}$     & $< 2   \times 10^{-10}$ \\
SiO $^{\rm g}$              &
 $< 5   \times 10^{12}$     & $< 3   \times 10^{-10}$ &
 $< 4   \times 10^{12}$     & $< 2   \times 10^{-10}$ &
 $< 3   \times 10^{12}$     & $< 2   \times 10^{-10}$ \\
CS                          &
 $2.0 (0.5) \times 10^{14}$ & $1.1 (0.4) \times 10^{-8}$ &
 $1.4 (0.3) \times 10^{14}$ & $7   (2)   \times 10^{-9}$ &
 $1.1 (0.3) \times 10^{14}$ & $6   (2)   \times 10^{-9}$ \\
SO                        &
 $7   (3)   \times 10^{13}$ & $4   (2)   \times 10^{-9}$ &
 $4   (2)   \times 10^{13}$ & $2   (1)   \times 10^{-9}$ &
 $3   (1)   \times 10^{13}$ & $1.8 (0.8) \times 10^{-9}$ \\
OCS $^{\rm g}$              &
 $< 5   \times 10^{13}$     & $< 3   \times 10^{-9}$ &
 $< 2   \times 10^{13}$     & $< 1   \times 10^{-9}$ &
 $< 2   \times 10^{13}$     & $< 8   \times 10^{-10}$ \\
HC$_3$N                     &
 $6   (3)   \times 10^{13}$ & $3   (2)   \times 10^{-9}$ &
 $2   (1)   \times 10^{13}$ & $1.2 (0.7) \times 10^{-9}$ &
 $1.3 (0.7) \times 10^{13}$ & $7   (4)   \times 10^{-10}$ \\
\tableline
  & \multicolumn{6}{c}{ $n_{\rm H_2} = 10^5$~cm$^{-3}$ $^{\rm b}$ } \\  \cline{2-7}
%         & $N^{\rm c}$ ($T_{\rm kin}$=10~K)$^{\rm d}$ & $X^{\rm e}$ ($T_{\rm kin}$=10~K)$^{\rm d}$ & $N^{\rm c}$ ($T_{\rm kin}$=15~K)$^{\rm d}$ & $X^{\rm e}$ ($T_{\rm kin}$=15~K)$^{\rm d}$ & $N^{\rm c}$ ($T_{\rm kin}$=20~K)$^{\rm d}$ & $X^{\rm e}$ ($T_{\rm kin}$=20~K)$^{\rm d}$ \\ 
%         & (cm$^{-2}$) &  & (cm$^{-2}$) &  & (cm$^{-2}$) & \\
\tableline
CCH                         &
 $1.5 (0.3) \times 10^{14}$ & $7   (2)   \times 10^{-8}$ & 
 $1.2 (0.3) \times 10^{14}$ & $5   (2)   \times 10^{-9}$ & 
 $1.2 (0.3) \times 10^{14}$ & $4   (1)   \times 10^{-9}$   \\
HCN $^{\rm f}$              &
 $9   (5)   \times 10^{13}$ & $5   (3)   \times 10^{-9}$ & 
 $6   (4)   \times 10^{13}$ & $3   (2)   \times 10^{-9}$ & 
 $5   (3)   \times 10^{13}$ & $2   (1)   \times 10^{-9}$   \\
H$^{13}$CN                  &
 $1.5 (0.9) \times 10^{12}$ & $8   (5)   \times 10^{-11}$ &
 $1.1 (0.6) \times 10^{12}$ & $5   (3)   \times 10^{-11}$ &
 $9   (5)   \times 10^{11}$ & $3   (2)   \times 10^{-11}$   \\
HC$^{15}$N $^{\rm g}$       &
 $< 6   \times 10^{11}$     & $< 3   \times 10^{-11}$    & 
 $< 4   \times 10^{11}$     & $< 2   \times 10^{-11}$    & 
 $< 4   \times 10^{11}$     & $< 1   \times 10^{-11}$      \\
HNC                         &
 $7   (2)   \times 10^{12}$ & $3   (1)   \times 10^{-10}$ &
 $6   (1)   \times 10^{12}$ & $2.4 (0.7) \times 10^{-10}$ &
 $5   (1)   \times 10^{12}$ & $1.9 (0.6) \times 10^{-10}$   \\
$^{13}$CO                   &
 $3.4 (0.8) \times 10^{16}$ & $1.7 (0.5) \times 10^{-6}$ & 
 $3.7 (0.8) \times 10^{16}$ & $1.6 (0.5) \times 10^{-6}$ & 
 $4.2 (0.9) \times 10^{16}$ & $1.6 (0.5) \times 10^{-6}$   \\
C$^{17}$O                   &
 $1.0 (0.5) \times 10^{15}$ & $5   (3)   \times 10^{-8}$ & 
 $1.2 (0.6) \times 10^{15}$ & $5   (3)   \times 10^{-8}$ & 
 $1.3 (0.6) \times 10^{15}$ & $5   (3)   \times 10^{-8}$   \\
C$^{18}$O                   &
 $3.4 (0.7) \times 10^{15}$ & \nodata &
 $3.9 (0.8) \times 10^{15}$ & \nodata &
 $4   (1)   \times 10^{15}$ & \nodata \\
HCO$^+$                     &
 $5   (1)   \times 10^{12}$ & $2.4 (0.7) \times 10^{-10}$ & 
 $4.1 (0.9) \times 10^{12}$ & $1.8 (0.5) \times 10^{-10}$ & 
 $3.8 (0.8) \times 10^{12}$ & $1.5 (0.4) \times 10^{-10}$   \\
H$^{13}$CO$^+$ $^{\rm g}$   &
 $< 1   \times 10^{11}$     & $< 7   \times 10^{-11}$ &
 $< 1   \times 10^{11}$     & $< 5   \times 10^{-11}$ &
 $< 1   \times 10^{11}$     & $< 4   \times 10^{-11}$   \\
CH$_3$OH                    &
 $2.4 (0.8) \times 10^{13}$ & $1.2 (0.5) \times 10^{-10}$ &
 $2.3 (0.8) \times 10^{13}$ & $1.0 (0.4) \times 10^{-9}$ &
 $2.4 (0.8) \times 10^{13}$ & $9.3 (0.4) \times 10^{-10}$  \\
N$_2$H$^+$                  &
 $2.2 (0.5) \times 10^{12}$ & $1.1 (0.3) \times 10^{-10}$ &
 $1.9 (0.4) \times 10^{12}$ & $8   (3)   \times 10^{-11}$ &
 $1.7 (0.4) \times 10^{12}$ & $7   (2)   \times 10^{-11}$   \\
c-C$_3$H$_2$ $^{\rm h}$     &
 $8   (3)   \times 10^{12}$ & $4   (2)   \times 10^{-10}$ &
 $5   (2)   \times 10^{12}$ & $2   (1)   \times 10^{-10}$ &
 $4   (2)   \times 10^{12}$ & $1.7 (0.7) \times 10^{-10}$ \\
CH$_3$CN $^{\rm g}$         &
 $< 2   \times 10^{12}$     & $< 9   \times 10^{-10}$ &
 $< 1   \times 10^{12}$     & $< 6   \times 10^{-11}$ &
 $< 1   \times 10^{12}$     & $< 5   \times 10^{-11}$   \\
HNCO $^{\rm g}$             &
 $< 2   \times 10^{12}$     & $< 1   \times 10^{-10}$ &
 $< 2   \times 10^{12}$     & $< 8   \times 10^{-11}$ &
 $< 2   \times 10^{12}$     & $< 7   \times 10^{-11}$   \\
SiO $^{\rm g}$              &
 $< 8   \times 10^{11}$     & $< 4   \times 10^{-11}$ &
 $< 6   \times 10^{11}$     & $< 3   \times 10^{-11}$ &
 $< 5   \times 10^{11}$     & $< 2   \times 10^{-11}$   \\
CS                          &
 $3.0 (0.6) \times 10^{13}$ & $1.5 (0.4) \times 10^{-9}$ &
 $2.3 (0.5) \times 10^{13}$ & $1.0 (0.3) \times 10^{-9}$ &
 $2.1 (0.4) \times 10^{13}$ & $8   (2)   \times 10^{-10}$ \\
SO                        &
 $1.4 (0.6) \times 10^{13}$ & $7   (3)   \times 10^{-10}$ &
 $1.1 (0.4) \times 10^{13}$ & $5   (2)   \times 10^{-10}$ &
 $1.0 (0.4) \times 10^{13}$ & $4   (2)   \times 10^{-10}$ \\
OCS $^{\rm g}$              &
 $< 3   \times 10^{13}$     & $< 1   \times 10^{-9}$ &
 $< 2   \times 10^{13}$     & $< 8   \times 10^{-10}$ &
 $< 1   \times 10^{13}$     & $< 5   \times 10^{-10}$ \\
HC$_3$N                     &
 $5   (3)   \times 10^{12}$ & $3   (2)   \times 10^{-10}$ &
 $2   (1)   \times 10^{12}$ & $1.1 (0.6) \times 10^{-10}$ &
 $2   (1)   \times 10^{12}$ & $7   (4)   \times 10^{-11}$   \\
\end{tabular}

\tablenotetext{a}{Errors of the column densities are estimated by taking into account the rms noise and calibration uncertainties of the chopper-wheel method (20~\%). }
\tablenotetext{b}{Assumed number density of molecular hydrogen.}
\tablenotetext{c}{Column density.}
\tablenotetext{d}{Assumed kinematic temperatures.}
\tablenotetext{e}{Fractional abundance relative to the H$_2$.  The column density of H$_2$ is calculated from the column density of C$^{18}$O, where the $N({\rm C^{18}O})/N({\rm H_2}) = 1.7 \times 10^{-7}$ is assumed.}
\tablenotetext{f}{Obtained from the H$^{13}$CN data assuming the $^{12}$C/$^{13}$C ratio of 60.}
\tablenotetext{g}{The upper limit to the column density is estimated from the 3$\sigma$ upper limit of the integrated intensity assuming the line width of 40~km/s.}
\tablenotetext{h}{An ortho-to-para ratio of 3 is assumed.}
\end{table*}

\clearpage
\appendix
\section{Appendix : Spectrum and Line Parameters of the Hot Cores e1/e2}
In this appendix, we summarize the spectrum and the line parameters of the hot cores e1/e2, observed as a calibration source in this observation.  The position is ($l$, $b$) = ($49.^{\circ}4898$, $-0.^{\circ}3874$), which corresponds to ($\alpha$ (J2000), $\delta$ (J2000)) = (19$^{\rm h}$ 23$^{\rm m}$ 43.9$^{\rm s}$, +14$^{\circ}$ 30$'$ 35.0$''$).  \citet{Kalenskii2010} also reported a spectral line survey toward the same position in the 3~mm band.  In the hot core, we detected 234 emission lines and identified 31 molecular species, 18 isotopologues, and hydrogen recombination lines on the basis of the spectral line databases, the Cologne Database for Molecular Spectroscopy (CDMS) managed by University of Cologne \citep{muller01,muller05} and the Submillimeter, Millimeter, and Microwave Spectral Line Catalog provided by Jet Propulsion Laboratory \citep{pickett98}.  Figure~\ref{fig01a} and Table~\ref{tab_a01} show the expanded spectrum and the line parameters, respectively.  The LSR velocity adopted for e1/e2 is 55~km~s$^{-1}$.

\clearpage
\begin{figure}
\epsscale{0.99}
%\plotone{fig_a01a.eps}
\includegraphics[angle=90,scale=0.8]{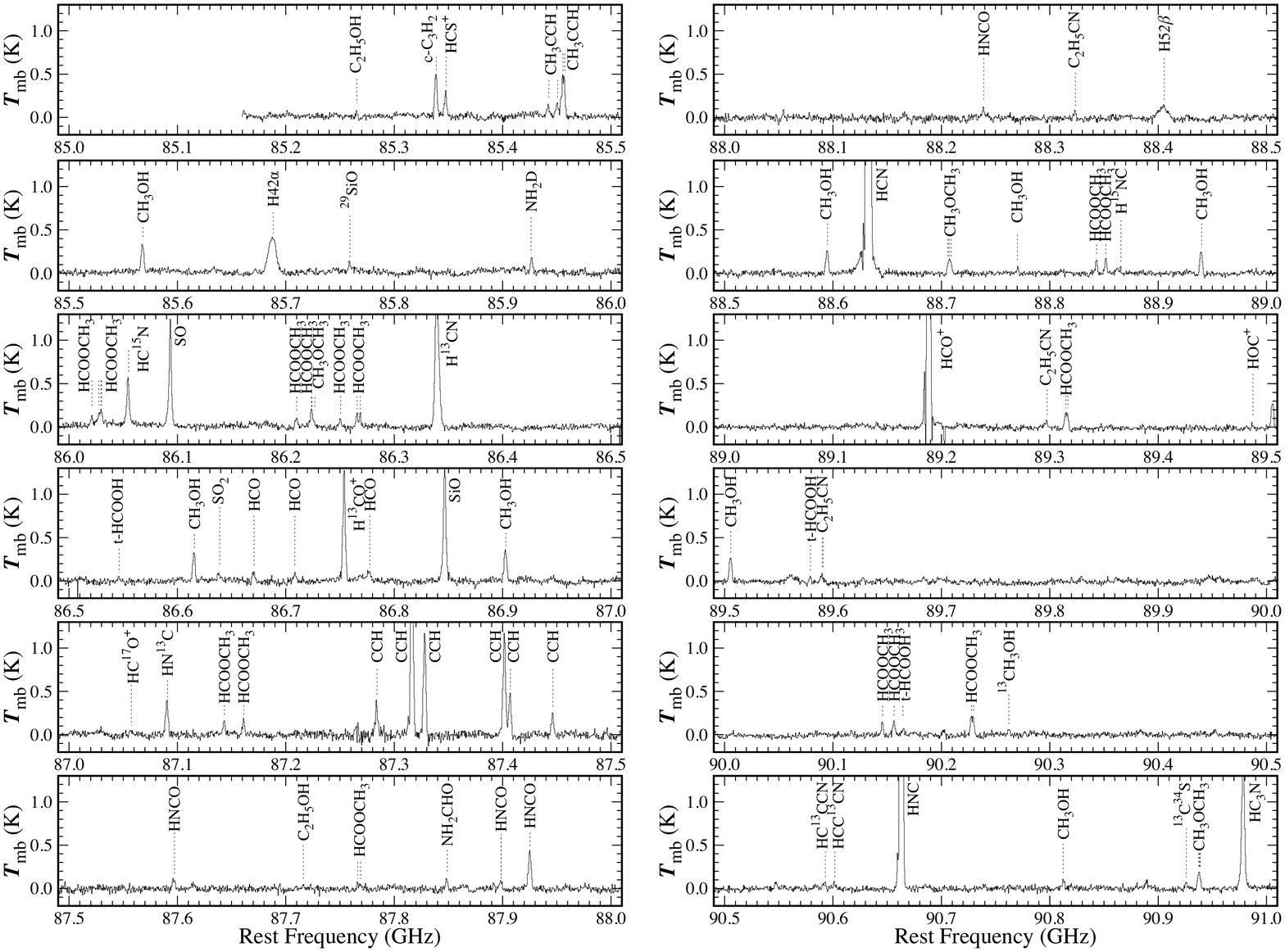}
\caption{Spectra of the hot cores e1/e2 in the 3~mm band observed with the Mopra~22~m telescope.  The frequency resolution is 0.5~MHz. The $V_{\rm LSR}$ of 55~km~s$^{-1}$ is assumed.  The position is ($l$, $b$) = ($49.^{\circ}4898$, $-0.^{\circ}3874$), which corresponds to ($\alpha$ (J2000), $\delta$ (J2000)) = (19$^{\rm h}$ 23$^{\rm m}$ 43.9$^{\rm s}$, +14$^{\circ}$ 30$'$ 35.0$''$).  [Image] indicates the contamination from the imaginary side band.}
\label{fig01a}
\end{figure}
\addtocounter{figure}{-1}

\clearpage
\begin{figure}
\epsscale{0.99}
%\plotone{fig_a01a.eps}
\includegraphics[angle=90,scale=0.8]{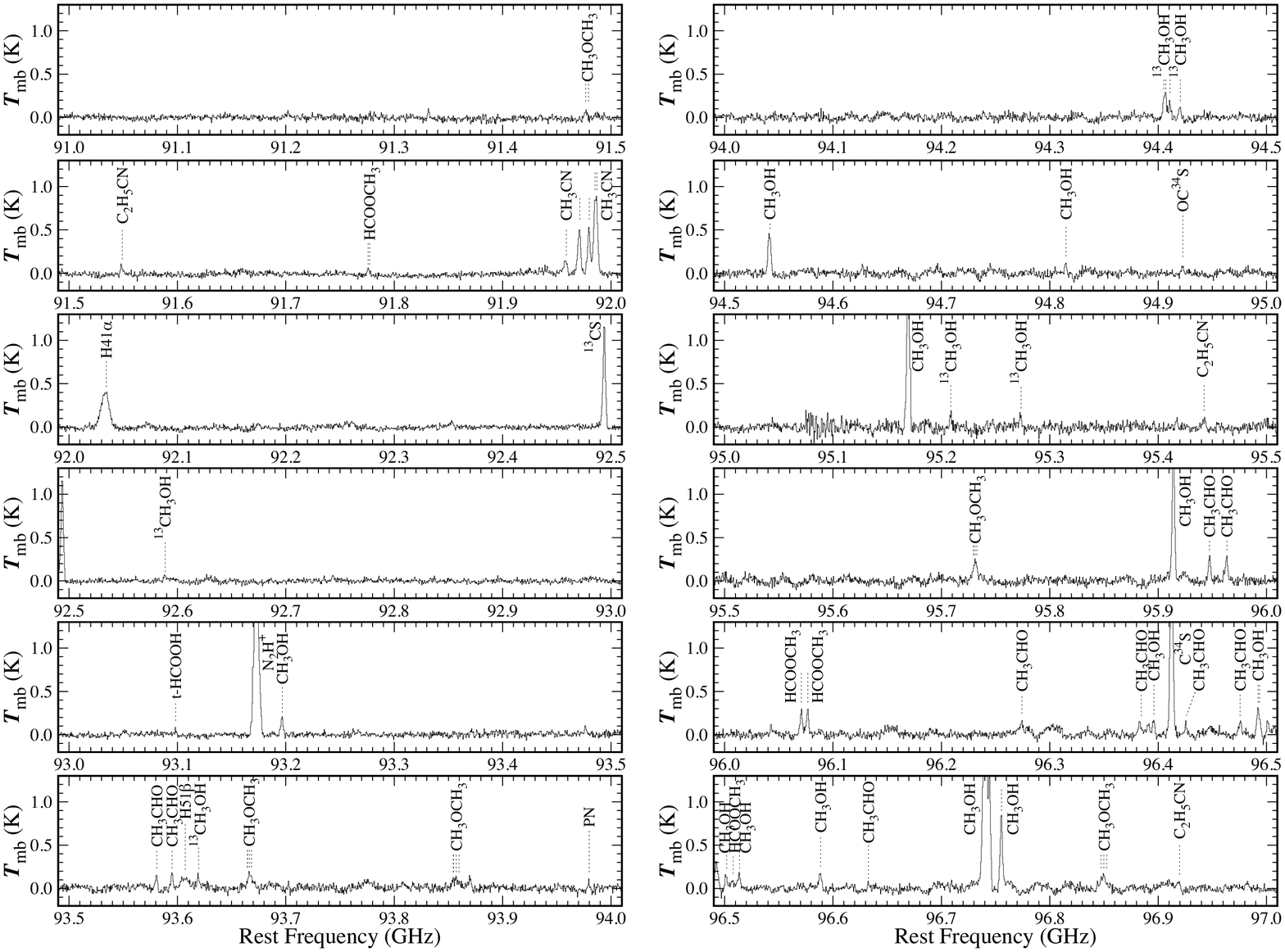}
\caption{\textit{Continued}}
\end{figure}
\addtocounter{figure}{-1}

\clearpage
\begin{figure}
\epsscale{0.99}
%\plotone{fig_a01a.eps}
\includegraphics[angle=90,scale=0.8]{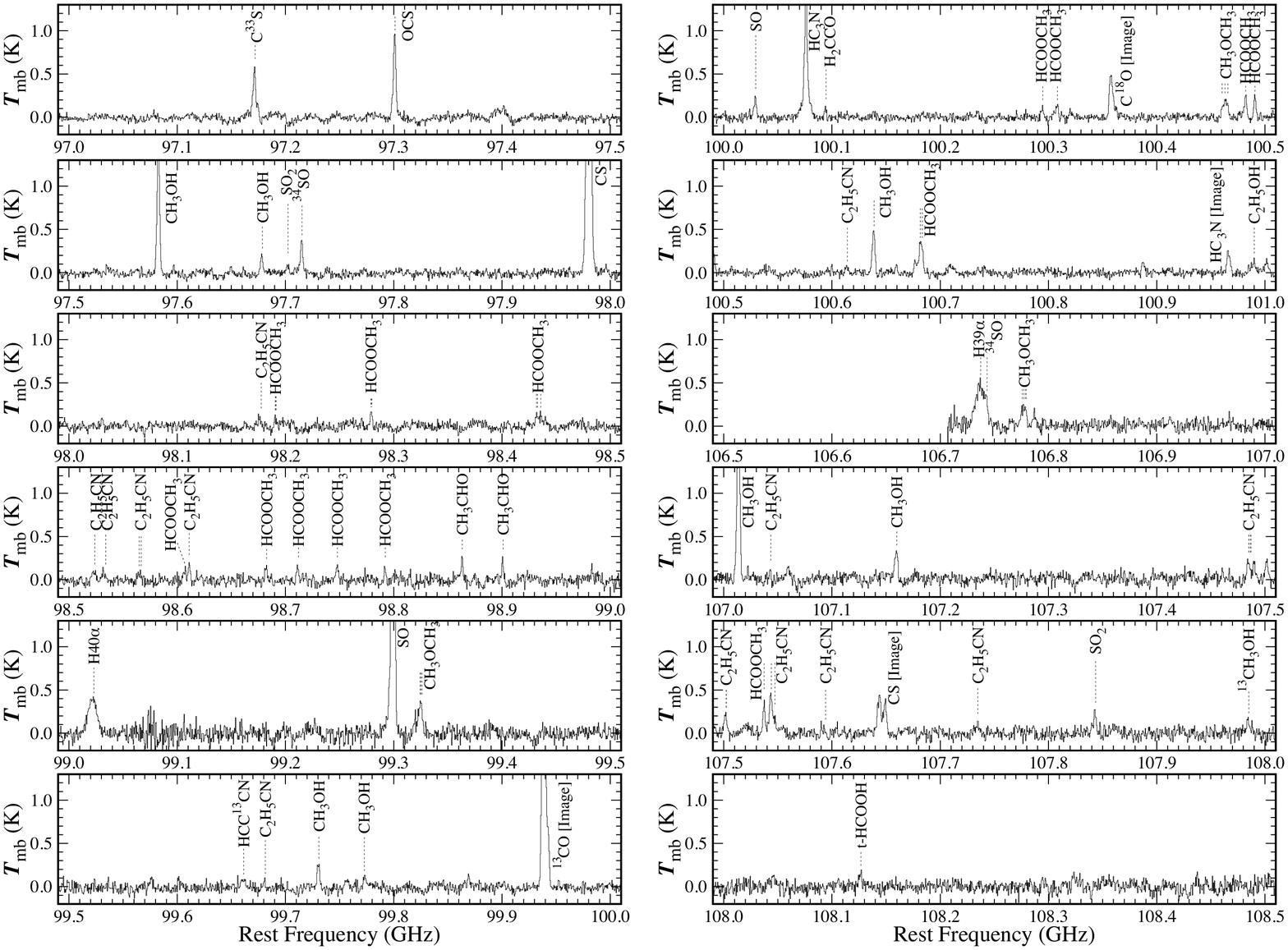}
\caption{\textit{Continued}}
\end{figure}
\addtocounter{figure}{-1}

\clearpage
\begin{figure}
\epsscale{0.99}
%\plotone{fig_a01a.eps}
\includegraphics[angle=90,scale=0.8]{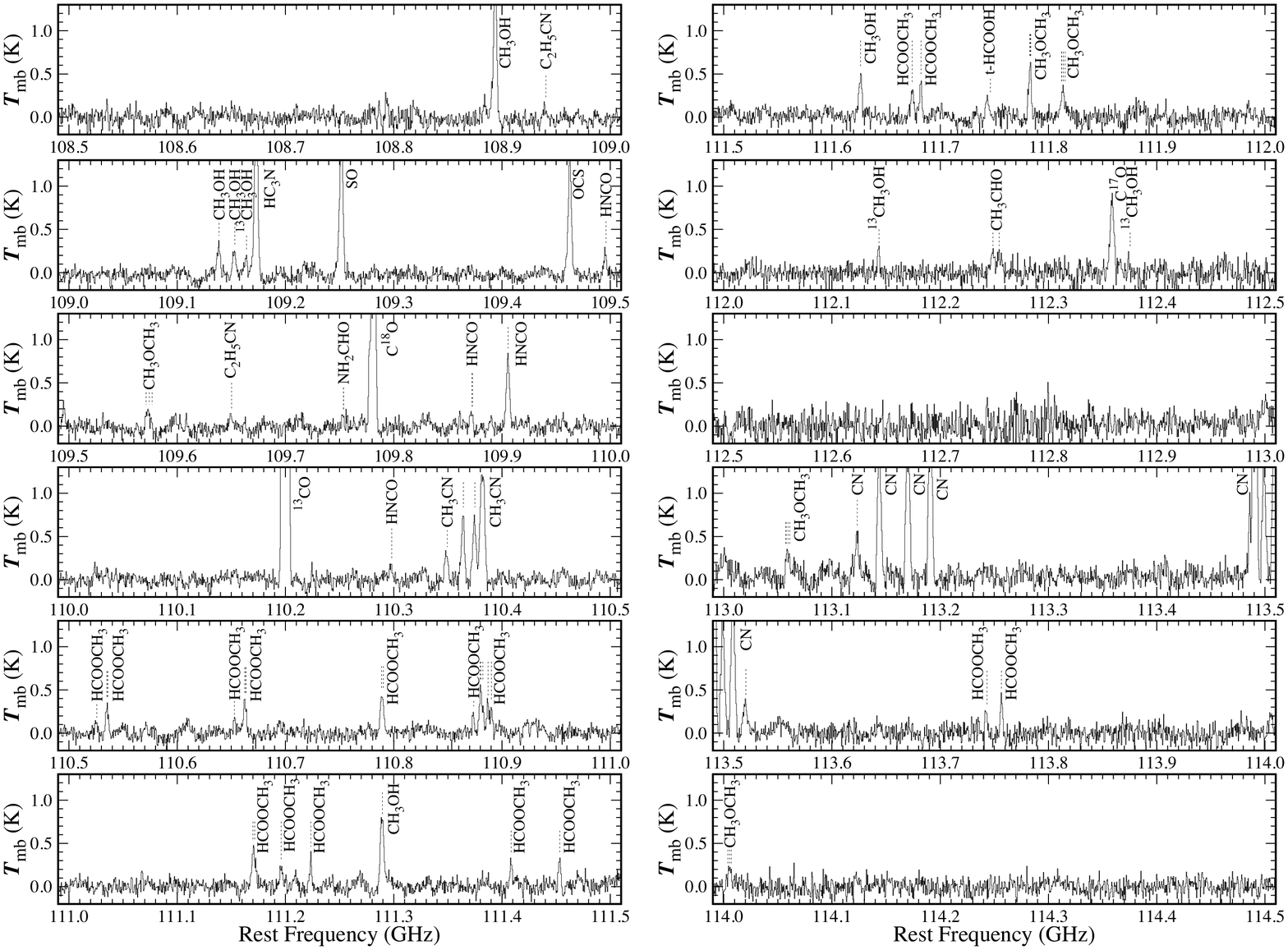}
\caption{\textit{Continued}}
\end{figure}
\addtocounter{figure}{-1}

\begin{deluxetable}{lrlrrrrrr}
%\rotate
\tabletypesize{\tiny}
\tablecolumns{7}
\tablewidth{0pt}
\tablecaption{Line parameters of W51~e1/e2 \tablenotemark{a} \label{tab_a01}}
\tablehead{
\colhead{Name} & \colhead{Frequency} & \colhead{Transition} & \colhead{$E_{\rm u}$} & \colhead{S$\mu^2$} & \colhead{$T_{\rm mb}$ Peak \tablenotemark{b}} & \colhead{$\int T_{\rm mb} dv$ \tablenotemark{b}}  \\
\colhead{} & \colhead{(GHz)} & \colhead{} & \colhead{(K)} & \colhead{(Debye$^2$)} & \colhead{(K)} & \colhead{(K~km~s$^{-1}$)} 
}

\startdata
C$_2$H$_5$OH     &  85.265503 & $6_{0\,6}-5_{1\,5}$                             &  17.5 &   5.34 &  0.08 (0.06) &  0.5 (0.3)   \\
c-C$_3$H$_2$     &  85.338894 & $2_{1\,2} - 1_{0\,1}$                           &   6.4 &  16.05 &  0.50 (0.07) &  4.9 (0.4)   \\
HCS$^+$          &  85.347890 & $J=2-1$                                         &   6.1 &   7.67 &  0.31 (0.07) &  2.7 (0.4)   \\
CH$_3$CCH        &  85.442601 & $5_3-4_3$                                       &  77.3 &   7.87 &  0.15 (0.07) &  1.6 (0.4)   \\
CH$_3$CCH        &  85.450766 & $5_2-4_2$                                       &  41.2 &   5.16 &  0.17 (0.07) &  1.8 (0.4)   \\
CH$_3$CCH \tablenotemark{c}       &  85.455667 & $5_1-4_1$                      &  19.5 &   5.90 &  0.49 (0.07) &  6.3 (0.5)   \\
CH$_3$CCH \tablenotemark{c}       &  85.457300 & $5_0-4_0$                      &  12.3 &   6.14 &              &              \\
CH$_3$OH \tablenotemark{d}        &  85.568084 & $6_{-2}-7_{-1}$,E              &  66.8 &   2.01 &  0.33 (0.06) &  3.8 (0.4)   \\
H42$\alpha$      &  85.688390 &                                                 &       &        &  0.42 (0.07) & 12.8 (0.7)   \\
$^{29}$SiO       &  85.759194 & $J=2-1$                                         &   6.2 &  19.20 &  0.14 (0.07) &  1.5 (0.4)   \\
NH$_2$D          &  85.926278 & $1_{1\,1}-1_{0\,1}$                             &  20.7 &  28.60 &  0.18 (0.07) &  1.7 (0.5)   \\
HCOOCH$_3$       &  86.021124 & $7_{5\,2} - 6_{5\,1}$, E                        &  33.1 &   9.12 &  0.14 (0.07) &  1.2 (0.4)   \\
HCOOCH$_3$ \tablenotemark{c}      &  86.027723 & $7_{5\,3} - 6_{5\,2}$, E       &  33.1 &   9.12 &  0.21 (0.07) &  3.9 (0.6)   \\
HCOOCH$_3$ \tablenotemark{c}      &  86.029442 & $7_{5\,3} - 6_{5\,2}$, A       &  33.1 &   9.13 &              &              \\
HCOOCH$_3$ \tablenotemark{c}      &  86.030186 & $7_{5\,2} - 6_{5\,1}$, A       &  33.1 &   9.13 &              &              \\
HC$^{15}$N       &  86.054966 & $J=1-0$                                         &   4.1 &   8.91 &  0.57 (0.07) &  6.6 (0.5)   \\
SO               &  86.093950 & $J_N=2_2-1_1$                                   &  19.3 &   3.53 &  1.24 (0.07) & 11.9 (0.5)   \\
HCOOCH$_3$       &  86.210057 & $7_{4\,4} - 6_{4\,3}$, A                        &  27.1 &  12.54 &  0.11 (0.08) &  1.0 (0.4)   \\
HCOOCH$_3$ \tablenotemark{c}      &  86.223655 & $7_{4\,3} - 6_{4\,2}$, E       &  27.2 &  12.47 &  0.21 (0.08) &  2.3 (0.5)   \\
HCOOCH$_3$ \tablenotemark{c}      &  86.224160 & $7_{4\,4} - 6_{4\,3}$, E       &  27.2 &  12.47 &              &              \\
CH$_3$OCH$_3$ \tablenotemark{c}   &  86.226733 & $2_{2\,0} - 2_{1\,1}$, EE      &   8.4 &  23.65 &              &              \\
HCOOCH$_3$       &  86.250552 & $7_{4\,3} - 6_{4\,2}$, A                        &  27.2 &  12.54 &  0.10 (0.08) &  0.8 (0.4)   \\
HCOOCH$_3$       &  86.265796 & $7_{3\,5} - 6_{3\,4}$, A                        &  22.5 &  15.19 &  0.16 (0.08) &  1.1 (0.4)   \\
HCOOCH$_3$       &  86.268739 & $7_{3\,5} - 6_{3\,4}$, E                        &  22.5 &  15.06 &  0.17 (0.08) &  1.2 (0.4)   \\
H$^{13}$CN       &  86.339921 & $J=1-0$                                         &   4.1 &  26.73 &  1.67 (0.06) & 23.7 (0.5)   \\
t-HCOOH          &  86.546189 & $4_{1\,4}-3_{1\,3}$                             &  13.6 &   7.58 &  0.1 (0.1)   &  0.3 (0.6)   \\
CH$_3$OH         &  86.615600 & $7_2-6_3$, A$^{-}$                              & 102.7 &   1.36 &  0.3 (0.1)   &  3.0 (0.6)   \\
SO$_2$           &  86.639088 & $8_{3\,5}-9_{2\,8}$                             &  55.2 &   3.02 &  0.09 (0.07) &  1.0 (0.4)   \\
HCO              &  86.670760 & $1_{0\,1}-0_{0\,0}, J=3/2-1/2, F=2-1$           &   4.2 &   3.09 &  0.11 (0.08) &  0.9 (0.5)   \\
HCO              &  86.708360 & $1_{0\,1}-0_{0\,0}, J=3/2-1/2, F=1-0$           &   4.2 &   1.82 &  0.10 (0.07) &  0.8 (0.4)   \\
HC$^{13}$O$^+$   &  86.754288 & $J=1-0$                                         &   4.2 &  15.21 &  1.27 (0.07) & 11.7 (0.5)   \\
HCO              &  86.777460 & $1_{0\,1}-0_{0\,0}, J=1/2-1/2, F=1-1$           &   4.2 &   1.82 &  0.12 (0.07) &  1.3 (0.4)   \\
SiO              &  86.846985 & $J=2-1$                                         &   6.3 &  19.20 &  1.40 (0.07) & 13.9 (0.6)   \\
CH$_3$OH         &  86.902949 & $7_2-6_3$, A$^{+}$                              & 102.7 &   1.36 &  0.36 (0.07) &  3.7 (0.5)   \\
HN$^{13}$C       &  87.090825 & $J=1-0$                                         &   4.2 &   9.30 &  0.40 (0.08) &  3.7 (0.5)   \\
HCOOCH$_3$       &  87.143282 & $7_{3\,4} - 6_{3\,3}$, E                        &  22.6 &  15.06 &  0.16 (0.07) &  1.4 (0.4)   \\
HCOOCH$_3$       &  87.161285 & $7_{3\,4} - 6_{3\,3}$, A                        &  22.6 &  15.20 &  0.19 (0.07) &  1.3 (0.4)   \\
CCH              &  87.284105 & $N=1-0, J=3/2-1/2, F=1-1$                       &   4.2 &   0.10 &  0.4 (0.1)   &  3.1 (0.7)   \\
CCH              &  87.316898 & $N=1-0, J=3/2-1/2, F=2-1$                       &   4.2 &   0.99 &  2.4 (0.1)   & 23.0 (0.7)   \\
CCH              &  87.328585 & $N=1-0, J=3/2-1/2, F=1-0$                       &   4.2 &   0.49 &  1.2 (0.1)   & 10.7 (0.7)   \\
CCH              &  87.401989 & $N=1-0, J=1/2-1/2, F=1-1$                       &   4.2 &   0.49 &  1.17 (0.08) & 11.2 (0.5)   \\
CCH              &  87.407165 & $N=1-0, J=1/2-1/2, F=0-1$                       &   4.2 &   0.20 &  0.48 (0.08) &  4.2 (0.4)   \\
CCH              &  87.446470 & $N=1-0, J=1/2-1/2, F=1-0$                       &   4.2 &   0.10 &  0.25 (0.08) &  2.1 (0.5)   \\
HNCO             &  87.597330 & $4_{1\,4}-3_{1\,3}$                             &  53.8 &   9.25 &  0.12 (0.07) &  1.0 (0.4)   \\
HCOOCH$_3$ \tablenotemark{c}      &  87.766382 & $8_{0\,8} - 7_{1\,7}$, E       &  20.1 &   2.88 &  0.08 (0.07) &  0.7 (0.5)   \\
HCOOCH$_3$ \tablenotemark{c}      &  87.769035 & $8_{0\,8} - 7_{1\,7}$, A       &  20.1 &   2.87 &              &              \\
NH$_2$CHO        &  87.848874 & $4_{1\,3}-3_{1\,2}$                             &  13.5 &  49.03 &  0.12 (0.07) &  0.8 (0.4)   \\
HNCO \tablenotemark{c}            &  87.898425 & $4_{2\,3}-3_{2\,2}$            & 180.8 &   7.15 &  0.09 (0.07) &  0.8 (0.4)   \\
HNCO \tablenotemark{c}            &  87.898628 & $4_{2\,2}-3_{2\,1}$            & 180.8 &   7.15 &              &              \\
HNCO             &  87.925237 & $4_{0\,4}-3_{0\,3}$                             &  10.5 &   9.99 &  0.44 (0.07) &  4.4 (0.4)   \\
HNCO             &  88.239020 & $4_{1\,3}-3_{1\,2}$                             &  53.9 &   9.26 &  0.12 (0.07) &  1.0 (0.5)   \\
C$_2$H$_5$CN     &  88.323735 & $10_{0\,10}-9_{0\,9}$                           &  23.4 & 147.87 &  0.08 (0.07) &  0.5 (0.4)   \\
H52$\beta$       &  88.405690 &                                                 &       &        &  0.15 (0.07) &  3.9 (0.7)   \\
CH$_3$OH         &  88.594960 & $15_3-14_4$, A$^{+}$                            & 328.3 &   4.21 &  0.26 (0.06) &  2.6 (0.4)   \\
HCN              &  88.631602 & $J=1-0$                                         &   4.3 &  26.74 &  5.27 (0.06) & 77.2 (0.5)   \\
CH$_3$OCH$_3$ \tablenotemark{c}   &  88.706231 & $15_{2\,13} - 15_{1\,14}$, EA  & 116.9 &  78.84 &  0.17 (0.05) &  2.1 (0.3)   \\
CH$_3$OCH$_3$ \tablenotemark{c}   &  88.706231 & $15_{2\,13} - 15_{1\,14}$, AE  & 116.9 & 118.27 &              &              \\
CH$_3$OCH$_3$ \tablenotemark{c}   &  88.707704 & $15_{2\,13} - 15_{1\,14}$, EE  & 116.9 & 315.39 &              &              \\
CH$_3$OCH$_3$ \tablenotemark{c}   &  88.709177 & $15_{2\,13} - 15_{1\,14}$, AA  & 116.9 & 197.12 &              &              \\
CH$_3$OH \tablenotemark{d}   &  88.769652 & $v_{\rm t} = 1, 22_{10}-22_{11}$, E &1490.1 &  11.85 &  0.08 (0.07) &  0.5 (0.3)   \\
HCOOCH$_3$       &  88.843187 & $7_{1\,6} - 6_{1\,5}$, E                        &  18.0 &  18.04 &  0.15 (0.08) &  1.1 (0.5)   \\
HCOOCH$_3$       &  88.851607 & $7_{1\,6} - 6_{1\,5}$, A                        &  17.9 &  18.04 &  0.17 (0.08) &  1.3 (0.4)   \\
H$^{15}$NC       &  88.865715 & $J=1-0$                                         &   4.3 &   7.29 &  0.08 (0.05) &  0.3 (0.2)   \\
CH$_3$OH         &  88.940090 & $15_3-14_4$, A$^{-}$                            & 328.3 &   4.21 &  0.25 (0.05) &  2.3 (0.3)   \\
HCO$^+$          &  89.188525 & $J=1-0$                                         &   4.3 &  15.21 &  6.74 (0.06) & 75.2 (0.5)   \\
C$_2$H$_5$CN     &  89.297660 & $10_{2\,9} - 9_{2\,8}$                          &  28.0 & 142.26 &  0.08 (0.06) &  0.7 (0.4)   \\
HCOOCH$_3$ \tablenotemark{c}      &  89.314657 & $8_{1\,8} - 7_{1\,7}$, E       &  20.2 &  20.83 &  0.17 (0.06) &  1.9 (0.4)   \\
HCOOCH$_3$ \tablenotemark{c}      &  89.316642 & $8_{1\,8} - 7_{1\,7}$, A       &  20.1 &  20.84 &              &              \\
HOC$^+$          &  89.487414 & $J  = 1 - 0$                                    &   4.3 &   7.68 &  0.06 (0.04) &  0.2 (0,2)   \\
CH$_3$OH \tablenotemark{d}        &  89.505808 & $8_{-4}-9_{-3}$, E             & 163.6 &  1.56  &  0.27 (0.05) &  2.5 (0.3)   \\
t-HCOOH          &  89.579178 & $4_{0\,4}-3_{0\,3}$                             &  10.8 &   8.08 &  0.06 (0.07) &  0.2 (0.3)   \\
C$_2$H$_5$CN \tablenotemark{c}    &  89.590028 & $10_{4\,7} - 9_{4\,6}$         &  41.4 & 124.52 &  0.09 (0.07) &  0.5 (0.4)   \\
C$_2$H$_5$CN \tablenotemark{c}    &  89.591013 & $10_{4\,6} - 9_{4\,5}$         &  41.4 & 124.51 &              &              \\
HCOOCH$_3$       &  90.145723 & $7_{2\,5} - 6_{2\,4}$, E                        &  19.7 &  17.14 &  0.15 (0.06) &  1.0 (0.3)   \\
HCOOCH$_3$       &  90.156473 & $7_{2\,5} - 6_{2\,4}$, A                        &  19.7 &  17.14 &  0.16 (0.06) &  1.3 (0.4)   \\
t-HCOOH          &  90.164630 & $4_{2\,2}-3_{2\,1}$                             &  23.5 &   6.06 &  0.07 (0.06) &  0.5 (0.3)   \\
HCOOCH$_3$ \tablenotemark{c}      &  90.227659 & $8_{0\,8} - 7_{0\,7}$, E       &  20.1 &  20.89 &  0.21 (0.06) &  2.6 (0.3)   \\
HCOOCH$_3$ \tablenotemark{c}      &  90.229624 & $8_{0\,8} - 7_{0\,7}$, A       &  20.1 &  20.90 &              &              \\
$^{13}$CH$_3$OH  &  90.262310 & $11_{-1}-10_{-2}$, E                            & 155.0 &   3.45 &  0.06 (0.06) &  0.5 (0.3)   \\
HC$^{13}$CCN     &  90.593059 & $J = 10 - 9$                                    &  23.9 & 139.26 &  0.07 (0.06) &  0.7 (0.3)   \\
HCC$^{13}$CN     &  90.601777 & $J = 10 - 9$                                    &  23.9 & 139.25 &  0.08 (0.06) &  0.5 (0.3)   \\
HNC              &  90.663568 & $J=1-0$                                         &   4.4 &   7.97 &  4.79 (0.06) & 50.4 (0.4)   \\
CH$_3$OH \tablenotemark{d}   &  90.812387 & $v_{\rm t} = 1, 20_{-3}-19_{-2}$, E & 800.6 &  13.20 &  0.11 (0.07) &  0.9 (0.3)   \\
$^{13}$C$^{34}$S &  90.926026 &                                                 &   6.5 &   7.67 &  0.07 (0.05) &  0.6 (0.3)   \\
CH$_3$OCH$_3$ \tablenotemark{c}   &  90.937508 & $6_{0\,6} - 5_{1\,5}$, AA      &  19.0 &  32.31 &  0.19 (0.05) &  1.9 (0.3)   \\
CH$_3$OCH$_3$ \tablenotemark{c}   &  90.938107 & $6_{0\,6} - 5_{1\,5}$, EE      &  19.0 &  86.16 &              &              \\
CH$_3$OCH$_3$ \tablenotemark{c}   &  90.938705 & $6_{0\,6} - 5_{1\,5}$, AE      &  19.0 &  10.77 &              &              \\
CH$_3$OCH$_3$ \tablenotemark{c}   &  90.938707 & $6_{0\,6} - 5_{1\,5}$, EA      &  19.0 &  21.54 &              &              \\
HC$_3$N          &  90.979023 & $J=10-9$                                        &  24.0 & 139.25 &  1.90 (0.06) & 18.2 (0.4)   \\
CH$_3$OCH$_3$ \tablenotemark{c}   &  91.476607 & $3_{2\,2} - 3_{1\,3}$, EE      &  11.1 &  38.95 &  0.09 (0.06) &  0.6 (0.3)   \\
CH$_3$OCH$_3$ \tablenotemark{c}   &  91.479263 & $3_{2\,2} - 3_{1\,3}$, AA      &  11.1 &  14.61 &              &              \\
C$_2$H$_5$CN     &  91.549112 & $10_{1\,9} - 9_{1\,8}$                          &  25.3 & 146.67 &  0.11 (0.05) &  0.7 (0.3)   \\
HCOOCH$_3$ \tablenotemark{c}      &  91.775935 & $8_{1\,8} - 7_{0\,7}$, E       &  20.2 &   2.91 &  0.06 (0.06) &  0.4 (0.3)   \\
HCOOCH$_3$ \tablenotemark{c}      &  91.777230 & $8_{1\,8} - 7_{0\,7}$, A       &  20.1 &   2.91 &              &              \\
CH$_3$CN         &  91.958726 & $5_4-4_4$                                       & 127.5 &  55.37 &  0.15 (0.05) &  1.8 (0.3)   \\
CH$_3$CN         &  91.971130 & $5_3-4_3$                                       &  77.5 & 196.88 &  0.50 (0.05) &  5.4 (0.4)   \\
CH$_3$CN         &  91.979994 & $5_2-4_2$                                       &  41.8 & 129.18 &  0.53 (0.05) &  5.1 (0.3)   \\
CH$_3$CN \tablenotemark{c}        &  91.985314 & $5_1-4_1$                      &  20.4 & 147.65 &  0.89 (0.05) & 12.5 (0.4)   \\
CH$_3$CN \tablenotemark{c}        &  91.987088 & $5_0-4_0$                      &  13.2 & 153.80 &              &              \\
H41$\alpha$      &  92.034430 &                                                 &       &        &  0.40 (0.05) & 11.5 (0.5)   \\
$^{13}$CS        &  92.494308 & $J=2-1$                                         &   6.7 &  15.34 &  1.16 (0.05) & 10.4 (0.3)   \\
$^{13}$CH$_3$OH  &  92.588704 & $7_2-8_1$, A$^{-}$                              & 101.2 &   2.51 &  0.07 (0.06) &  0.7 (0.4)   \\
t-HCOOH          &  93.098363 & $4_{1\,3}-3_{1\,2}$                             &  14.4 &   7.58 &  0.08 (0.06) &  0.4 (0.3)   \\
N$_2$H$^+$       &  93.173398 & $J=1-0$                                         &   4.5 & 104.04 &  2.71 (0.05) & 39.0 (0.4)   \\
CH$_3$OH \tablenotemark{d}        &  93.196673 & $v_{\rm t} = 1, 1_0-2_1$, E    & 295.0 &   1.34 &  0.21 (0.05) &  1.7 (0.3)   \\
CH$_3$CHO        &  93.580909 & $5_{1\,5}-4_{1\,4}$, A                          &  15.7 &  60.70 &  0.15 (0.08) &  1.1 (0.4)   \\
CH$_3$CHO        &  93.595235 & $5_{1\,5}-4_{1\,4}$, E                          &  15.8 &  60.70 &  0.18 (0.08) &  1.3 (0.4)   \\
H51$\beta$       &  93.607320 &                                                 &       &        &  0.14 (0.08) &  3.2 (0.6)   \\
$^{13}$CH$_3$OH  &  93.619460 & $2_1-1_1$, A$^{+}$                              &  21.3 &   1.21 &  0.18 (0.08) &  1.4 (0.5)   \\
CH$_3$OCH$_3$ \tablenotemark{c}   &  93.664597 & $12_{1\,11} - 12_{0\,12}$, AE  &  74.0 &  18.78 &  0.20 (0.08) &  3.0 (0.5)   \\
CH$_3$OCH$_3$ \tablenotemark{c}   &  93.664597 & $12_{1\,11} - 12_{0\,12}$, EA  &  74.0 &  37.55 &              &              \\
CH$_3$OCH$_3$ \tablenotemark{c}   &  93.666463 & $12_{1\,11} - 12_{0\,12}$, EE  &  74.0 & 150.21 &              &              \\
CH$_3$OCH$_3$ \tablenotemark{c}   &  93.668329 & $12_{1\,11} - 12_{0\,12}$, AA  &  74.0 &  56.34 &              &              \\
CH$_3$OCH$_3$ \tablenotemark{c}   &  93.854438 & $4_{2\,3} - 4_{1\,4}$, EA      &  14.7 &  13.26 &  0.14 (0.08) &  2.5 (0.6)   \\
CH$_3$OCH$_3$ \tablenotemark{c}   &  93.854560 & $4_{2\,3} - 4_{1\,4}$, AE      &  14.7 &  19.90 &              &              \\
CH$_3$OCH$_3$ \tablenotemark{c}   &  93.857113 & $4_{2\,3} - 4_{1\,4}$, EE      &  14.7 &  53.05 &              &              \\
CH$_3$OCH$_3$ \tablenotemark{c}   &  93.859727 & $4_{2\,3} - 4_{1\,4}$, AA      &  14.7 &  33.16 &              &              \\
PN               &  93.979770 & $J = 2 - 1$                                     &   6.8 &  15.09 &  0.11 (0.08) &  0.4 (0.4)   \\
$^{13}$CH$_3$OH \tablenotemark{c} &  94.405163 & $2_{-1}-1_{-1}$, E             &  12.4 &   1.21 &  0.29 (0.08) &  3.3 (0.5)   \\
$^{13}$CH$_3$OH \tablenotemark{c} &  94.407129 & $2_0-1_0$, A$^{+}$             &   6.8 &   1.62 &              &              \\
$^{13}$CH$_3$OH  &  94.411016 & $2_0-1_0$, E                                    &  19.9 &   1.62 &  0.20 (0.08) &  1.2 (0.4)   \\
$^{13}$CH$_3$OH  &  94.420449 & $2_1-1_1$, E                                    &  27.9 &   1.25 &  0.12 (0.08) &  0.9 (0.5)   \\
CH$_3$OH \tablenotemark{d}        &  94.541765 & $8_3-9_2$,E                    & 123.4 &   2.24 &  0.46 (0.08) &  4.6 (0.5)   \\
CH$_3$OH \tablenotemark{c}        &  94.814987 & $19_7 - 20_6$, A$^{+}$         & 685.2 &   4.55 &  0.12 (0.08) &  0.8 (0.3)   \\
CH$_3$OH \tablenotemark{c}        &  94.814987 & $19_7 - 20_6$, A$^{-}$         & 685.2 &   4.55 &              &              \\
OC$^{34}$S       &  94.922799 & $J = 8 - 7$                                     &  20.5 &   4.09 &  0.08 (0.07) &  0.6 (0.4)   \\
CH$_3$OH         &  95.169463 & $8_0-7_1$, A$^{+}$                              &  83.5 &   7.22 &  2.8 (0.1)   & 21.6 (0.8)   \\
$^{13}$CH$_3$OH  &  95.208660 & $2_1-1_1$, A$^{-}$                              &  21.4 &   1.21 &  0.2 (0.1)   &  1.0 (0.6)   \\
$^{13}$CH$_3$OH  &  95.273440 & $6_{-2}-7_{-1}$, E                              &  73.6 &   2.01 &  0.2 (0.1)   &  1.1 (0.7)   \\
C$_2$H$_5$CN     &  95.442479 & $11_{1\,11} - 10_{1\,10}$                       &  28.6 & 161.63 &  0.1 (0.1)   &  0.8 (0.7)   \\
CH$_3$OCH$_3$ \tablenotemark{c}   &  95.729780 & $16_{2\,14} - 16_{1\,15}$, EA  & 131.8 &  79.74 &  0.3 (0.1)   &  4.0 (0.8)   \\
CH$_3$OCH$_3$ \tablenotemark{c}   &  95.729781 & $16_{2\,14} - 16_{1\,15}$, AE  & 131.8 &  39.87 &              &              \\
CH$_3$OCH$_3$ \tablenotemark{c}   &  95.731253 & $16_{2\,14} - 16_{1\,15}$, EE  & 131.8 & 318.96 &              &              \\
CH$_3$OCH$_3$ \tablenotemark{c}   &  95.732726 & $16_{2\,14} - 16_{1\,15}$, AA  & 131.8 & 119.60 &              &              \\
CH$_3$OH         &  95.914309 & $2_1-1_1$, A$^{+}$                              &  21.4 &   1.21 &  1.7 (0.1)   & 14.0 (0.6)   \\
CH$_3$CHO        &  95.947437 & $5_{0\,5}-4_{0\,4}$, E                          &  13.9 &  63.23 &  0.3 (0.1)   &  2.2 (0.6)   \\
CH$_3$CHO        &  95.963459 & $5_{0\,5}-4_{0\,4}$, A                          &  13.8 &  63.19 &  0.3 (0.1)   &  2.6 (0.6)   \\
HCOOCH$_3$       &  96.070725 & $8_{2\,7} - 7_{2\,6}$, E                        &  23.6 &  19.84 &  0.30 (0.09) &  2.2 (0.5)   \\
HCOOCH$_3$       &  96.076845 & $8_{2\,7} - 7_{2\,6}$, A                        &  23.6 &  19.85 &  0.30 (0.09) &  2.5 (0.5)   \\
CH$_3$CHO        &  96.274252 & $5_{2\,4}-4_{2\,3}$, A                          &  22.9 &  53.13 &  0.2 (0.1)   &  2.1 (0.7)   \\
CH$_3$CHO        &  96.384409 & $5_{3\,3}-4_{3\,2}$, E                          &  34.2 &  40.48 &  0.15 (0.09) &  1.6 (0.5)   \\
CH$_3$OH         &  96.396040 & $v_{\rm t}=1, 2_1-1_1$, A$^{+}$                 & 332.2 &   1.21 &  0.17 (0.09) &  1.4 (0.5)   \\
C$^{34}$S        &  96.412950 & $J=2-1$                                         &   6.9 &   7.67 &  2.74 (0.09) & 24.6 (0.6)   \\
CH$_3$CHO        &  96.425614 & $5_{2\,4}-4_{2\,3}$, E                          &  22.9 &  52.83 &  0.16 (0.09) &  0.8 (0.4)   \\
CH$_3$CHO        &  96.475524 & $5_{2\,3}-4_{2\,2}$, E                          &  23.0 &  52.84 &  0.16 (0.09) &  1.4 (0.5)   \\
CH$_3$OH \tablenotemark{cd}        &  96.492152 & $v_{\rm t}=1, 2_1-1_1$, E     & 290.6 &   1.21 &  0.31 (0.09) &  3.1 (0.6)   \\
CH$_3$OH \tablenotemark{cd}        &  96.493540 & $v_{\rm t}=1, 2_0-1_0$, E     & 299.7 &   1.62 &              &              \\
CH$_3$OH \tablenotemark{d}     &  96.501705 & $v_{\rm t} = 1, 2_{-1}-1_{-1}$, E & 412.5 &   1.21 &  0.16 (0.07) &  1.2 (0.3)   \\
HCOOCH$_3$       &  96.507882 & $7_{4\,4} - 7_{3\,5}$, E                        &  27.2 &   1.00 &  0.09 (0.09) &  0.8 (0.4)   \\
CH$_3$OH         &  96.513675 & $v_{\rm t} = 1, 2_0-1_0$, A$^{+}$               & 430.6 &   1.62 &  0.19 (0.07) &  1.6 (0.4)   \\
CH$_3$OH         &  96.588582 & $v_{\rm t}=1, 2_1-1_1$, A$^{-}$                 & 332.2 &   1.21 &  0.17 (0.08) &  1.6 (0.4)   \\
CH$_3$CHO        &  96.632663 & $5_{2\,2}-4_{2\,2}$, A                          &  23.0 &  53.13 &  0.08 (0.09) &  0.5 (0.4)   \\
CH$_3$OH \tablenotemark{cd}        &  96.739362 & $2_{-1}-1_{-1}$, E            &   4.6 &   1.21 &  2.6 (0.1)   & 46.1 (0.9)   \\
CH$_3$OH \tablenotemark{c}        &  96.741375 & $2_0-1_0$, A$^{+}$             &   7.0 &   1.62 &              &              \\
CH$_3$OH \tablenotemark{cd}        &  96.744550 & $2_0-1_0$, E                  &  12.1 &   1.62 &              &              \\
CH$_3$OH \tablenotemark{d}        &  96.755511 & $2_{1}-1_{1}$, E               &  20.1 &   1.24 &  0.8 (0.1)   &  7.2 (0.6)   \\
CH$_3$OCH$_3$ \tablenotemark{c}   &  96.847241 & $5_{2\,4} - 5_{1\,5}$, EA      &  19.3 &  16.41 &  0.17 (0.09) &  2.8 (0.7)   \\
CH$_3$OCH$_3$ \tablenotemark{c}   &  96.849292 & $5_{2\,4} - 5_{1\,5}$, AE      &  19.3 &   8.20 &              &              \\
CH$_3$OCH$_3$ \tablenotemark{c}   &  96.849890 & $5_{2\,4} - 5_{1\,5}$, EE      &  19.3 &  65.63 &              &              \\
CH$_3$OCH$_3$ \tablenotemark{c}   &  96.852514 & $5_{2\,4} - 5_{1\,5}$, AA      &  19.3 &  24.61 &              &              \\
C$_2$H$_5$CN     &  96.919762 & $11_{0\,11} - 10_{0\,10}$                       &  28.1 & 162.61 &  0.07 (0.09) &  0.5 (0.4)   \\
C$^{33}$S        &  97.172064 & $J=2-1$                                         &   7.0 &  30.67 &  0.58 (0.09) &  6.0 (0.6)   \\
OCS              &  97.301209 & $J=8-7$                                         &  21.0 &   4.09 &  1.0 (0.1)   &  8.9 (0.6)   \\
CH$_3$OH         &  97.582804 & $2_1-1_1$, A$^{-}$                              &  21.6 &   1.21 &  1.7 (0.1)   & 14.0 (0.7)   \\
CH$_3$OH \tablenotemark{c}        &  97.677738 & $21_6-22_5$, A$^{-}$           & 729.3 &   5.75 &  0.2 (0.1)   &  1.6 (0.5)   \\
CH$_3$OH \tablenotemark{c}        &  97.678852 & $21_6-22_5$, A$^{+}$           & 729.3 &   5.75 &              &              \\
SO$_2$           &  97.702333 & $7_{3\,5}-8_{2\,6}$                             &  47.8 &   2.51 &  0.1 (0.1)   &  0.8 (0.6)   \\
$^{34}$SO        &  97.715317 & $J_N=3_2-2_1$                                   &   9.1 &   6.92 &  0.4 (0.1)   &  3.3 (0.7)   \\
CS               &  97.980953 & $J=2-1$                                         &   7.1 &   7.67 & 12.3 (0.1)   & 135.4 (0.7)  \\
C$_2$H$_5$CN     &  98.177574 & $11_{2\,10} - 10_{2\,9}$                        &  32.8 & 157.60 &  0.1 (0.1)   &  1.0 (0.6)   \\
HCOOCH$_3$ \tablenotemark{c}      &  98.190658 & $8_{7\,1} - 7_{7\,0}$, A       &  53.8 &   4.99 &  0.1 (0.1)   &  0.8 (0.5)   \\
HCOOCH$_3$ \tablenotemark{c}      &  98.190658 & $8_{7\,2} - 7_{7\,1}$, A       &  53.8 &   4.99 &              &              \\
HCOOCH$_3$ \tablenotemark{c}      &  98.191460 & $8_{7\,2} - 7_{7\,1}$, E       &  53.8 &   4.99 &              &              \\
HCOOCH$_3$ \tablenotemark{c}      &  98.278921 & $8_{6\,3} - 7_{6\,2}$, E       &  45.1 &   9.32 &  0.2 (0.1)   &  1.2 (0.5)   \\
HCOOCH$_3$ \tablenotemark{c}      &  98.279762 & $8_{6\,2} - 7_{6\,1}$, A       &  45.1 &   9.32 &              &              \\
HCOOCH$_3$ \tablenotemark{c}      &  98.279762 & $8_{6\,3} - 7_{6\,2}$, A       &  45.1 &   9.32 &              &              \\
HCOOCH$_3$ \tablenotemark{c}      &  98.431803 & $8_{5\,4} - 7_{5\,3}$, E       &  37.8 &  12.97 &  0.2 (0.1)   &  3.0 (0.9)   \\
HCOOCH$_3$ \tablenotemark{c}      &  98.432760 & $8_{5\,4} - 7_{5\,3}$, A       &  37.8 &  12.98 &              &              \\
HCOOCH$_3$ \tablenotemark{c}      &  98.435802 & $8_{5\,3} - 7_{5\,2}$, A       &  37.8 &  12.97 &              &              \\
C$_2$H$_5$CN \tablenotemark{c}    &  98.523872 & $11_{6\,5} - 10_{6\,4}$        &  68.4 & 114.54 &  0.1 (0.1)   &  1.0 (0.6)   \\
C$_2$H$_5$CN \tablenotemark{c}    &  98.523872 & $11_{6\,6} - 10_{6\,5}$        &  68.4 & 114.54 &              &              \\
C$_2$H$_5$CN \tablenotemark{c}    &  98.533987 & $11_{5\,6} - 10_{5\,5}$        &  56.2 & 129.36 &  0.1 (0.1)   &  0.9 (0.6)   \\
C$_2$H$_5$CN \tablenotemark{c}    &  98.533987 & $11_{5\,7} - 10_{5\,6}$        &  56.2 & 129.36 &              &              \\
C$_2$H$_5$CN \tablenotemark{c}    &  98.564827 & $11_{4\,8} - 10_{4\,7}$        &  46.2 & 141.49 &  0.1 (0.1)   &  0.8 (0.5)   \\
C$_2$H$_5$CN \tablenotemark{c}    &  98.566792 & $11_{4\,7} - 10_{4\,6}$        &  46.2 & 141.49 &              &              \\
HCOOCH$_3$       &  98.606856 & $8_{3\,6} - 7_{3\,5}$, E                        &  27.3 &  18.25 &  0.2 (0.1)   &  0.9 (0.5)   \\
HCOOCH$_3$       &  98.611163 & $8_{3\,6} - 7_{3\,5}$, A                        &  27.2 &  18.27 &  0.2 (0.1)   &  1.2 (0.5)   \\
HCOOCH$_3$       &  98.682615 & $8_{4\,5} - 7_{4\,4}$, A                        &  31.9 &  15.97 &  0.2 (0.1)   &  1.1 (0.5)   \\
HCOOCH$_3$       &  98.712001 & $8_{4\,5} - 7_{4\,4}$, E                        &  31.9 &  15.44 &  0.2 (0.1)   &  1.1 (0.6)   \\
HCOOCH$_3$       &  98.747906 & $8_{4\,4} - 7_{4\,3}$, E                        &  31.9 &  15.44 &  0.2 (0.1)   &  1.3 (0.6)   \\
HCOOCH$_3$       &  98.792289 & $8_{4\,4} - 7_{4\,3}$, A                        &  31.9 &  15.96 &  0.2 (0.1)   &  0.8 (0.5)   \\
CH$_3$CHO        &  98.863314 & $5_{1\,4}-4_{1\,3}$, E                          &  16.6 &  60.69 &  0.3 (0.2)   &  1.7 (0.7)   \\
CH$_3$CHO        &  98.900944 & $5_{1\,4}-4_{1\,3}$, A                          &  16.5 &  60.70 &  0.3 (0.2)   &  1.4 (0.7)   \\
H40$\alpha$      &  99.022950 &                                                 &       &        &  0.4 (0.2)   & 12 (2)       \\
SO               &  99.299870 & $J_N=3_2-2_1$                                   &   9.2 &   6.91 &  5.6 (0.2)   & 56 (1)       \\
CH$_3$OCH$_3$ \tablenotemark{c}   &  99.324362 & $4_{1\,4} - 3_{0\,3}$, EA      &  10.2 &  17.45 &  0.4 (0.2)   &  5 (1)       \\
CH$_3$OCH$_3$ \tablenotemark{c}   &  99.324364 & $4_{1\,4} - 3_{0\,3}$, AE      &  10.2 &  26.18 &              &              \\
CH$_3$OCH$_3$ \tablenotemark{c}   &  99.325217 & $4_{1\,4} - 3_{0\,3}$, EE      &  10.2 &  69.82 &              &              \\
CH$_3$OCH$_3$ \tablenotemark{c}   &  99.326072 & $4_{1\,4} - 3_{0\,3}$, AA      &  10.2 &  43.64 &              &              \\
HCC$^{13}$CN     &  99.661467 & $J = 11 - 10$                                   &  28.7 & 153.19 &  0.1 (0.1)   &  1.0 (0.5)   \\
CH$_3$OH \tablenotemark{d}        &  99.730920 & $v_{\rm t} = 1, 6_1-5_0$, E    & 332.3 &   3.14 &  0.3 (0.1)   &  2.3 (0.7)   \\
CH$_3$OH \tablenotemark{d}        &  99.772834 & $v_{\rm t} = 1, 20_3-21_4$, E  & 894.6 &  17.63 &  0.1 (0.1)   &  1.0 (0.4)   \\
SO               & 100.029640 & $J_N=4_5-4_4$                                   &  38.6 &   0.84 &  0.25 (0.09) &  1.8 (0.4)   \\
HC$_3$N          & 100.076392 & $J=11-10$                                       &  28.8 & 153.17 &  1.51 (0.09) & 16.8 (0.7)   \\
H$_2$CCO         & 100.094510 & $5_{1\,5}-4_{1\,4}$                             &  27.5 &  29.04 &  0.13 (0.08) &  0.6 (0.3) ) \\
HCOOCH$_3$       & 100.294604 & $8_{3\,5} - 7_{3\,4}$, E                        &  27.4 &  18.26 &  0.14 (0.07) &  0.8 (0.3)   \\
HCOOCH$_3$       & 100.308179 & $8_{3\,5} - 7_{3\,4}$, A                        &  27.4 &  18.28 &  0.15 (0.07) &  1.5 (0.4)   \\
CH$_3$OCH$_3$ \tablenotemark{c}   & 100.460413 & $6_{2\,5} - 6_{1\,6}$, EA      &  24.7 &  19.23 &  0.21 (0.07) &  3.2 (0.5)   \\
CH$_3$OCH$_3$ \tablenotemark{c}   & 100.460437 & $6_{2\,5} - 6_{1\,6}$, AE      &  24.7 &  28.84 &              &              \\
CH$_3$OCH$_3$ \tablenotemark{c}   & 100.463075 & $6_{2\,5} - 6_{1\,6}$, EE      &  24.7 &  76.92 &              &              \\
CH$_3$OCH$_3$ \tablenotemark{c}   & 100.465726 & $6_{2\,5} - 6_{1\,6}$, AA      &  24.7 &  48.07 &              &              \\
HCOOCH$_3$       & 100.482241 & $8_{1\,7} - 7_{1\,6}$, E                        &  22.8 &  20.63 &  0.26 (0.07) &  2.1 (0.4)   \\
HCOOCH$_3$       & 100.490682 & $8_{1\,7} - 7_{1\,6}$, A                        &  22.8 &  20.64 &  0.26 (0.07) &  1.8 (0.4)   \\
C$_2$H$_5$CN     & 100.614281 & $11_{1\,10} - 10_{1\,9}$                        &  30.1 & 161.58 &  0.1 (0.1)   &  0.6 (0.5)   \\
CH$_3$OH \tablenotemark{d}        & 100.638900 & $13_2-12_3$, E                 & 225.7 &   3.84 &  0.5 (0.1)   &  4.4 (0.6)   \\
HCOOCH$_3$ \tablenotemark{c}      & 100.681545 & $9_{0\,9} - 8_{0\,8}$, E       &  24.9 &  23.53 &  0.4 (0.1)   &  5.5 (0.7)   \\
HCOOCH$_3$ \tablenotemark{c}      & 100.683368 & $9_{0\,9} - 8_{0\,8}$, A       &  24.9 &  23.54 &              &              \\
C$_2$H$_5$OH     & 100.990102 & $8_{2\,7} - 8_{1\,8}$                           &  35.2 &   8.83 &  0.17 (0.09) &  1.5 (0.5)   \\
H39$\alpha$ \tablenotemark{c}     & 106.737360 &                                &       &        &  0.6 (0.2)   & 15 (1)     ) \\
$^{34}$SO \tablenotemark{c}       & 106.743244 & $J_N=2_3-1_2$                  &  20.9 &   3.56 &              &              \\
CH$_3$OCH$_3$ \tablenotemark{c}   & 106.775679 & $9_{1\,8} - 8_{2\,7}$, AA      &  43.4 &  36.60 &  0.3 (0.1)   &  3.6 (0.8)   \\
CH$_3$OCH$_3$ \tablenotemark{c}   & 106.777372 & $9_{1\,8} - 8_{2\,7}$, EE      &  43.4 &  58.57 &              &              \\
CH$_3$OCH$_3$ \tablenotemark{c}   & 106.779061 & $9_{1\,8} - 8_{2\,7}$, AE      &  43.4 &  21.96 &              &              \\
CH$_3$OCH$_3$ \tablenotemark{c}   & 106.779069 & $9_{1\,8} - 8_{2\,7}$, EA      &  43.4 &  14.64 &              &              \\
CH$_3$OH         & 107.013803 & $3_1-4_0$, A$^{+}$                              &  28.3 &   3.01 &  2.0 (0.1)   & 17.4 (0.7)   \\
CH$_3$OH \tablenotemark{d}        & 107.159820 & $15_{-2}-15_{1}$, E            & 296.9 &   2.60 &  0.3 (0.1)   &  3.4 (0.7)   \\
C$_2$H$_5$CN \tablenotemark{c}    & 107.485160 & $12_{7\,5} - 11_{7\,4}$        &  88.0 & 117.36 &  0.2 (0.1)   &  2.1 (0.7)   \\
C$_2$H$_5$CN \tablenotemark{c}    & 107.485160 & $12_{7\,6} - 11_{7\,5}$        &  88.0 & 117.36 &              &              \\
C$_2$H$_5$CN \tablenotemark{c}    & 107.486949 & $12_{6\,6} - 11_{6\,5}$        &  73.6 & 133.41 &              &              \\
C$_2$H$_5$CN \tablenotemark{c}    & 107.486949 & $12_{6\,7} - 11_{6\,6}$        &  73.6 & 133.41 &              &              \\
C$_2$H$_5$CN \tablenotemark{c}    & 107.502432 & $12_{5\,7} - 11_{5\,6}$        &  61.3 & 146.99 &  0.2 (0.1)   &  2.0 (0.7)   \\
C$_2$H$_5$CN \tablenotemark{c}    & 107.502432 & $12_{5\,8} - 11_{5\,7}$        &  61.3 & 146.99 &              &              \\
HCOOCH$_3$       & 107.537258 & $9_{2\,8} - 8_{2\,7}$, E                        &  28.8 &  22.61 &  0.4 (0.1)   &  2.4 (0.7)   \\
HCOOCH$_3$       & 107.543711 & $9_{2\,8} - 8_{2\,7}$, A                        &  28.8 &  22.61 &  0.5 (0.1)   &  5.5 (0.8)   \\
C$_2$H$_5$CN     & 107.734723 & $12_{3\,9} - 11_{3\,8}$                         &  43.6 & 166.76 &  0.1 (0.1)   &  1.1 (0.7)   \\
SO$_2$           & 107.843470 & $12_{4\,8}-13_{3\,11}$                          & 111.0 &   4.54 &  0.3 (0.1)   &  2.3 (0.7)   \\
$^{13}$CH$_3$OH  & 107.984970 & $8_3-9_2$, E                                    & 129.5 &   2.24 &  0.2 (0.1)   &  1.4 (0.7)   \\
t-HCOOH          & 108.126720 & $5_{1\,4}-4_{1\,3}$                             &  18.8 &   9.70 &  0.2 (0.1)   &  1.2 (0.6)   \\
CH$_3$OH \tablenotemark{d}        & 108.893963 & $0_0-1_{-1}$, E                &   5.2 &   0.98 &  1.8 (0.1)   & 17.6 (0.9)   \\
C$_2$H$_5$CN     & 108.940590 & $12_{2\,10} - 11_{2\,9}$                        &  38.2 & 172.92 &  0.2 (0.1)   &  1.0 (0.6)   \\
CH$_3$OH \tablenotemark{d}        & 109.138710 & $14_5-15_4$, E                 & 371.8 &   3.41 &  0.4 (0.2)   &  2.9 (0.7)   \\
CH$_3$OH \tablenotemark{d}        & 109.153107 & $16_{-2}-16_{1}$, E            & 334.1 &   3.68 &  0.3 (0.2)   &  2.2 (0.8)   \\
$^{13}$CH$_3$OH  & 109.164120 & $0_0-1_{-1}$, E                                 &  13.1 &   0.98 &  0.2 (0.2)   &  1.5 (0.7)   \\
HC$_3$N          & 109.173634 & $J=12-11$                                       &  34.1 & 167.09 &  2.5 (0.2)   & 23 (1)       \\
SO               & 109.252220 & $J_N=2_3-1_2$                                   &  21.1 &   3.56 &  2.2 (0.2)   & 1 (1)        \\
OCS              & 109.463063 & $J=9-8$                                         &  26.3 &   4.60 &  1.4 (0.1)   & 13.2 (0.9)   \\
HNCO             & 109.495996 & $5_{1\,5}-4_{1\,4}$                             &  59.0 &  11.85 &  0.3 (0.1)   &  1.7 (0.6)   \\
CH$_3$OCH$_3$ \tablenotemark{c}   & 109.571391 & $8_{2\,7} - 8_{1\,8}$, EA      &  38.3 &  23.95 &  0.2 (0.1)   &  2.1 (0.7)   \\
CH$_3$OCH$_3$ \tablenotemark{c}   & 109.571398 & $8_{2\,7} - 8_{1\,8}$, AE      &  38.3 &  35.92 &              &              \\
CH$_3$OCH$_3$ \tablenotemark{c}   & 109.574127 & $8_{2\,7} - 8_{1\,8}$, EE      &  38.3 &  95.79 &              &              \\
CH$_3$OCH$_3$ \tablenotemark{c}   & 109.576860 & $8_{2\,7} - 8_{1\,8}$, AA      &  38.3 &  59.87 &              &              \\
C$_2$H$_5$CN     & 109.650295 & $12_{1\,11} - 11_{1\,10}$                       &  35.4 & 176.48 &  0.2 (0.1)   &  1.1 (0.8)   \\
C$^{18}$O        & 109.782173 & $J=1-0$                                         &   5.3 &   0.01 &  4.3 (0.2)   & 46 (1)       \\
HNCO \tablenotemark{c}            & 109.872337 & $5_{2\,4}-4_{2\,3}$            & 186.1 &  10.01 &  0.2 (0.2)   &  1.3 (0.7)   \\
HNCO \tablenotemark{c}            & 109.872765 & $5_{2\,3}-4_{2\,2}$            & 186.1 &  10.01 &              &              \\
HNCO             & 109.905749 & $5_{0\,5}-4_{0\,4}$                             &  15.8 &  12.48 &  0.8 (0.2)   &  7.8 (0.9)   \\
$^{13}$CO        & 110.201354 & $J=1-0$                                         &   5.3 &   0.01 & 20.3 (0.2)   & 250 (1)      \\
HNCO             & 110.298089 & $5_{1\,4}-4_{1\,3}$                             &  59.2 &  11.85 &  0.2 (0.2)   &  1.5 (0.8)   \\
CH$_3$CN         & 110.349470 & $6_4-5_4$                                       & 132.8 & 102.53 &  0.3 (0.2)   &  3.0 (0.8)   \\
CH$_3$CN         & 110.364354 & $6_3-5_3$                                       &  82.8 & 276.87 &  0.7 (0.2)   &  6.9 (0.9)   \\
CH$_3$CN         & 110.374989 & $6_2-5_2$                                       &  47.1 & 164.05 &  0.8 (0.2)   &  7.2 (0.9)   \\
CH$_3$CN \tablenotemark{c}        & 110.381372 & $6_1-5_1$                      &  25.7 & 179.43 &  1.2 (0.2)   & 17 (1)       \\
CH$_3$CN \tablenotemark{c}        & 110.383500 & $6_0-5_0$                      &  18.5 & 184.56 &              &              \\
HCOOCH$_3$       & 110.525741 & $9_{7\,2} - 8_{7\,1}$, E                        &  59.1 &   9.46 &  0.2 (0.1)   &  1.1 (0.6)   \\
HCOOCH$_3$ \tablenotemark{c}      & 110.535186 & $9_{7\,2} - 8_{7\,1}$, A       &  59.1 &   9.47 &  0.4 (0.1)   &  2.3 (0.7)   \\
HCOOCH$_3$ \tablenotemark{c}      & 110.535186 & $9_{7\,3} - 8_{7\,2}$, A       &  59.1 &   9.47 &              &              \\
HCOOCH$_3$ \tablenotemark{c}      & 110.536003 & $9_{7\,3} - 8_{7\,2}$, E       &  59.1 &   9.47 &              &              \\
HCOOCH$_3$       & 110.652813 & $9_{6\,3} - 8_{6\,2}$, E                        &  50.5 &  13.31 &  0.2 (0.1)   &  1.0 (0.6)   \\
HCOOCH$_3$ \tablenotemark{c}      & 110.662315 & $9_{6\,4} - 8_{6\,3}$, E       &  50.4 &  13.31 &  0.4 (0.1)   &  3.4 (0.7)   \\
HCOOCH$_3$ \tablenotemark{c}      & 110.663273 & $9_{6\,4} - 8_{6\,3}$, A       &  50.4 &  13.31 &              &              \\
HCOOCH$_3$ \tablenotemark{c}      & 110.663429 & $9_{6\,3} - 8_{6\,2}$, A       &  50.4 &  13.31 &              &              \\
HCOOCH$_3$ \tablenotemark{c}      & 110.788664 & $10_{1\,10} - 9_{1\,9}$, E     &  30.3 &  26.17 &  0.4 (0.2)   &  5 (1)       \\
HCOOCH$_3$ \tablenotemark{c}      & 110.790526 & $10_{1\,10} - 9_{1\,9}$, A     &  30.3 &  26.18 &              &              \\
HCOOCH$_3$       & 110.873955 & $9_{5\,4} - 8_{5\,3}$, E                        &  43.2 &  16.56 &  0.2 (0.2)   &  1.3 (0.5)   \\
HCOOCH$_3$ \tablenotemark{c}      & 110.879766 & $9_{3\,7} - 8_{3\,6}$, E       &  32.6 &  21.24 &  0.6 (0.2)   &  5.1 (0.8)   \\
HCOOCH$_3$ \tablenotemark{c}      & 110.880447 & $9_{5\,5} - 8_{5\,4}$, A       &  43.2 &  16.56 &              &              \\
HCOOCH$_3$ \tablenotemark{b}      & 110.882331 & $9_{5\,5} - 8_{5\,4}$, E       &  43.2 &  16.55 &              &              \\
HCOOCH$_3$       & 110.887092 & $9_{3\,7} - 8_{3\,6}$, A                        &  32.6 &  21.26 &  0.4 (0.2)   &  2.5 (0.6)   \\
HCOOCH$_3$       & 110.890256 & $9_{5\,4} - 8_{5\,3}$, A                        &  43.2 &  16.56 &  0.3 (0.2)   &  1.6 (0.5)   \\
HCOOCH$_3$ \tablenotemark{c}      & 111.169903 & $10_{0\,10} - 9_{0\,9}$, E     &  30.2 &  26.19 &  0.5 (0.2)   &  4.8 (0.8)   \\
HCOOCH$_3$ \tablenotemark{c}      & 111.171634 & $10_{0\,10} - 9_{0\,9}$, A     &  30.2 &  26.19 &              &              \\
HCOOCH$_3$       & 111.195962 & $9_{4\,6} - 8_{4\,5}$, A                        &  37.2 &  19.22 &  0.2 (0.2)   &  2.0 (0.7)   \\
HCOOCH$_3$       & 111.223491 & $9_{4\,6} - 8_{4\,5}$, E                        &  37.2 &  18.18 &  0.4 (0.2)   &  2.0 (0.7)   \\
CH$_3$OH         & 111.289550 & $7_2-8_1$, A$^{+}$                              & 102.7 &   2.34 &  0.8 (0.2)   &  9 (1)       \\
HCOOCH$_3$       & 111.408412 & $9_{4\,5} - 8_{4\,4}$, E                        &  37.3 &  18.19 &  0.3 (0.2)   &  1.9 (0.7)   \\
HCOOCH$_3$       & 111.453300 & $9_{4\,5} - 8_{4\,4}$, A                        &  37.2 &  19.22 &  0.3 (0.2)   &  2.1 (0.7)   \\
CH$_3$OH \tablenotemark{d}        & 111.626449 & $17_{-2}-17_{1}$, E            & 373.7 &   5.06 &  0.5 (0.2)   &  4.7 (0.9)   \\
HCOOCH$_3$       & 111.674131 & $9_{1\,8} - 8_{1\,7}$, E                        &  28.1 &  23.19 &  0.3 (0.2)   &  2.7 (0.8)   \\
HCOOCH$_3$       & 111.682189 & $9_{1\,8} - 8_{1\,7}$, A                        &  28.1 &  23.20 &  0.4 (0.2)   &  3.1 (0.8)   \\
t-HCOOH          & 111.746784 & $5_{0\,4}-4_{0\,3}$                             &  16.1 &  10.09 &  0.3 (0.2)   &  2.1 (0.9)   \\
CH$_3$OCH$_3$ \tablenotemark{c}   & 111.782600 & $7_{0\,7} - 6_{1\,6}$, AA      &  25.2 &  68.05 &  0.6 (0.2)   &  4.8 (0.9)   \\
CH$_3$OCH$_3$ \tablenotemark{c}   & 111.783117 & $7_{0\,7} - 6_{1\,6}$, EE      &  25.2 & 108.87 &              &              \\
CH$_3$OCH$_3$ \tablenotemark{c}   & 111.783633 & $7_{0\,7} - 6_{1\,6}$, AE      &  25.3 &  40.83 &              &              \\
CH$_3$OCH$_3$ \tablenotemark{c}   & 111.783634 & $7_{0\,7} - 6_{1\,6}$, EA      &  25.3 &  27.22 &              &              \\
CH$_3$OCH$_3$ \tablenotemark{c}   & 111.812252 & $18_{3\,15} - 18_{2\,16}$, AE  & 169.8 &  48.29 &  0.4 (0.2)   &  4 (1)       \\
CH$_3$OCH$_3$ \tablenotemark{c}   & 111.812253 & $18_{3\,15} - 18_{2\,16}$, EA  & 169.8 &  96.56 &              &              \\
CH$_3$OCH$_3$ \tablenotemark{c}   & 111.813812 & $18_{3\,15} - 18_{2\,16}$, EE  & 169.8 & 386.28 &              &              \\
CH$_3$OCH$_3$ \tablenotemark{c}   & 111.815372 & $18_{3\,15} - 18_{2\,16}$, AA  & 169.8 & 144.84 &              &              \\
$^{13}$CH$_3$OH  & 112.143546 & $3_1-4_0$, A$^{+}$                              &  28.0 &   3.01 &  0.3 (0.2)   &  1.6 (0.7)   \\
CH$_3$CHO        & 112.248716 & $6_{1\,6}-5_{1\,5}$, A                          &  21.1 &  73.77 &  0.3 (0.2)   &  3 (1)       \\
CH$_3$CHO        & 112.254508 & $6_{1\,6}-5_{1\,5}$, E                          &  21.2 &  73.80 &  0.3 (0.2)   &  3 (1)       \\
C$^{17}$O        & 112.359284 & $J=1-0$                                         &   5.4 &   0.01 &  0.9 (0.2)   & 10 (1)       \\
CH$_3$OCH$_3$ \tablenotemark{c}   & 113.057591 & $17_{3\,14} - 17_{2\,15}$, AE  & 153.1 & 132.79 &  0.4 (0.2)   &  4 (1)       \\
CH$_3$OCH$_3$ \tablenotemark{c}   & 113.057593 & $17_{3\,14} - 17_{2\,15}$, EA  & 153.1 &  88.52 &              &              \\
CH$_3$OCH$_3$ \tablenotemark{c}   & 113.059352 & $17_{3\,14} - 17_{2\,15}$, EE  & 153.1 & 354.11 &              &              \\
CH$_3$OCH$_3$ \tablenotemark{c}   & 113.061112 & $17_{3\,14} - 17_{2\,15}$, AA  & 153.1 & 221.32 &              &              \\
CN               & 113.123370 & $N=1-0, J=1/2-1/2, F=1/2-1/2$                   &   5.4 &   0.15 &  0.6 (0.2)   &  5 (1)       \\
CN               & 113.144157 & $N=1-0, J=1/2-1/2, F=1/2-3/2$                   &   5.4 &   1.25 &  1.7 (0.2)   & 16 (1)       \\
CN               & 113.170492 & $N=1-0, J=1/2-1/2, F=3/2-1/2$                   &   5.4 &   1.22 &  2.3 (0.2)   & 20 (1)       \\
CN               & 113.191279 & $N=1-0, J=1/2-1/2, F=3/2-3/2$                   &   5.4 &   1.58 &  2.2 (0.2)   & 21 (1)       \\
CN \tablenotemark{c}              & 113.488120 & $N=1-0, J=3/2-1/2, F=3/2-1/2$  &   5.4 &   1.58 &  5.0 (0.2)   & 58 (1)       \\
CN \tablenotemark{c}              & 113.490970 & $N=1-0, J=3/2-1/2, F=5/2-3/2$  &   5.4 &   4.20 &              &              \\
CN               & 113.499644 & $N=1-0, J=3/2-1/2, F=1/2-1/2$                   &   5.4 &   1.25 &  1.5 (0.2)   & 13 (1)       \\
CN               & 113.508907 & $N=1-0, J=3/2-1/2, F=3/2-3/2$                   &   5.4 &   1.22 &  1.8 (0.2)   & 17 (1)       \\
CN               & 113.520432 & $N=1-0, J=3/2-1/2, F=1/2-3/2$                   &   5.4 &   0.15 &  0.4 (0.2)   &  4 (1)       \\
HCOOCH$_3$       & 113.743107 & $9_{3\,6} - 8_{3\,5}$, E                        &  32.9 &  21.28 &  0.3 (0.2)   &  1 (1)       \\
HCOOCH$_3$       & 113.756610 & $9_{3\,6} - 8_{3\,5}$, A                        &  32.9 &  21.29 &  0.5 (0.2)   &  2 (1)       \\
CH$_3$OCH$_3$ \tablenotemark{c}   & 114.003844 & $18_{2\,16} - 18_{1\,17}$, AE  & 164.4 &  39.31 &  0.2 (0.2)   &  2 (1)       \\
CH$_3$OCH$_3$ \tablenotemark{c}   & 114.003844 & $18_{2\,16} - 18_{1\,17}$, EA  & 164.4 &  78.62 &              &              \\
CH$_3$OCH$_3$ \tablenotemark{c}   & 114.005401 & $18_{2\,16} - 18_{1\,17}$, EE  & 164.4 & 314.48 &              &              \\
CH$_3$OCH$_3$ \tablenotemark{c}   & 114.006957 & $18_{2\,16} - 18_{1\,17}$, AA  & 164.4 & 117.92 &              &              \\
\enddata

\tablenotetext{a}{The position is ($l$, $b$) = ($49.^{\circ}4898$, $-0.^{\circ}3874$), which corresponds to ($\alpha$ (J2000), $\delta$ (J2000)) = (19$^{\rm h}$ 23$^{\rm m}$ 43.9$^{\rm s}$, +14$^{\circ}$ 30$'$ 35.0$''$).}
\tablenotetext{b}{The numbers in parentheses represent $3\sigma$ errors.}
\tablenotetext{c}{The line is blended with other transition lines.}
\tablenotetext{d}{The upper state energy is calculated from an energy state of $1_{-1}$~E.}
\end{deluxetable}

\end{document}